\documentclass[aps,pra,twocolumn,floats,superscriptaddress,tighten,letterpaper,floatfix,eqsecnum]{revtex4-2}
\usepackage{bbm,mathtools}
\usepackage[utf8]{inputenc}
\usepackage{amssymb}
\usepackage{amsbsy}
\usepackage{amsmath}
\usepackage{graphicx}
\usepackage{graphics}
\usepackage{setspace}
\usepackage{array}
\usepackage{color}
\usepackage{xcolor}
\usepackage{fontenc}
\usepackage{textcomp}
\usepackage{rotating}
\usepackage{bm}
\usepackage{braket}
\usepackage[colorlinks=true,linkcolor=black,bookmarksopen=false,urlcolor=blue,citecolor=blue]{hyperref}
\hypersetup{pdfpagemode=UseNone}
\usepackage{chngcntr}
\usepackage{subfigure}

\usepackage{comment}
\usepackage{color}
\usepackage{slashed}
\usepackage[colorlinks=true,linkcolor=black,bookmarksopen=false,urlcolor=blue,citecolor=blue]{hyperref}
\usepackage{amsmath}
\usepackage{enumitem}
\usepackage{revsymb}

\let\normalint\int 
\def\nint{\displaystyle\normalint} 

\let\normalsum\sum 
\def\nsum{\displaystyle\normalsum} 

\newcommand{\calD}{{\cal D}}
\newcommand{\calM}{{\cal M}}
\newcommand{\calI}{{\cal I}}

\newcommand{\calR}{{\cal R}}

\newcommand{\tilX}{\tilde{X}}

\newcommand{\tilzeta}{\tilde{\zeta}}

\newcommand{\hSig}{\hat{\Sigma}}

\newcommand{\hG}{\hat{G}}

\newcommand{\hD}{\hat{D}}

\newcommand{\hq}{\hat{q}}
\newcommand{\hW}{\hat{W}}

\newcommand{\Wdag}{{W}^{\dagger}}
\newcommand{\hWdag}{\hat{W}^{\dagger}}

\newcommand{\hqsp}{\hat{q}_{\mathsf{sp}}}
\newcommand{\hU}{\hat{U}}

\newcommand{\homega}{\hat{\omega}}
\newcommand{\hM}{\hat{M}}
\newcommand{\hphi}{\hat{\phi}}
\newcommand{\hgamma}{\hat{\gamma}}

\newcommand{\hIdSUN}{\hat{1}_{\text{SU} (N) }}
\newcommand{\hPi}{\hat{\Pi}}

\newcommand{\fraka}{\mathfrak{a}}
\newcommand{\frakb}{\mathfrak{b}}

\newcommand{\sfT}{\mathsf{T}}
\newcommand{\sfTbar}{\bar{\mathsf{T}}}
\newcommand{\sfLO}{\mathsf{LO}}
\newcommand{\sfclean}{\mathsf{clean}}
\newcommand{\sfcl}{\mathsf{cl}}
\newcommand{\sfq}{\mathsf{q}}
\newcommand{\sfdis}{\mathsf{dis}}

\newcommand{\sfmax}{\mathsf{max}}
\newcommand{\sfFL}{\mathsf{FL}}
\newcommand{\sfAA}{\mathsf{AA}}
\newcommand{\sfdc}{\mathsf{dc}}
\newcommand{\sfsp}{\mathsf{sp}}
\newcommand{\sfdyn}{\mathsf{dyn}}
\newcommand{\sfscr}{\mathsf{scr}}
\newcommand{\sfint}{\mathsf{int}}
\newcommand{\sfDrude}{\mathsf{Drude}}
\newcommand{\sfintI}{\mathsf{int\,I}}
\newcommand{\sfintII}{\mathsf{int\,II}}

\newcommand{\vareps}{\varepsilon}
\newcommand{\gel}{\gamma_{\mathsf{el}}}

\newcommand{\gbar}{\bar{g}}
\newcommand{\lamphi}{\lambda_{\phi}}
\newcommand{\htau}{\hat{\tau}}

\newcommand{\gammael}{\gamma_{\mathsf{el}}}

\newcommand{\tauel}{\tau_{\mathsf{el}}}
\newcommand{\mb}{m_{\mathsf{b}}}
\newcommand{\varE}{\tilde{\varepsilon}}

\newcommand{\diag}{\mathsf{diag}}
\newcommand{\Tr}{\mathsf{Tr}\, }

\newcommand{\im}{\,\text{Im}\,}
\newcommand{\re}{\,\text{Re}\,}

\newcommand{\e}{\text{e}}

\newcommand{\vex}[1]{\bm{\mathrm{#1}}}
\newcommand{\Nabla}{\bm{\nabla}}
\newcommand{\drho}{\delta\rho_{\mathsf{in}}}

\begin{document}

\title{
Quantum Interference of Hydrodynamic Modes in a Dirty Marginal Fermi Liquid 
}

\author{Tsz Chun Wu}
\affiliation{Department of Physics and Astronomy, Rice University, Houston, Texas 77005, USA}
\author{Yunxiang Liao}
\affiliation{Joint Quantum Institute, University of Maryland, College Park, MD 20742, USA}
\affiliation{
Condensed Matter Theory Center, Department of Physics,
University of Maryland, College Park, MD 20742, USA
}
\author{Matthew S. Foster}
\affiliation{Department of Physics and Astronomy, Rice University, Houston, Texas 77005, USA}
\affiliation{Rice Center for Quantum Materials, Rice University, Houston, Texas 77005, USA}

\date{\today}

\begin{abstract}
We study the electrical transport of a two-dimensional non-Fermi liquid with disorder, and we determine the first quantum
correction to the semiclassical dc conductivity due to quantum interference.
We consider a system with $N$ flavors of fermions coupled 
to SU($N$) critical matrix bosons.  Motivated by the SYK model, we employ the bilocal field formalism and derive a set of 
finite-temperature saddle-point equations governing the fermionic and bosonic self-energies in the large-$N$ limit.  
Interestingly, disorder smearing induces a marginal Fermi liquid (MFL) self-energy for the fermions.  
We next consider fluctuations around the saddle points and derive a MFL-Finkel'stein nonlinear sigma model.
We find that the Altshuler-Aronov quantum conductance correction gives linear-$T$ resistivity 
that can dominate over the Drude result at low temperature. 
The strong temperature dependence of the quantum correction arises due to rapid relaxation of the mediating quantum-critical 
bosons. We verify that our calculations explicitly satisfy the Ward identity at the semiclassical and quantum levels. 
Our results establish that quantum interference persists in two-particle hydrodynamic modes, even when quasiparticles are subject to 
strong (Planckian) dissipation. 
\end{abstract}

\maketitle
\tableofcontents

\section{Introduction}

There has been a recent surge of interest in non-Fermi liquids (NFLs), 
which cannot be described by the Landau quasiparticle paradigm.
NFL properties are observed in a variety of correlated electron systems, including 
the cuprates \cite{MFL_Varma_CuO_PRL_89,NFL_CuO_review_Kivelson_Nat_15}, 
iron-based superconductors \cite{NFL_FeHTSC_Shibauchi_AnnuRevCMP_14,NFL_FeHTSC_Fisher_Science_16}, 
ruthenates \cite{NFL_ruthenates1_Hussey_PRB_98,NFL_ruthenates2_Schneider_PRL_14,NFL_ruthenates3_Klein_PRL_96,NFL_ruthenates4_Allen_PRB_96}, 
twisted bilayer graphene \cite{NFL_TBLG_Cao_PRL_20,NFL_TBLG_AYoung_NatPhy_19}, 
and 
heavy fermion materials \cite{NFL_heavy_fermion_Stewart_RMP_01,NFL_heavy_fermion_Lohneysen_RMP_07,NFL_heavy_fermion_Brando_Science_13} tuned to quantum criticality. 
Some of these systems demonstrate a resistivity that increases linearly with temperature down to the low-temperature regime, 
in sharp contrast with ordinary metals in which the resistivity varies quadratically with temperature \cite{FL_book_Lifshitz_81}. 
This ``strange metal'' phase has been ubiquitously observed in the normal state of high-temperature superconductors 
\cite{MFL_Varma_CuO_PRL_89,NFL_CuO_review_Kivelson_Nat_15,NFL_FeHTSC_Shibauchi_AnnuRevCMP_14,NFL_FeHTSC_Fisher_Science_16}. 

The anomalous properties observed in strange metals might indicate ultrafast ``Planckian'' dissipation, 
due to the extremely rapid collision dynamics between charge carriers \cite{Patel2019,SYK_review_Sachdev_arxiv_21}. 
Arguably the most important question is why superconductivity, a macroscopic, coherent many-body quantum state, can paradoxically arise 
\emph{at anomalously high temperatures}
from a collision-dominated, seemingly incoherent strange metal. In this work, we address a precursor to this question:
does some form of quantum coherence survive in a strange metal that simultaneously exhibits Planckian dissipation? 

We consider the effects of both interactions and quenched disorder; the latter is certainly present in 
most quantum materials.  Strange metals are typically quasi-two-dimensional, where quantum interference
effects are strong and can induce Anderson localization for arbitrarily weak disorder \cite{disO_review_PALee_85,disO_review_Mirlin_RMP_08}. 
From a semiclassical (hydrodynamic or kinetic-theory) perspective, strong carrier-carrier collisions might 
be expected to wipe out interference effects like weak localization 
\cite{MFL_Galitski_PRB_05,Ludwig08}, 
particularly if the \emph{electron transport dephasing time} $\tau_\phi$ \cite{AAG99} due
to inelastic carrier-carrier scattering becomes comparable to the elastic lifetime $\tauel$. 

On the other hand, interactions
can mediate other types of quantum interference: Altshuler-Aronov (AA) corrections to the conductivity \cite{AA4_Altshuler_Review_85,AA3_Aleiner_PRB_01}. 
AA corrections can appear in the particle-hole channel, where a conservation law (the Ward identity) prevents the dephasing rate from 
directly modifying the hydrodynamic response function \cite{Ward3_PALee_PRB_86}. 
The coherent scattering of interacting electrons off the self-consistent potential due to impurity-induced Friedel oscillations 
results in AA corrections in Fermi liquids \cite{AA3_Aleiner_PRB_01}. 
We show in this work that AA corrections can
arise in a theory of a marginal Fermi liquid \cite{MFL_Varma_CuO_PRL_89}, and produce surprising effects. 
In our theory, interactions are mediated by quantum-critical bosons in the finite-temperature 
(quantum relaxational \cite{quantum_relax_Sachdev_PRB_94,quantum_relax_Sachdev_PRB_99,NFL_Subir_book_CUP_11}) regime. 
As a result of the \emph{rapid relaxation of the bosons}, 
the AA correction 
can give linear-in-temperature (linear-$T$) resistivity at low temperatures $T \lesssim 1/\tauel$.
Here $\tauel$ is the elastic lifetime due to impurity scattering that determines the semiclassical Drude conductivity.

Our most important takeaway is this: although individual quasiparticles rapidly decay in a strongly correlated marginal Fermi liquid,
two-particle hydrodynamic modes can retain spatial quantum coherence, leading to virtual interference effects that modulate low-temperature
transport. The key question for future work is to determine what implication this coherence may have for unconventional
superconductivity \cite{Feigelman07,Feigelman10,Burmistrov12,Mayoh15,Fan20}.

\begin{figure}[t]
\centering
\includegraphics[width=0.495\textwidth]{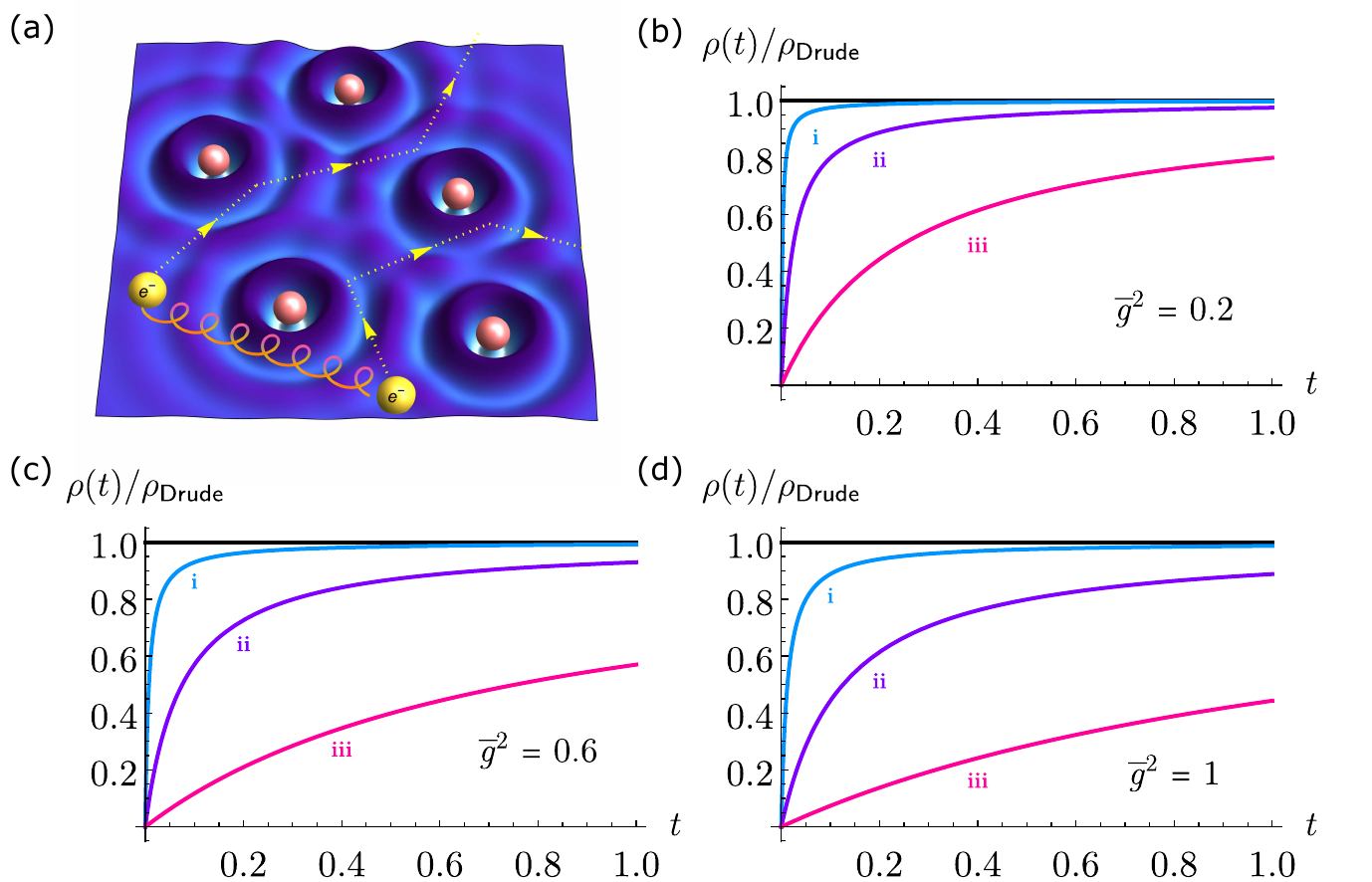}
\caption{
(a) An impurity potential induces inhomogeneity in the single-particle wave functions, resulting in density Friedel oscillations. 
In a Fermi liquid, interacting electrons scatter coherently off of the Friedel oscillations, 
resulting in the Altshuler-Aronov (AA) quantum correction to conductance \cite{AA4_Altshuler_Review_85,AA3_Aleiner_PRB_01}.
In two-dimensional (2D) diffusive Fermi liquids, AA corrections are a source of logarithmic temperature $T$-dependence. 
By contrast, in this paper, we show that AA corrections can be a source for linear-$T$ resistivity 
(``strange metal'' behavior), in a consistent theory of quantum transport for a 2D marginal Fermi liquid. 
(b)--(d) Plots of the dc resistivity $\rho(t)/\rho_{\sfDrude}$ versus the temperature $t = T / \gammael$, from Eq.~(\ref{eq:resist_dc}). 
We ignore the Fermi-liquid-like inelastic scattering contributions $\delta \rho_{\mathsf{in}}$ in all plots.
Panels (b), (c) and (d) depict results for three different values of the effective dimensionless squared-Yukawa coupling $\gbar^2 = \{0.2,0.6,1\}$, respectively. 
The black curve in each panel has $\bar{\rho}\,\mathcal{G}_{AA} = 0$, which neglects the AA corrections. 
The other curves include the AA corrections with $\bar{\rho}\,\mathcal{G}_{AA} \in \{0.001,0.01,0.1\}$ (blue to red, top to bottom).
The antilocalizing AA correction drives $\rho(t)$ towards zero as $t \rightarrow 0 $ and effectively shifts the resistivity downward. 
}
\label{fig:rho_dc_schematic}
\end{figure}


\subsection{Summary of main results \label{Sec:Results}}

Much recent theoretical work incorporates multicomponent randomness (quenched ``exchange'' disorder) into the strange metal phase 
\cite{SYK_Balents_PRL_17,SYK_Chowdhury_Berg_AnnPhy_20,SYK_Patel_PRB_21,SYK_Patel_arxiv_20,SYK_Senthil_PRX_18,SYK_YBKim_arxiv_21,SYK_review_Sachdev_arxiv_21,SYK_Patel_linearT_arxiv_22}.
These works draw upon or are inspired by progress in understanding of the Sachdev-Ye-Kitaev (SYK) model, as reviewed below in Sec.~\ref{Sec:SMTheory}.

In this paper, we uncover an important role played by simple potential disorder in a model of a strange metal. 
We show that 
a two-dimensional (2D) system of fermions coupled to quantum-critical bosons, at finite temperature in the quantum
relaxational regime of the latter \cite{quantum_relax_Sachdev_PRB_94,quantum_relax_Sachdev_PRB_99,NFL_Subir_book_CUP_11},
exhibits 
a singular quantum interference correction that gives linear-$T$ resistivity at low temperature, where it can dominate
over the semiclassical Drude result.  
Specifically, we consider $N$ flavors of fermions with finite density coupled to $N \times N$ matrix bosons representing magnetic collective 
fluctuations \cite{NFL_SU_N_Raghu_PRL_19,NFL_SU_N_disO_Raghu_PRL_20}. 
The mass of the bosons is fine-tuned to zero at a zero-temperature quantum critical point (QCP), 
so that the interaction mediated by them becomes long-ranged in the temperature $T \rightarrow 0$ limit. 
In contrast to the SYK class of models 
\cite{SYK_Patel_arxiv_20,NFL_SYK_largeN_Subir_PRB_21,SYK_Patel_PRB_21,SYK_Patel_arxiv_20,SYK_YBKim_arxiv_21,SYK_Chowdhury_Berg_AnnPhy_20,SYK_Patel_linearT_arxiv_22}, 
the Yukawa coupling between the fermions and bosons here is uniform, not random. 
As a result, for every fixed realization of disorder, our system is manifestly SU($N$) invariant.

The starting point in this paper is the same large-$N$ microscopic model considered in Ref.~\cite{NFL_SU_N_disO_Raghu_PRL_20}.
Different from that work, we construct a nonlinear sigma model based upon a distinct, finite-temperature saddle point. 
Our saddle point obtains from 
disorder-smearing of the clean NFL physics \cite{NFL_SU_N_Raghu_PRL_19} 
and
the generation of a thermal mass for the quantum relaxational bosons \cite{NFL_Subir_book_CUP_11},
leading to a fermion self-energy of exactly the marginal Fermi liquid type \cite{MFL_Varma_CuO_PRL_89}. 
Our primary interest in strange-metal physics is as a platform for nucleating exotic correlated phases at finite temperature;
we therefore do not consider the stability of our theory in the zero-temperature limit. 

We derive a set of saddle point equations governing the self-energies in the large-$N$ limit, 
treating non-perturbatively the effects of both disorder and interactions.
We work in the Keldysh formalism at finite temperature \cite{Keldysh_conv4_Kamenev_AdvPhy_09,NLsM6_Kamenev_CUP_11,NLsM2_Matt_Yun_Ann_17}. 
Importantly, the singularity in the Landau-damped bosonic self-energy of the clean theory 
is smeared out due to disorder \cite{SYK_Patel_PRB_21}, leading to the effective retarded boson propagator 
\begin{align}\label{DR--intro}
	D^{R}_{\sfsp}(\omega,\textbf{k})
	=&\,	
	-
	\frac{1}{2}
	\left[
		\vex{k}^2
		+
		\mb^2
		-
		i \, \alpha \, \omega
	\right]^{-1}.
\end{align}
Here 
\begin{align}\label{thmass--intro}
	\mb^2 \simeq \alpha_m \, T
\end{align}
is the ``thermal mass'' \cite{NFL_Subir_book_CUP_11,thermal_mass_Bellac}
of the critical matrix bosons in the quantum relaxational regime; such a mass generically arises due
to interactions above a quantum critical point. 
In Eq.~(\ref{thmass--intro}), $T$ denotes the temperature, while $\alpha_m$ carries units of inverse diffusion constant
and is independent of temperature in the leading approximation.  
We note that SU($N$) symmetry does \emph{not} 
prevent the appearance of the thermal mass in the boson propagator; instead, the SU($N$) Ward identity 
constrains only correlation functions of the associated Noether currents (these are bosonic bilinears). 
The coefficient $\alpha$ in Eq.~(\ref{DR--intro}) carries units of inverse diffusion constant,
and is independent of temperature. It takes a value $\alpha = 2 \pi^2 \gbar^2 \nu_0 / N$.
Here $\nu_0$ is the bare density of states per fermion species, and $\gbar^2 = g^2 / [(2 \pi)^2 \gammael]$,
with $g$ the bare fermion-boson (Yukawa) coupling and $\gammael$ the elastic impurity scattering rate. 

At the same semiclassical saddle-point level, we obtain the retarded fermion self-energy
\begin{align}
\label{eq:Sig_MFL_intro}
	\Sigma^{R}_{\mathsf{fermion}}(\omega,\textbf{k})
	=&\,
	-
	i 
	\gammael 
	-
	\gbar^2
	\left[
		\omega 
		\ln 
		\left(\frac{\omega_c}{x}\right)
		+
		i
		\frac{\pi}{2}
		x
	\right],
\end{align}
where $x = \max(|\omega|,J \, T)$ and $\omega_c$ is an ultraviolet cutoff. 
The second term in Eq.~(\ref{eq:Sig_MFL_intro}) takes exactly the form 
predicted 30 years ago in the phenomenological strange-metal theory of a ``marginal Fermi liquid'' (MFL)
\cite{MFL_Varma_CuO_PRL_89}. 
In Eq.~(\ref{eq:Sig_MFL_intro}), the constant $J = J(\alpha/\alpha_m)$ [Eq.~(\ref{eq:MFL_J})]. 
We note that $J \rightarrow 2 \ln(\alpha/\alpha_m)$ logarithmically diverges in the (artificial) 
limit of vanishing thermal mass, $\alpha_m \rightarrow 0$.    

We next go beyond the saddle-point level, and derive a modified Finkel'stein nonlinear
sigma model \cite{NLsM5_Finkelshtein_83,disO_review_Kirkpatrick_94,NLsM2_Matt_Yun_Ann_17,NLsM10_Burmistrov_review_19}
that governs quantum interference corrections on top of the MFL saddle-point. 
The sigma model incorporates the MFL self-energy [Eq.~(\ref{eq:Sig_MFL_intro})] as well as an additional
effective interaction channel \cite{SintII1_Altland_PRL_03,SintII2_Altland_IJMP_04} that is induced relative to the Fermi-liquid case. 
We verify the consistency of our theory by explicitly confirming particle conservation (the Ward identity)
for the density-density correlation function, at both 
the semiclassical and quantum levels 
through explicit calculations. 

Our main physics result is the prediction of the temperature-dependent dc electrical resistivity, 
which takes the form
\begin{align}\label{eq:resist_dc}
	\frac{\rho(t)}{\rho_{\sfDrude}}
	=
	\left[	1
		+
		\bar{\rho}
		\,
		\mathcal{G}_{AA}
		\left(
		\frac{4 \pi \gbar^2}{t}
		\right)
	\right]^{-1}
	+
	\frac{\drho(t)}{\rho_{\sfDrude}},
\end{align}
where $t \equiv T / \gammael$ is the dimensionless temperature measured
relative to the elastic scattering rate. 
Here $\rho_{\sfDrude}$ denotes the $t$-independent semiclassical
Drude conductivity.
The second term in the square brackets of Eq.~(\ref{eq:resist_dc})
is the AA correction; $\bar{\rho} = (N e^2 / h) \rho_{\sfDrude}$ 
denotes
the dimensionless
Drude resistance per channel. 
Eq.~(\ref{eq:resist_dc}) holds for $ \gbar^2/N \lesssim t \lesssim 1$.
The key observation is that the AA correction in Eq.~(\ref{eq:resist_dc}) 
can give linear-$T$ resistivity for low $t$.
In this equation, the parameter $\mathcal{G}_{AA}$ is dimensionless and order-one. 
The AA coefficient $\mathcal{G}_{AA}$ becomes particularly simple in the limit of a static boson 
[$\alpha \rightarrow 0$ in Eq.~(\ref{DR--intro})]
and Fermi-liquid self-energy [$\gbar^2 = 0$ in Eq.~(\ref{eq:Sig_MFL_intro})]. In that case, $\mathcal{G}_{AA}$ 
is a logarithmic function of $D \alpha_m$, independent of $N$ and temperature. 
(This is the usual 2D AA logarithm \cite{AA4_Altshuler_Review_85,AAG99}, 
automatically cut off here because we retain the irrelevant squared-momentum of the boson). 
In the general case of dynamical bosons and MFL fermions, 
$\mathcal{G}_{AA}$ acquires weak temperature dependence, but remains
order one for $D \alpha \gtrsim 1$ and $D \alpha_m \gtrsim 1$, see Sec.~\ref{Sec:AAEval} 
and Figs.~\ref{fig:GAA_z_plot1}--\ref{fig:GAA_T_plot}
for details.  

The residual term $\drho(t)$ in Eq.~(\ref{eq:resist_dc}) 
vanishes at zero temperature,
and describes real inelastic scattering between fermions and the quantum-relaxational
bosons. In the absence of strong fermion-boson drag, which can suppress the inelastic scattering 
contribution to the conductivity \cite{NEQM_Levchenko_AnnOfPhys_20},
$\drho(t)$ should be nonzero for any $t > 0$. In this paper, we do not 
compute $\drho(t)$, but we note that $t^2$ dependence was predicted for a similar model 
\cite{SYK_Patel_linearT_arxiv_22}. Consistent with that work
and required here by the Ward identity, we find a perfect
cancelation between the MFL self-energy and boson vertex corrections that 
eliminates the naively expected linear-$T$ resistivity behavior of the \emph{semiclassical} limit.
The key conclusion is that the MFL self-energy alone does not determine the transport lifetime.

The linear-$T$ behavior implied by Eq.~(\ref{eq:resist_dc}) becomes apparent 
when the AA correction becomes comparable to one,
as shown in Figs.~\ref{fig:rho_dc_schematic}(b)--(d). How reliable is this prediction? Usually,
when a perturbative quantum correction overwhelms a tree-level result (here the Drude conductivity), 
one has to either include higher-order corrections and/or resum the perturbation series \`a la
renormalization-group (RG) improved perturbation theory \cite{DoubleLogRG_Peskin}. Moreover, the form of the AA correction in 
Eq.~(\ref{eq:resist_dc}), proportional to $g^2/T$, seemingly indicates a strongly relevant interaction,
mediated by the bosons.

We emphasize two important points to keep in mind that distinguish the situation here. 
(1) The AA correction in Eq.~(\ref{eq:resist_dc}) is \emph{anti-localizing}, similar to weak anti-localization predicted for
a spin-orbit coupled, noninteracting 2D electron gas \cite{disO_review_Mirlin_RMP_08}. In that case, the conductance flows to ever larger values
due to constructive quantum interference. This corrresponds to \emph{weaker} coupling in the 
field theory: the disorder-induced distribution functions for physical observables such as the local density of states or conductance
fluctuations become \emph{narrower} as the conductance is enhanced. Moreover, because the average conductance
is a coupling constant in the sigma model, RG-improved and ordinary perturbation theory are identical. 
(2) As usual in the interacting sigma model  \cite{NLsM5_Finkelshtein_83,disO_review_Kirkpatrick_94,NLsM2_Matt_Yun_Ann_17,NLsM10_Burmistrov_review_19},
interaction corrections are summed to all orders, with the inverse conductance 
(which becomes smaller and smaller here)
serving as the control parameter. 
The form in Eq.~(\ref{eq:resist_dc}) neglects the effects of dynamical screening, which play an important role at low temperatures $t \lesssim \gbar^2/N$. 
In the $t \rightarrow 0$ limit, we find that $1/t$ is replaced by $\log^{2}(t)$, analogous to that of the zero-bias anomaly in disordered metals 
\cite{AA1_Altshuler_JETP_79,AA4_Altshuler_Review_85}. 
The linear-$T$ due to the AA correction is predicted to hold for $\gbar^2/N \lesssim t \lesssim 1$. 
The relevance of the interaction coupling $g^2$ is key to understanding the zero-temperature
quantum phase diagram \cite{disO_review_Kirkpatrick_94,AA5_N_flav_Finkel_Sci_05}, but we defer this question to future work.

The strong temperature dependence of the AA correction in Eq.~(\ref{eq:resist_dc}) results
directly from the thermal mass of the mediating virtual quantum-relaxational boson
[Eqs.~(\ref{DR--intro}) and (\ref{thmass--intro})].
By contrast,
in the usual Fermi liquid case 
\cite{NLsM5_Finkelshtein_83,disO_review_Kirkpatrick_94,AA5_N_flav_Finkel_Sci_05},
one typically assumes short-ranged spin-exchange interactions; these give rise to 
a spin-triplet diffuson hydrodynamic mode that is prevented by the SU(2) Ward identity 
from acquiring a thermal mass. 

We emphasize that the AA correction arises here due to the fact that SU($N$) symmetry is \emph{preserved in every
realization of disorder}, and therefore flavor polarization forms a slow hydrodynamic
mode that survives on time and length scales larger than those set by the impurity scattering. 
Technically, this manifests through SU($N$) flavor degrees of freedom that appear in the matrix field of the sigma model. 
If instead we were to include random flavor-dependent Yukawa couplings 
(as in SYK-type models 
\cite{SYK_Balents_PRL_17,SYK_Chowdhury_Berg_AnnPhy_20,SYK_Patel_PRB_21,SYK_Patel_arxiv_20,SYK_Senthil_PRX_18,SYK_YBKim_arxiv_21,SYK_review_Sachdev_arxiv_21,SYK_Patel_linearT_arxiv_22}),
the SU($N$) mode and associated AA correction 
would be suppressed by the interflavor-impurity scattering rate 
\cite{disO_review_Kirkpatrick_94,DellAnna1,DellAnna2,NLsM2_Matt_Yun_Ann_17,NFL_SU_N_disO_Raghu_PRL_20}.

Our results suggest that the combined effects of forces mediated by quantum-relaxational bosons
and spatial quantum coherence in the particle-hole channel can induce strange metal behavior, 
offering a possible microscopic origin for MFL phenomenology.


\subsection{A brief survey of strange-metal theory \label{Sec:SMTheory}}

There have been a number of theoretical frameworks proposed to understand strange-metal behavior.  
In the cuprates, an early attempt was to employ MFL phenomenology in which quasiparticles are only marginally defined \cite{MFL_Varma_CuO_PRL_89}. 
However, the underlying microscopic origin remains unclear, despite recent developments (see e.g.\ Refs.~\cite{SYK_Patel_PRB_21,SYK_Senthil_PRX_18,SYK_Patel_linearT_arxiv_22}). 
On the other hand, it has been suspected that the strange-metal physics is associated with an underlying QCP \cite{NFL_HTC_review_PALee_RMP_06,NFL_QMC_review_Berg_AnnRevCMP_19}. 
Models with $N$ flavors of fermions coupled to quantum-critical bosons, which represent critical order-parameter fields, have been extensively 
studied in the large-$N$ limit and are known to result in a critical Fermi surface without well-defined quasiparticles (see e.g.\ 
Refs.~\cite{NFL_SSLee_Review_18,NFL_Subir_book_CUP_11,NFL_Chubukov_PRB_06,NFL_Hooley_PRL_15,NFL_Metlitski_Sachdev1_PRB_10,NFL_Metlitski_Sachdev2_PRB_10,NFL_Raghu_BCS_PRB_15,NFL_Raghu_NFL_fp_PRB_13,NFL_Raghu_QC_metal_PRB_13,NFL_SSLee_PRB_15,NFL_SSLee_PRB_09,NFL_SSLee_PRX_17,NFL_SSLee_Sur_PRB_14,NFL_SSLee_Sur_PRB_16,NFL_Senthil_PRB_10,NFL_Mandal_BCS_PRB_16} for recent developments).
Lee \cite{NFL_SSLee_PRB_09} showed that these models cannot however be controlled by large-$N$ and remain strongly coupled in 2D, 
due to an extra enhancement factor of $N$ in higher-loop diagrams. 
An alternative approach is to consider the matrix large-$N$ limit, in which $N$ fermion flavors are coupled to $N \times N$ matrix bosons, as recently proposed by 
Damia \textit{et al.} \cite{NFL_SU_N_Raghu_PRL_19} and employed in this work. It was shown that this model yields a tractable large-$N$ limit and gives the same non-analytical form of self-energies as in the vector-boson model. 
Nevertheless, a fully quantum treatment of the transport properties in these models remains elusive given the complication in satisfying the Ward identity \cite{Ward1_Chubukov_PRB_05,Ward4_Varma_PRB_09}. 
Linear-$T$ resistivity has been observed in quantum Monte Carlo simulations of an fermions undergoing an Ising-nematic transition \cite{NFL_NQCP_QMC_Kivelson_PNAS_17}.

Another playground for exploring systems without quasiparticles is offered by the Sachdev-Ye-Kitaev
(SYK) class of models \cite{SYK_Balents_PRL_17,SYK_Chowdhury_Berg_AnnPhy_20,SYK_HYao_PRB_18,SYK_PRX_15,SYK_Patel_PRB_21,SYK_Patel_linearT_arxiv_22,SYK_Patel_arxiv_20,SYK_Sachdev_PRB_17,SYK_Sachdev_Ye_PRL_93,SYK_Senthil_PRX_18,SYK_Stanford_JHEP_17,SYK_YBKim_arxiv_21,SYK_review_Sachdev_arxiv_21,SYK_e_ph_SC_Schmalian_PRB_19,SYK_quantum_dot_YWang_PRL_20}. 
The SYK model is an exactly solvable (0+1) dimensional model involving a large number of fermions coupled via randomized all-to-all interactions \cite{SYK_Sachdev_Ye_PRL_93,SYK_PRX_15}.
Upon averaging over the randomness, the system exhibits a number of appealing features, such as emergent conformal invariance and NFL behavior.
In particular, transport properties have been investigated in higher-dimensional generalizations of the SYK model 
\cite{SYK_Balents_PRL_17,SYK_Stanford_JHEP_17,SYK_Sachdev_PRB_17,Patel2019,SYK_review_Sachdev_arxiv_21}.
For example, linear-$T$ resistivity and anomalous Lorenz number are obtained in lattice models of SYK quantum dots with random intersite hopping \cite{SYK_Balents_PRL_17}. 
Translationally invariant versions of the lattice model have also been considered and found to display local quantum critically and MFL behaviors in a certain 
temperature range \cite{SYK_Senthil_PRX_18}.   

The idea of systematically controlling models of strongly correlated systems with random interaction couplings has ignited new theories 
for realizing linear-$T$ resistivity by revisiting more conventional physical systems 
\cite{SYK_Patel_arxiv_20,NFL_SYK_largeN_Subir_PRB_21,SYK_Patel_PRB_21,SYK_Patel_arxiv_20,SYK_YBKim_arxiv_21,SYK_Chowdhury_Berg_AnnPhy_20,SYK_Patel_linearT_arxiv_22}. 
In Refs.~\cite{SYK_Patel_PRB_21,SYK_Patel_linearT_arxiv_22}, a model with a critical Fermi surface with random Yukawa-like couplings was constructed. 
Specifically, when the couplings are random in both flavor and position space, the system demonstrates Planckian transport.  
On the other hand, in Ref.~\cite{SYK_Patel_arxiv_20}, an effective theory of heavy fermions with random couplings in the flavor space near 
the critical point was studied and found to realize robust MFL behavior at the strong-coupling metallic fixed point. 
Pairing instabilities in these kinds of models have also been explored recently \cite{SYK_Patel_PRB_21,SYK_Patel_arxiv_20,SYK_YBKim_arxiv_21,SYK_Chowdhury_Berg_AnnPhy_20}. 
Since these models often assume Gaussian distributed random flavor couplings with zero mean, consensus on the applicability of them to real materials is yet to be reached.
Nevertheless, they successfully capture some signatures of the strange metal phase. 

Recently a model-independent theory for the origin of strange metals 
was proposed \cite{NFL_ErsatzFL_Senthil_PRL_21}.  
It was shown that linear-$T$ resistivity down to zero temperature in a clean and translationally invariant metal requires a divergent 
susceptibility for an observable that is odd under inversion/time reversal symmetries, and has zero crystal momentum.


\subsection{Disordered, interacting Fermi liquids}

The long-standing problem of disordered and interacting electrons in Fermi liquids has been well-studied over the past 4 decades 
\cite{disO_review_PALee_85,disO_review_Kirkpatrick_94,NLsM6_Kamenev_CUP_11,NLsM2_Matt_Yun_Ann_17,NLsM10_Burmistrov_review_19}. 
Both electron-electron interactions and quenched disorder can be crucial in describing low-temperature metal-insulator transitions (MITs). 
The effect of interactions is nontrivial, contributing both indirectly and directly to quantum corrections to the dc electrical conductivity. 
Interactions dephase pure quantum interference corrections such as weak (anti)localization (WL,WAL), and separately induce additional
AA corrections (logarithmic in 2D).
An AA correction can appear with sign opposite to that of WL or WAL, and can therefore alter the localization physics
relative to the single-particle problem
\cite{AA1_Altshuler_JETP_79,AA2_Altshuler_PRL_80,AA3_Aleiner_PRB_01,AA4_Altshuler_Review_85,disO_review_PALee_85}. 
Disorder can also enhance interaction matrix elements, through the combination of wave-function multifractality and
Chalker scaling \cite{Feigelman07,Feigelman10,Burmistrov12,Foster12,Foster14,Mayoh15}, and this can induce 
or augment interaction-driven instabilities relative to the clean
Fermi liquid \cite{disO_review_Kirkpatrick_94,DellAnna1,DellAnna2,NLsM10_Burmistrov_review_19}.

In experiments, a MIT was observed in semiconductor inversion layers \cite{disO_2DInvLayer_Kravchenko_RPP_04},
challenging the conventional view from the noninteracting picture that all states localize 2D. 
Theoretically analyzing systems with both interactions and disorder in a controlled manner is typically challenging. 
The standard tool for summing quantum corrections to transport that can incorporate all of the aforementioned effects of interactions 
is the Finkel'stein nonlinear sigma model \cite{NLsM5_Finkelshtein_83,disO_review_Kirkpatrick_94,DellAnna1,DellAnna2,NLsM2_Matt_Yun_Ann_17,NLsM10_Burmistrov_review_19}. 
This enabled a scaling theory for the MIT with interactions, using the renormalization group. In work designed to address the 2D MIT, 
the sigma model was combined with a large-$N$ analysis in order to cope with a potential magnetic instability \cite{AA5_N_flav_Finkel_Sci_05,Finkel_Rev_2010}.

The problem studied in this paper of fermions with a Fermi surface, interacting through critical collective modes in 2D and subject to quenched disorder, has 
recently attracted significant attention \cite{NFL_SU_N_disO_Raghu_PRL_20,NFL_disO_Halbinger_PRB_21,NFL_disO_Joseph_PRB_18,SYK_Patel_linearT_arxiv_22,SYK_Patel_PRB_21}. 
Unlike the Fermi liquid case, in which the interactions are dynamically screened and thus effectively short-ranged, interactions mediated by critical bosons are manifestly
long-ranged in proximity to the QCP. 
In particular, the stability of the zero-temperature conducting phase of the model studied in this paper was addressed in Ref.~\cite{NFL_SU_N_disO_Raghu_PRL_20},
where a modified sigma model incorporating true non-Fermi liquid exponents was considered. Our focus here is instead on finite-temperature physics
that gives rise to the MFL phenomenology discussed above in Sec.~\ref{Sec:Results}.


\subsection{Technical approach of this work}

We employ the $G$-$\Sigma$-$D$-$\Pi$ formalism \cite{SYK_PRX_15,SYK_Senthil_PRX_18} to derive a set of saddle-point equations 
governing the Green's functions and self-energies for large $N$, incorporating both disorder and interactions on the same footing. 
The saddle point solutions yield the quantum-relaxational propagator for the bosons and the MFL self-energy for the fermions
in Eqs.~(\ref{DR--intro}) and (\ref{eq:Sig_MFL_intro}), above. 

We then examine fluctuations around the saddle points to derive the MFL
Finkel'stein nonlinear sigma model (MFL-FNLsM) --- a field theory describing a system of interacting diffusons. 
We restrict our attention in this paper to the particle-hole channel, formally assuming unitary (class A \cite{disO_review_Mirlin_RMP_08}) symmetry, as in 
Refs.~\cite{NLsM5_Finkelshtein_83,NFL_SU_N_disO_Raghu_PRL_20}.  
The derivation is carried out in the Keldysh framework 
\cite{NLsM1_Ludwig_PRB_99,NLsM2_Matt_Yun_Ann_17,NLsM3_Matt_PRB_08,NLsM4_Matt_PRB_06,NLsM5_Finkelshtein_83,NLsM6_Kamenev_CUP_11,NLsM7_Kane_Stone_AoP_81,NLsM8_Keldysh_SC_PRB_20,NLsM9_Burmistrov_PRB_18}, 
which naturally avoids the subtlety of analytical continuation from imaginary time
and allows extension to explore dephasing 
\cite{NLsM2_Matt_Yun_Ann_17,Dephasing_AAK_PhysC_82,Dephasing_Seth_PRB_20,disO_Liao_Matt_dephasing_PRL_18} 
and 
nonequilibrium phenomena 
\cite{NLsM1_Ludwig_PRB_99,NLsM2_Matt_Yun_Ann_17,NLsM6_Kamenev_CUP_11}. 

Armed with this powerful analytical tool, we evaluate the density (linear) response function and demonstrate how the Ward identity for 
particle conservation can be satisfied at both semiclassical and quantum levels.  
We assume that the bosons are at thermal equilibrium and therefore neglect their kinetics. 
Satisfying the Ward identity in this system is a nontrivial task because of the anomalous diffuson propagator that originates from the MFL self-energy. 
We show that it is necessary to incorporate a new term in the sigma model action, 
describing the interaction ladder \cite{SintII1_Altland_PRL_03,SintII2_Altland_IJMP_04}, 
which is absent in ordinary Fermi liquid case \cite{NLsM1_Ludwig_PRB_99,NLsM2_Matt_Yun_Ann_17,NLsM5_Finkelshtein_83,NLsM6_Kamenev_CUP_11}. 
This term introduces new vertex corrections for the Feynman diagrams contributing to the density response, necessary to ensure 
particle conservation.   

We derive the lowest-order quantum AA correction, taking into account various vertex corrections and the effects of the boson thermal mass 
[Eq.~(\ref{thmass--intro})] and dynamical screening. 
Upon summing over all Feynman diagrams to leading order in the dimensionless Drude resistivity per channel 
$\bar{\rho}$ and $1/N$, we recover particle conservation. The final formula for the AA correction is relatively simple 
and can be reduced to the one for disordered metals in an appropriate limit 
\cite{AA1_Altshuler_JETP_79,AA2_Altshuler_PRL_80,NLsM6_Kamenev_CUP_11,AA3_Aleiner_PRB_01},
see Sec.~\ref{Sec:AAEval}.


\subsection{Outline}

The rest of the paper is organized as follows. 
In Sec.~\ref{sec:model}, we present the ingredients of our model and our convention for the Keldysh path integral. 
We then introduce disorder and perform disorder averaging in Sec.~\ref{sec:disO_avg}, 
and we derive and solve
a set of saddle-point equations for the self-energies in Sec.~\ref{sec:biloca_field}.
In Sec.~\ref{sec:NLsM}, we consider the effects of fluctuations around the saddle point and derive the MFL-FNLsM. 
Using the MFL-FNLsM, we compute the semiclassical result and the AA correction to the density response function in 
Secs.~\ref{sec:semi_classic} and \ref{sec:interaction_AA_corr}, respectively. 
Finally, we discuss and conclude our results in Sec.~\ref{sec:discuss_conclu}. 
Various technical details are relegated to the appendices.

\section{The model}\label{sec:model}

We consider a system of 
fermions $\psi$ at finite density with $N$ flavors coupled to critical SU($N$) matrix bosons $\hat{\phi}$. 
For simplicity, we assume time-reversal symmetry is broken so that the system belongs to unitary class A according 
to the ten-fold classification scheme \cite{tenfold1_Ryu_NewJPhy_13,tenfold2_Ryu_RMP_16,tenfold3_Ludwig_PRB_08}, 
as in the early seminal work by Finkel'stein \cite{NLsM5_Finkelshtein_83}. 
We study this system using the finite-temperature Keldysh formalism 
\cite{
Keldysh_conv5_Kamenev_PRB_99,
NLsM6_Kamenev_CUP_11,
NLsM2_Matt_Yun_Ann_17,
NLsM8_Keldysh_SC_PRB_20,
Keldysh_conv2_Liao_PRB_20,
Keldysh_conv1_Wu_PRB_21,
Keldysh_conv3_Liao_arxiv_21}. 
In the presence of external source field $V$, the generating function for the closed Keldysh contour going from 
$t = -\infty$ to $t = \infty$ and then back to $t = -\infty$ is given by
\begin{widetext}
\begin{equation}
\label{eq:Z1}
\!\!\!
	Z[V]
	=
	\int  
	\calD \bar{\psi}  
	\calD \psi 
	\calD \hphi
	\;
	\exp
	\begin{Bmatrix}
		i \nint\limits_{\omega,\textbf{x},\textbf{x}'}
		\bar{\psi}_{i}^a(\omega,\textbf{x})
		{G}^{-1}_{0,ab}(\omega;\textbf{x},\textbf{x}')
		\psi_{i}^b(\omega,\textbf{x}')
%
	+
		i
		\dfrac{1}{2}
		\nint\limits_{t,\textbf{x}}
		{\phi}_{ij}^a(t,\textbf{x})
		\left[
		(i\partial_t)^2 + c^2 \partial_x^2
		\right]
		{\phi}_{ji}^a(t,\textbf{x})
		\tau^3_{a,a}
\\
	+
		i 
		\dfrac{g}{\sqrt{N}}
		\nint\limits_{t,\textbf{x}}
		\tau^3_{a,a}
		\phi_{jk}^a(t,\textbf{x})
		\bar{\psi}_j^a(t,\textbf{x})
		\psi_k^a(t,\textbf{x})
	-
		i
		\dfrac{\lamphi}{2N^2}
		\nint\limits_{t,\textbf{x}}
		\left[
		\phi^a_{ij}(t,\textbf{x}) 
		\phi^a_{ji}(t,\textbf{x})
		\right]^2
		\tau^3_{a,a}
\\
	-
		i
		\nint\limits_{t,\textbf{x}}
		\frac{1}{\sqrt{N}}
		u(\textbf{x})
		\bar{\psi}_i^a (t,\textbf{x})
		\psi_i^a (t,\textbf{x})
		\tau^3_{a,a}
	-
		i 
		\nint\limits_{t,\textbf{x}}
		V^a(t,\textbf{x})
		\bar{\psi}^a (t,\textbf{x})
		\psi^a (t,\textbf{x})
		\tau^3_{a,a}
\end{Bmatrix}
\!\!,
\!\!
\end{equation}
where 
$c$ is the speed of the bosons, 
$g$ is the ``Yukawa'' coupling constant, 
$\lamphi$ is the coupling constant for quartic boson-boson interactions, 
and 
$u(\textbf{x})$ is the onsite impurity potential. 
Meanwhile, 
$i=1,2,...,N$ is the SU($N$) flavor index, 
$a = 1,2$ (which corresponds to forward or backward time-ordering) is the Keldysh label, and 
$\tau^{s}$ are the Pauli matrices acting in the Keldysh space. 
We also introduced the shorthand $\int_{t,\textbf{x}} = \int dt \int d^2 \textbf{x} $. 
Throughout this paper, summation over repeated indices is assumed unless otherwise specified and 
we work with the unit $\hbar = k_B = c = 1$, where $\hbar$ is the Planck constant and $k_B$ is the Boltzmann constant.   
The fermionic field $\psi_i^a$ transforms in the defining representation of SU($N$), 
while the matrix bosonic field $\hphi \rightarrow \phi_{ij}^a$ transforms in the adjoint representation.
The quartic bosonic interaction is symmetry-allowed and analogous to the 
biquadratic 
spin-spin interaction 
$\sim(\textbf{S}_{\text{site $A$}}\cdot \textbf{S}_{\text{site $B$}})^2$ in lattice systems.
The source field $V$ is incorporated to facilitate the calculation of the density response function. 

Our Keldysh conventions and approach mainly follow 
Refs.~\cite{Keldysh_conv1_Wu_PRB_21,Keldysh_conv2_Liao_PRB_20,NLsM2_Matt_Yun_Ann_17}. 
Detailed reviews on the Keldysh formalism can be found in (e.g.) 
Refs.~\cite{NLsM6_Kamenev_CUP_11,Keldysh_conv4_Kamenev_AdvPhy_09,Keldysh_conv5_Kamenev_PRB_99}.

The clean, non-interacting Green's function for the fermions and bosons are respectively denoted by $\hG_0$ and $\hD_0$. 
In the space-time basis, they are given respectively by
\begin{equation}
	i\hG_0(t,t';\textbf{x},\textbf{x}')
	=
	\begin{bmatrix}
	i \hG_{0,\sfT} & i \hG_{0,<}
	\\
	i \hG_{0,>} & i\hG_{0,\sfTbar}
	\end{bmatrix}
	=
	\begin{bmatrix}
	\left\langle 
	\sfT \psi_i(\textbf{x},t) \bar{\psi}_i(\textbf{x}',t')
	\right\rangle_0
	&
	-
	\left\langle 
	\bar{\psi}_i(\textbf{x}',t') {\psi}_i(\textbf{x},t)
	\right\rangle_0
	\\
	\left\langle 
	{\psi}_i(\textbf{x},t) \bar{\psi}_i(\textbf{x}',t')
	\right\rangle_0
	&
	\left\langle 
	\sfTbar \psi_i(\textbf{x},t) \bar{\psi}_i(\textbf{x}',t')
	\right\rangle_0
	\end{bmatrix},
\end{equation}
where $i$ is not summed over, and
\begin{equation}
	i\hD_0(t,t';\textbf{x},\textbf{x}')
	=
	\begin{bmatrix}
	i \hD_{0,\sfT} & i \hD_{0,<}
	\\
	i \hD_{0,>} & i\hD_{0,\sfTbar}
	\end{bmatrix}
	=
	\begin{bmatrix}
	\left\langle 
	\sfT \phi_{ij}(\textbf{x},t) \phi_{ji}(\textbf{x}',t')
	\right\rangle_0
	&
	\left\langle 
	\phi_{ij}(\textbf{x}',t') 
	\phi_{ji}(\textbf{x},t)
	\right\rangle_0
	\\
	\left\langle 
	\phi_{ij}(\textbf{x},t) 
	\phi_{ji}(\textbf{x}',t')
	\right\rangle_0
	&
	\left\langle 
	\sfTbar \phi_{ij}(\textbf{x},t) \phi_{ji}(\textbf{x}',t')
	\right\rangle_0
	\end{bmatrix},
\end{equation}
where $i,j$ are not summed. The symbol $\sfT$ ($\sfTbar$) denotes the time-ordering (anti-time-ordering) operator. 

We now decouple the quartic bosonic interaction with the Hubbard-Stratonovich (H.S.) fields $X_{1,2}$ and express
\begin{equation}
\label{eq:phi4_b4_Keldysh_rot}
	\e^{
	-
	i
	\frac{\lamphi}{2N^2}
	\int\limits_{t,\textbf{x}}
	\tau^3_{a,a}
	[
	\phi^a_{ij}(t,\textbf{x}) 
	\phi^a_{ji}(t,\textbf{x})
	]^2
	}
\\
	=
	\int \calD X
	\;
	\e^{
	-
	\int\limits_{t,\textbf{x}}
	\left\lbrace 
	\frac{N^2}{2i \lamphi}
	(
	X_1^2
	+
	X_2^2
	)
	+
	i
	\Tr
	[
	(\hphi^1)^2
	]
	X_1
	+
	\Tr
	[
	(\hphi^2)^2
	]
	X_2
	\right\rbrace 
	},
\end{equation}
\end{widetext}
where $\int \calD X = \int \calD X_1 \calD X_2$. 
We further introduce the classical component $V_{\sfcl}$ and quantum component $V_{\sfq}$ of the external scalar potential. They are related to the forward (backward) component $V_1$ ($V_2$) by
\begin{equation}
V_{\sfcl} = {\textstyle{\frac{1}{2}}}(V_1 + V_2)
,
\qquad
V_{\sfq} = {\textstyle{\frac{1}{2}}}(V_1 - V_2)
.
\end{equation}
The coupling between the density and the external potential [the last term in the exponent of Eq.~(\ref{eq:Z1})] can then be written as
\begin{equation}
	-i
	\int\limits_{t,\textbf{x}}
	\left[
		V^{\sfcl}
		\bar{\psi}_i
		\htau^3
		\psi_i
		+
		V^{\sfq}
		\bar{\psi}_i
		\psi_i
	\right].
\end{equation}
Similarly, we introduce
\begin{equation}
\hat{\phi}^{\sfcl} = {\textstyle{\frac{1}{2}}}(\hat{\phi}^1 + \hat{\phi}^2)
,
\qquad
\hat{\phi}^{\sfq} = {\textstyle{\frac{1}{2}}}(\hat{\phi}^1 - \hat{\phi}^2)
,
\end{equation}
so that the free bosonic action [second term in the exponent of Eq. (\ref{eq:Z1})] becomes
\begin{equation}
\begin{aligned}
&
\frac{i}{2}
\int\limits_{t,\textbf{x}}
\phi^{\alpha}_{ij}
[D_{0}^{-1}]^{\alpha \beta}
\phi_{ji}^{\beta}
\\
&\equiv
\frac{i}{2}
\int\limits_{t,\textbf{x}}
\begin{bmatrix}
\phi^{\sfcl}_{ij} & \phi^{\sfq}_{ij} 
\end{bmatrix}
\begin{bmatrix}
0 & (D_0^A)^{-1}
\\
(D_0^R)^{-1} & (D_0^K)^{-1}
\end{bmatrix}
\begin{bmatrix}
\phi^{\sfcl}_{ji} 
\\ 
\phi^{\sfq}_{ji} 
\end{bmatrix}
,
\end{aligned}
\end{equation}
where the retarded (R), advanced (A) and Keldysh (K) components of the bosonic propagator respectively take the following form in Fourier space
\begin{equation}
D^{R/A}_0(\Omega,\textbf{q})
=
\frac{1}{2}
\frac{1}{
(\Omega \pm i\eta)^2 - \textbf{q}^2
}
,
\end{equation}
\begin{equation}
D^K_0(\Omega,\textbf{q})
=
\left[
D^R_0(\Omega,\textbf{q})
-
D^A_0(\Omega,\textbf{q})
\right]
F_B(\Omega)
,
\end{equation}
where $\eta \rightarrow 0^+$ and $F_B(\varepsilon) = \coth (\varepsilon/2T)$ is the generalized Bose distribution function at temperature $T$. 
By defining 
\begin{equation}
\tilX_1
=
X_1 - i X_2
,
\qquad
\tilX_2
=
X_1 + i X_2,
\end{equation}
Eq. (\ref{eq:phi4_b4_Keldysh_rot}) can be rewritten as 
\begin{equation}
\begin{aligned}
&
\e^{
-
i
\frac{\lamphi}{2N^2}
\int\limits_{t,\textbf{x}}
\tau^3_{a,a}
[
\phi^a_{ij}(t,\textbf{x}) 
\phi^a_{ji}(t,\textbf{x})
]^2
}
\\
&=
\int \calD \tilX
\exp
\begin{Bmatrix}
-i
\nint\limits_{t,\textbf{x}}
\begin{bmatrix}
\phi^{\sfcl}_{ij} 
& 
\phi^{\sfq}_{ij}
\end{bmatrix}
\begin{bmatrix}
\tilX_1 & \tilX_2
\\
\tilX_2 & \tilX_1
\end{bmatrix}
\begin{bmatrix}
\phi^{\sfcl}_{ji} 
\\ 
\phi^{\sfq}_{ji}
\end{bmatrix}
\\\\
-
\nint\limits_{t,\textbf{x}}
\frac{N^2}{2i \lamphi}
(
\tilX_1 \tilX_2
)
\end{Bmatrix}
,
\end{aligned}
\end{equation}
where 
$\int \calD \tilX = \int \calD \tilX_1 \calD \tilX_2$.

For the fermions, we switch to the physical representation via the following non-unitary transformation
\begin{equation}
\psi
\rightarrow
\htau^3 \hU_{\sfLO} \psi
,
\quad
\bar{\psi}
\rightarrow
\bar{\psi}
\hU_{\sfLO}^{\dagger}
,
\end{equation}
where
\begin{equation}
\hU_{\sfLO}
=
\frac{1}{\sqrt{2}}
(
1 + i\htau^2
).
\end{equation}
The fermionic Green's function then turns into 
\begin{equation}
\hG_0
\rightarrow
\hU_{\sfLO}^{\dagger} \htau^3 \hG_0 \hU_{\sfLO}
=
\begin{bmatrix}
G^R_{0} & G^K_{0}
\\
0 & G^A_{0}
\end{bmatrix}_{\tau}
,
\end{equation}
where the retarded (R), advanced (A) and Keldysh (K) components are respectively given by
\begin{eqnarray}
	G^{R/A}_{0}(\omega,\textbf{k})
	&=&
	\frac{1}{
	\omega \pm i\eta  - \varE_{\textbf{k}}
	},
\label{GRABareDef}
\\
	G^{K}_{0}(\omega,\textbf{k})
	&=&
	[
		G^R_0(\omega,\textbf{k})
		-
		G^A_0(\omega,\textbf{k})
	]
	F(\omega),
\\
	\varE_{\textbf{k}} 
	&=&
	\frac{\textbf{k}^2}{2m} - \mu.
\end{eqnarray}
Here, 
$m$ is the mass of fermions, 
$\mu$ is the chemical potential, and 
$F(\omega) = \tanh (\omega/2T)$ is the generalized Fermi distribution function at temperature $T$. 
The subscript $\tau$ indicates that the matrix is expressed in the Keldysh space. 
We further perform a thermal rotation
\begin{equation}\label{Ferm_Therm_Rot}
\psi \rightarrow \hM_F \psi
,
\quad
\bar{\psi} \rightarrow \bar{\psi} \hM_F
,
\end{equation}
where 
\begin{equation}
\hM_F(\hat{\omega})
=
\begin{bmatrix}
1 & F(\homega)
\\
0 & -1
\end{bmatrix}_{\tau}
=
\hM_F^{-1}(\hat{\omega}).
\end{equation} 
Under this transformation, the fermionic Green's function acquires a diagonal form
\begin{equation}\label{hG0_Diag}
\hG_0
\rightarrow 
\begin{bmatrix}
G^R_{0} & 0 
\\
0 & G^A_{0}
\end{bmatrix}_{\tau}
,
\end{equation}
which is convenient for subsequent analysis. 

After the above transformations, the partition function and action can be expressed as
\begin{equation}
Z 
=
\int 
\calD \bar{\psi} 
\,
\calD \psi 
\,
\calD \hphi 
\,
\calD \tilX
\;
\e^{-S},
\end{equation}
\begin{equation}
S 
=
S_{\psi} + S_{\phi} + S_{\psi \phi} + S_{\phi X} + S_{\psi V} + S_{\sfdis} + S_X ,
\end{equation}
where
\begin{align}
	S_{\psi}	
	=&\,
	-
	i
	\int\limits_{\omega,\textbf{k}}
	\bar{\psi}_{i}(\omega,\textbf{k})
	\hG_{0}^{-1}(\omega,\textbf{k})
	\psi_{i}(\omega,\textbf{k}),
\\
	S_{\phi}
	=&\,
	-\frac{i}{2}
	\int\limits_{\Omega,\textbf{q}}
	\Tr
	[
	\hat{\phi}(\Omega,\textbf{q}) 
	\hD^{-1}_0(\Omega,\textbf{q}) 
	\hat{\phi}(-\Omega,-\textbf{q})
	],
\\
\!\!\!\!
	S_{\psi \phi}
	=&\,
	-i
	\frac{g}{\sqrt{N}}
	\int\limits_{\omega,\omega',\textbf{x}}
\!\!
	\bar{\psi}_i(\omega,\textbf{x})
	\hat{\Phi}_{ij}(\omega,\omega',\textbf{x})
	\psi_{j}(\omega',\textbf{x}),
\end{align}
\begin{align}
	S_{\phi X}
	=&\,
	i
	\int\limits_{t,\textbf{x}}
	\Tr
	\left\{
	\hphi(t,\textbf{x})
	\left[
	\tilX_1(t,\textbf{x}) + \tilX_2 (t,\textbf{x}) \, \htau^1
	\right]
	\hphi(t,\textbf{x})
	\right\},
\end{align}
\begin{align}
	S_{\psi V}
	=&\,
	i
	\int\limits_{\omega,\omega',\textbf{x}}
	\bar{\psi}_i(\omega,\textbf{x})  
	\hat{\cal V}(\omega,\omega',\textbf{x})
	\psi_{i}(\omega',\textbf{x}),
\end{align}
\begin{align}
	S_{\sfdis}
	=&\,
	i
	\int\limits_{t,\textbf{x}}
	\frac{1}{\sqrt{N}}
	u(\textbf{x})
	\bar{\psi}_i (t,\textbf{x})
	\psi_i(t,\textbf{x}),
\end{align}
\begin{align}
	S_X
	=&\,
	\int\limits_{t,\textbf{x}}
	\frac{N^2}{2i \lamphi}
	\tilX_1 (t,\textbf{x}) 
	\tilX_2 (t,\textbf{x}),
\end{align}
and where 
$\int_{\omega} = \int \frac{d\omega}{2\pi}$, 
$\int_{\textbf{k}} = \int \frac{d^2\textbf{k}}{(2\pi)^2}$. 
Here, we defined the shorthand notations
\begin{eqnarray}
\hat{\cal V}(\omega,\omega',\textbf{x})
&=&
\sum_{s =\sfcl,\sfq}
\hgamma^{s}_{\omega,\omega'}V^{s}_{\omega - \omega'}(\textbf{x})
,
\\
\hat{\Phi}_{ij}(\omega,\omega',\textbf{x})
&=&
\sum_{s =\sfcl,\sfq}
\hgamma^{s}_{\omega,\omega'}\phi^{s}_{ij,\omega - \omega'}(\textbf{x}), 
\end{eqnarray}
with
\begin{equation}
	\hgamma^{\sfcl}_{\omega,\omega'}
	=
	\hM_F(\omega) \hM_F(\omega'),
	\qquad
	\hgamma^{\sfq}_{\omega,\omega'}
	=
	\hM_F(\omega) \htau^1 \hM_F(\omega').
\end{equation}
Now the fermion distribution function appears only in 
the coupling matrices $\hgamma^{s}_{\omega,\omega'}$.
This convention allows a clear separation between purely virtual 
quantum-interference modes on one hand, 
and 
on the other 
hydrodynamic fluctuations of conserved currents \cite{NLsM2_Matt_Yun_Ann_17}.

\begin{widetext}

\section{Disorder averaging}
\label{sec:disO_avg}

Assume the impurity potential $u(\textbf{x})$ is described by a Gaussian distribution function $P[u]$ given by
\begin{equation}
P[u]
=
\e^{
-\pi \nu_0 \tauel \int\limits_{\textbf{x}} u^2(\textbf{x})
},
\end{equation}
where $\tauel$ is the elastic scattering time due to impurities.
We perform disorder averaging by integrating out the potential $u(\textbf{x})$
\begin{equation}
\begin{aligned}
	\int 
	\calD u 
	\;
	P[u]
	\;
	\e^{-S_{\sfdis}}
	=&\,
	\int 
	\calD u 
	\;
	\e^{
	-\pi \nu_0 \tauel 
	\int\limits_{\textbf{x}} u^2(\textbf{x})
	}
	\e^{
	i
	\int\limits_{t,\textbf{x}}
	\frac{1}{\sqrt{N}}
	u(\textbf{x})
	\bar{\psi}_i^a (t,\textbf{x})
	\psi_i^{a}(t,\textbf{x})
	}
	=
	\e^{
	\frac{1}{4N \pi \nu_0 \tauel}
	\int\limits_{t,t',\textbf{x}}
	\bar{\psi}_i^a
	(t,\textbf{x})
	\psi_j^{b}
	(t',\textbf{x})
	\bar{\psi}_j^b
	(t',\textbf{x})
	\psi_i^{a}
	(t,\textbf{x})
	}
\\
	=&\,
	\int
	\calD \hq
	\;
	\e^{
	- 
	\frac{\pi \nu_0 }{4\tauel N}
	\int\limits_{\textbf{x}}
	\Tr
	\left[
	\hat{q}^2(\textbf{x})
	\right]
	-
	\frac{1}{2\tauel N}
	\int\limits_{t,t',\textbf{x}}
	q_{ij;tt'}^{ab}(\textbf{x})
	\bar{\psi}^a_i(t,\textbf{x})
	\psi^b_j(t',\textbf{x})
	}.
\end{aligned}
\end{equation} 
In the last step, we decoupled the disordered induced four-fermion interaction using the H.S. field,  
$\hat{q}(\textbf{x}) \rightarrow q_{ij,tt'}^{ab}(\textbf{x})$, which is a Hermitian matrix containing 
time indices $t, t'$, 
flavor indices $i,j$, 
and 
Keldysh indices $a,b$. 
In this work, we focus on the unitary class A in which time-reversal symmetry (TRS) is broken (e.g. by a weak external magnetic field), 
so that the Cooperon contribution to transport is suppressed \cite{NLsM2_Matt_Yun_Ann_17,NLsM6_Kamenev_CUP_11}. 
Hence, we restrict our attention to the particle-hole channel
in the above decoupling. 
We note, however, that restoring TRS is essential for exploring the interplay of dephasing and weak localization  
\cite{NLsM6_Kamenev_CUP_11,NLsM2_Matt_Yun_Ann_17,disO_review_PALee_85}, as well as pairing instabilities to 
superconductivity. We leave these for a separate study.

\section{Bilocal field formulation}
\label{sec:biloca_field}

In this section, we derive a set of saddle-point equations governing fermionic and bosonic self-energies 
arising due to disorder and interactions in the large-$N$ limit. 
Analogous to the treatment for the SYK model \cite{SYK_PRX_15,SYK_Senthil_PRX_18}, 
we introduce the bilocal auxiliary fields $\hG$ and $\hD$ defined by
\begin{eqnarray}
\label{eq:def_G}
G^{ab}_{ij}(x',x)
&=&
-
i
\psi_i^a (x')
\bar{\psi}_j^b (x) 
,
\\
\label{eq:def_D}
D^{ss'}(x',x)
&=&
-\frac{i}{N^2}
\sum_{ij}
\phi_{ji}^{s}(x')
\phi_{ij}^{s'}(x)
,
\end{eqnarray}
where $x = (t,\textbf{x})$, 
$a,b \in \{R,A\}$ (retarded, advanced) are fermion Keldysh indices, 
and 
$s,s' \in \{\sfcl,\sfq\}$ (classical, quantum) are boson Keldysh indices. 
[Recall that the bare fermion Green's function is diagonal after the transformation in 
Eq.~(\ref{Ferm_Therm_Rot}), possessing only retarded and advanced nonzero components,
Eq.~(\ref{hG0_Diag})].
We impose the constraints in Eqs.~(\ref{eq:def_G}) and (\ref{eq:def_D}) using the identities
\begin{equation}
\begin{aligned}
	1 
	&=
	\int 
	\calD \hG
	\;
	\delta\left[
		G^{ba}_{ji}(x',x) 
		-
		i
		\bar{\psi}_i^a (x) \psi_j^b (x')
	\right]
	=
	\int 
	\calD \hG
	\,
	\calD \hSig
	\;
	\e^{
	\sum\limits_{ij}
	\int\limits_{x,x'}
	\Sigma^{ab}_{ij}(x,x')
	\left[
		G^{ba}_{ji}(x',x)
		-
		i
		\bar{\psi}_i^a(x) \psi_j^b (x')
	\right]
	},
\end{aligned}
\end{equation}
\begin{equation}
\begin{aligned}
	1
	&=
	\int 
	\calD \hD
	\;
	\delta\left[
		D^{s' s}(x',x) 
		+
		\frac{i}{N^2}
		\sum_{ij} \phi_{ij}^{s} (x) \phi_{ji}^{s'} (x')
	\right]
	=
	\int 
	\calD \hD
	\,
	\calD \hat{\Pi}
	\;
	\e^{
	-\frac{1}{2}
	\int\limits_{x,x'}
	\Pi^{s s'}(x,x')
	\left[
		N^2 D^{s' s}(x',x)
		+
		i
		\sum\limits_{ij} \phi_{ij}^{s} (x) \phi_{ji}^{s'} (x')
	\right]
}.
\end{aligned}
\end{equation}
Physically, the Lagrange multipliers $\hSig$ and $\hat{\Pi}$ describe respectively the fermionic and bosonic self-energies 
due to the Yukawa interaction. 

Upon integrating out the fermions and bosons, we obtain the following disorder-averaged large-$N$ partition function for the 
$\hq$ matrix field (bilocal in time, but local in position),
the boson-boson interaction-mediating Hubbard-Stratonovich field $\tilX$,
and the spacetime bilocal fields 
$\hG$, $\hSig$, $\hD$, and $\hat{\Pi}$: 
\begin{equation}
	Z 
	=
	\int 
	\calD \hq \, 
	\calD \tilX \, 
	\calD \hG\, 
	\calD \hSig \, 
	\calD \hD \, 
	\calD \hat{\Pi}
	\;
	\e^{-S},
\end{equation}
where the effective action is
\begin{equation}
\label{eq:S_GDPiSig}
\begin{aligned}
	S
	&=
	\frac{N^2}{2}
	\Tr 
	\ln
	\left(
		\hD_0^{-1} -\hPi - 2\tilX_1  - 2 \tilX_2 \, \htau^1
	\right)
	-
	\Tr
	\ln
	\left(
		\hG_{0}^{-1}  - \hat{\cal  V}
		-
		\hSig
		+
		\frac{i}{2\tauel N}
		\hq\
	\right)
	-
	\Tr[
		\hSig
		\,
		\hG
	]
	+
	\frac{N^2}{2}
	\Tr[
		\hPi
		\,
		\hD
	]
\\
&
+
i\frac{g^2}{2N}
\int\limits_{\omega_1,\omega_2,\omega_3,\omega_4,\textbf{x},\textbf{x}'}
[
\hgamma^s_{\omega_1,\omega_2}
]^{ab}
[
\hgamma^{s'}_{\omega_3,\omega_4}
]^{cd}
G^{bc}_{jj}(\omega_2,\textbf{x};\omega_3,\textbf{x}')
G^{da}_{ii}(\omega_4,\textbf{x}';\omega_1,\textbf{x})
D^{s s'}(\omega_1- \omega_2, \textbf{x} ; \omega_3 - \omega_4,\textbf{x}')
\\
&
+
\frac{\pi \nu_0 }{4\tauel N}
\int\limits_{\textbf{x}}
\Tr
\left[
\hat{q}(\textbf{x})^2
\right]
+
\int\limits_{t,\textbf{x}}
\frac{N^2}{2i \lamphi}
\tilX_1 (t,\textbf{x}) 
\tilX_2 (t,\textbf{x}) 
.
\end{aligned}
\end{equation}
In the second last line, we performed a second-order cumulant expansion to write
\begin{equation}
\begin{aligned}
&
\e^{
i \frac{g}{\sqrt{N}}
\int\limits_{\omega,\omega',\textbf{x}}
\bar{\psi}_i(\omega,\textbf{x})
\hat{\Phi}_{ij}(\textbf{x},\omega,\omega')
\psi_j(\omega',\textbf{x})
}
\\
&\rightarrow
\e^{
-
i\frac{g^2}{2N}
\int\limits_{\omega_1,\omega_2,\omega_3,\omega_4,\textbf{x},\textbf{x}'}
[
\hgamma^s_{\omega_1,\omega_2}
]^{ab}
[
\hgamma^{s'}_{\omega_3,\omega_4}
]^{cd}
(-i)
\psi_j^b(\omega_2,\textbf{x})
\bar{\psi}_l^c(\omega_3,\textbf{x}')
(-i)
\psi_m^d(\omega_4,\textbf{x}')
\bar{\psi}_i^a(\omega_1,\textbf{x})
(-i)
\phi^s_{ij}(\omega_1 - \omega_2,\textbf{x})
\phi^{s'}_{lm}(\omega_3 - \omega_4,\textbf{x}')
}
\\
&=
\e^{
-
i\frac{g^2}{2N}
\int\limits_{\omega_1,\omega_2,\omega_3,\omega_4,\textbf{x},\textbf{x}'}
[
\hgamma^s_{\omega_1,\omega_2}
]^{ab}
[
\hgamma^{s'}_{\omega_3,\omega_4}
]^{cd}
G^{bc}_{jj}(\omega_2,\textbf{x};\omega_3,\textbf{x}')
G^{da}_{ii}(\omega_4,\textbf{x}';\omega_1,\textbf{x})
D^{s s'}(\omega_1- \omega_2, \textbf{x} ; \omega_3 - \omega_4,\textbf{x}')
}
,
\end{aligned}
\end{equation}
where we used the definitions in Eqs. (\ref{eq:def_G}) and (\ref{eq:def_D}) in the last step.

\subsection{Saddle-point equations and solutions}
\label{sec:saddle_pt_eqn}

We now look for spacetime-translationally invariant and flavor-space SU($N$)-invariant saddle-point solutions. 
In particular, we let the saddle points of $\hG$ and $\hSig$ to be respectively $\hG_{\sfsp} \otimes \hIdSUN$ and $\hSig_{\sfsp} \otimes \hIdSUN$, 
where  $\hIdSUN$ is the identity matrix of size $N \times N$ in the flavor space. 
Also, we assume that the saddle points for $\tilX_1$, $\tilX_2$ and $\hq_{\sfsp}$ are spatially uniform and time-independent.  
Setting the external source fields $V^{\sfcl}, V^{\sfq}$ to zero and minimizing with respect to 
$\{\hSig,\hat{\Pi},\hG,\hD,\tilX_1,\tilX_2,\hq\}$,
we obtain the following set of saddle-point equations:
\begin{eqnarray}
\label{eq:saddle_pt_G}
	\hG^{-1}_{\sfsp}
	&=&
	\hG_{0}^{-1} 
	- 
	\hSig_{\sfsp}
	+ 
	i \gammael \, \hqsp,
\\
	\hD^{-1}_{\sfsp}
	&=&
	\hD_0^{-1} - \hat{\Pi}_{\sfsp} - 2\tilX_{1,\sfsp}  - 2 \tilX_{2,\sfsp} \, \htau^1,
\end{eqnarray}
\begin{equation}
\label{eq:Sig_sp}
	\Sigma^{ab}_{\sfsp}(\omega,\textbf{k})
	=
	ig^2
	\int\limits_{\Omega,\textbf{q}}
	\left[
		\hgamma^{s'}_{\omega,\Omega}
		\hG_{\sfsp}(\Omega,\textbf{q})
		\hgamma^{s}_{\Omega,\omega}
	\right]^{ab}
	D^{ss'}_{\sfsp}(\Omega-\omega, \textbf{q} - \textbf{k}),
\end{equation}
\begin{equation}
\label{eq:Pi_sp}
	\Pi^{ss'}_{\sfsp}(\Omega,\textbf{q})
	=
	-
	i \frac{g^2}{N} 
	\int\limits_{\omega,\textbf{k}}
	\Tr 
	\left[
		\hgamma^{s'}_{\omega,\omega-\Omega} 
		\hG_{\sfsp}(\omega-\Omega,\textbf{k}) 
		\hgamma^{s}_{\omega-\Omega,\omega} 
		\hG_{\sfsp}(\omega,\textbf{k} + \textbf{q})
	\right],
\end{equation}
\end{widetext}
\begin{equation}
\label{eq:saddle_X1}
\tilX_{1,\sfsp}
=
2i\lamphi
\int\limits_{\Omega,\textbf{q}}
\Tr
\left[
\hD_{\sfsp}(\Omega,\textbf{q})
\htau^1
\right]
,
\end{equation}
\begin{equation}
\label{eq:saddle_X2}
\tilX_{2,\sfsp}
=
2i\lamphi
\int\limits_{\Omega,\textbf{q}}
\Tr
\left[
\hD_{\sfsp}(\Omega,\textbf{q})
\right],
\end{equation}
and
\begin{equation}
\label{eq:q_sp}
\hqsp
=
\frac{i}{\pi \nu_0 }
\int\limits_{\textbf{k}}
\frac{1}{
\hG_{0}^{-1}(\omega,\textbf{k}) - \hSig_{\sfsp}(\omega,\textbf{k})
+
i
\gel
\,
\hqsp
}
.
\end{equation}
Here, 
$x = (t,\textbf{x})$,  
$a,b \in \left\lbrace R,A\right\rbrace$ 
and
$s,s' \in \left\lbrace \sfcl,\sfq\right\rbrace$ 
are Keldysh labels, 
and 
\begin{align}
\label{gammaelDef}
	\gammael 
	\equiv
	\frac{1}{2 N \tauel}
\end{align} 
is the elastic scattering rate for fermions in each flavor channel. 
Our approach automatically generates the Schwinger-Dyson equations for the self-energies $\hSig_{\sfsp}$ and $\hat{\Pi}_{\sfsp}$. 
Note that interaction-vertex corrections to the bosonic self-energy are absent due to large-$N$ suppression. 
The last equation (\ref{eq:q_sp}) means that disorder is treated under the self-consistent Born approximation at the saddle-point level. 

\begin{figure}
	\centering
	{\includegraphics[width=0.3\textwidth]{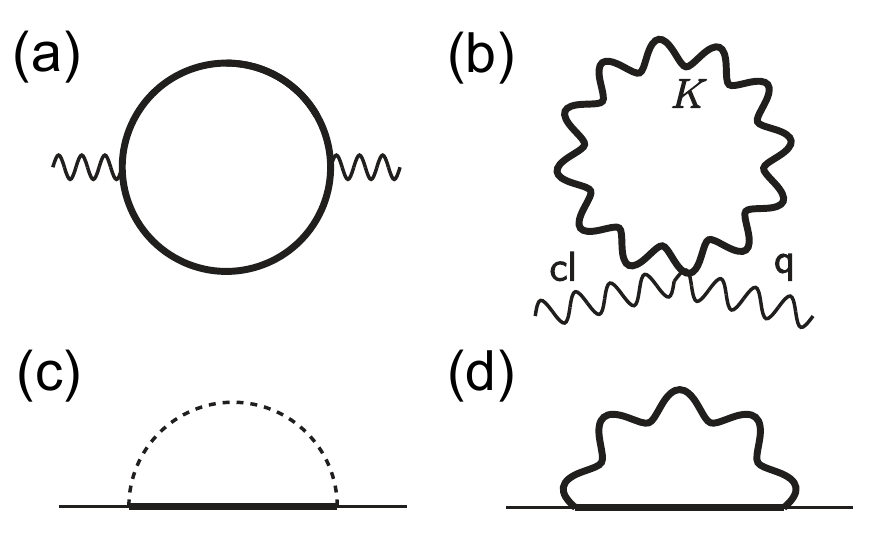} }%
	\caption{
	Diagrams for the saddle-point solution to 
	(a) the bosonic self-energy $\hat{\Pi}_{\sfsp}$ due to the Yukawa interaction, 
	(b) the bosonic thermal mass $\mb^2$ due to the quartic bosonic ($\tilX$-mediated) interaction, 
	(c) the $q$-matrix $\hqsp$, 
	and  
	(d) the fermionic self-energy $\hSig_{\sfsp}$ due to the Yukawa interaction. 
	Straight, wavy and dotted lines respectively represent fermions, bosons and disorder. 
	The propagator lines in bold are fully dressed at the saddle-point level.  
	Vertex corrections due to interactions are suppressed by $1/N$.
	In diagram (b), the symbol $K$ labels the Keldysh component of the bosonic propagator, 
	while $\sfcl$ and $\sfq$ label respectively the classical and quantum component of the external legs.    
	}  	
	\label{fig:saddle_point_selfE}
\end{figure}

The self-energies in Eqs.~(\ref{eq:Sig_sp})--(\ref{eq:q_sp}) are diagrammatically depicted in Fig.~\ref{fig:saddle_point_selfE}. 
We use straight, wavy and dotted lines to respectively represent fermionic, bosonic, and disorder propagators. 
The saddle-point propagators are denoted by bold lines. 
The advantage of the current approach is that disorder and interactions are treated under equal footing self-consistently. 
In the absence of disorder and boson self-interactions, 
the above set of equations agrees with the Schwinger-Dyson equations in Refs.~\cite{NFL_SU_N_Raghu_PRL_19,NFL_SU_N_disO_Raghu_PRL_20}.

We now solve the saddle-point equations. 
For the $\hq$ matrix, 
we assume that the elastic scattering rate $\gammael$ due to the disorder is the largest energy scale, 
relative to the other symmetry-breaking terms (the frequency $\omega$ or the fermion self-energy $\hSig_{\sfsp}$).
We find the following time-translational invariant and spatially homogeneous saddle point solution:
\begin{equation}\label{qSPDef}
	\hqsp 
	= 
	\htau^3
	\otimes
	\hIdSUN.
\end{equation} 
The causality structure is consistent with the 
infinitesimal
``$\eta$ prescription'' 
in the clean, noninteracting fermionic Green's function
[Eq.~(\ref{GRABareDef})].

For the bosonic and fermionic self-energies generated by the Yukawa interaction 
[Eqs.~(\ref{eq:Sig_sp}) and (\ref{eq:Pi_sp})], 
their retarded component in Fourier space can be expressed respectively as
\begin{equation}
\label{eq:Pi_R_formal}
\begin{aligned}
\Pi^R_{\sfsp}(\Omega,\textbf{q})
=
-
i
\frac{g^2}{N}
\int\limits_{\omega,\textbf{k}}
\begin{bmatrix}
G^R_{\sfsp}(\omega + \Omega, \textbf{k} + \textbf{q})  
G^K_{\sfsp}(\omega,\textbf{k})
\\\\
+
G^K_{\sfsp}(\omega + \Omega, \textbf{k} + \textbf{q})  
G^A_{\sfsp} (\omega,\textbf{k})
\end{bmatrix}
\end{aligned}
\end{equation}
and
\begin{equation}
\label{eq:Sig_R_formal}
\begin{aligned}
\Sigma^R_{\sfsp}(\omega,\textbf{k})
=
ig^2
\int\limits_{\Omega,\textbf{q}}
\begin{bmatrix}
D^K_{\sfsp}(\Omega,\textbf{q})
G^R_{\sfsp}(\omega + \Omega,\textbf{k} +\textbf{q})
\\\\
+
D^A_{\sfsp}(\Omega,\textbf{q})
G^K_{\sfsp}(\omega + \Omega,\textbf{k} +\textbf{q})
\end{bmatrix}
.
\end{aligned}
\end{equation}
By linearizing the dispersion of the fermions around the Fermi surface as $\varE_{\textbf{k}} \simeq v_F \delta k$, 
where $\delta \textbf{k} = \textbf{k} - \textbf{k}_F$ and $v_F = k_F/m$ is the Fermi velocity, Eq. (\ref{eq:Pi_R_formal}) can be written as
\begin{equation}
\begin{aligned}
	\Pi^R_{\sfsp}(\Omega,\textbf{q})
	=&\,
	-i
	\frac{g^2}{N}
	\int_{-\infty}^{\infty}
	\frac{d\omega}{2\pi}
	\int_{-\infty}^{\infty}
	\frac{k_F d\delta k}{2\pi}
	\int_0^{2\pi} \frac{d\theta}{2\pi}
\\
&\times
	\frac{1}{
	\tilde{\Sigma}^R_{\omega + \Omega}  + i\gammael -v_F(\delta k + q\cos \theta)
	}
\\
&\times
	\frac{1}{
	\tilde{\Sigma}^A_{\omega} - i\gammael - v_F \delta k
	}
	(F_{\omega + \Omega} - F_{\omega}),
\end{aligned}
\end{equation}  
where $\tilde{\Sigma}^{R/A}_{\omega} = \omega - \Sigma^{R/A}_{\sfsp}(\omega)$ and $\theta$ is the angle between $\textbf{k}_F$ and $\textbf{q}$. 
We assumed for the fermionic self-energy 
$\Sigma^R_{\sfsp}(\omega,\textbf{k}) \simeq \Sigma^R_{\sfsp}(\omega,\textbf{k}_F) \equiv \Sigma^R_{\sfsp,\omega}$. 
Throughout this work, we use $\Sigma^R_{\sfsp,\omega}$ and $\Sigma^R_{\sfsp} (\omega)$ interchangeably. 
Performing the momentum and angular integral, we have
\begin{equation}
\begin{aligned}
	\Pi^R_{\sfsp}(\Omega,\textbf{q})
	=&\,
	-i
	\frac{mg^2}{2\pi N}
	\int_{-\infty}^{\infty}
	d\omega 
\\
&\times 
	\frac{
	F_{\omega + \Omega} - F_{\omega}
	}{
	\sqrt{
	(v_F q)^2
	-
	(
	\tilde{\Sigma}_{\omega + \Omega}^R
	-
	\tilde{\Sigma}_{\omega}^A
	+
	2i\gammael
	)^2
	}
	}.
\end{aligned}
\end{equation}
To leading order in $1/\gammael$, this evaluates to 
\begin{equation}
\label{eq:PiR_saddle_pt}
	\Pi^R_{\sfsp}(\Omega,\textbf{q})
	\simeq
	- 2 i \alpha \Omega,
\quad
	\alpha 
	\equiv 
	\frac{g^2 \nu_0}{2 N \gammael}, 
\end{equation}
where $\nu_0 = m / 2 \pi$ is the bare density of states per fermion flavor. 
The parameter $\alpha$ carries units of inverse diffusion constant, 
is independent of temperature, 
and gives rise to diffusive dynamics at zero-temperature for the quantum-critical bosons.
In evaluating Eq.~(\ref{eq:PiR_saddle_pt}), we have neglected the 
fermion-fermion contribution to the boson mass $\mb^2$ [Eq.~(\ref{DR--intro})],
which is negative, proportional to $\nu_0$, and independent of temperature to leading order. 
We assume that the bosons are tuned to a zero-temperature QCP, with 
paramagnetic critical fluctuations at $T > 0$. These generically induce
a positive $\mb^2$; we consider an explicit calculation next.

We examine the saddle-point equations for $\tilX_{1,2,\sfsp}$ 
[Eqs.~(\ref{eq:saddle_X1}) and (\ref{eq:saddle_X2})]. 
Since $\tilX_{1,\sfsp}$ is proportional to an integral over purely retarded or advanced components of the bosonic propagator, 
it vanishes 
\begin{equation}
	\tilX_{1,\sfsp} = 0.
\end{equation}
This is consistent with the Keldysh structure of $\hD_{\sfsp}$. 
For $\tilX_{2,\sfsp}$, we have 
\begin{equation}
	\tilX_{2,\sfsp}
	=
	2i\lamphi
	\int\limits_{\Omega,\textbf{q}}
	D^K_{\sfsp}(\Omega,\textbf{q}). 
\vspace{2pt}
\end{equation} 
By performing the frequency $\Omega$ integral, followed by the momentum $\textbf{q}$ integral as outlined in Appendix \ref{app:thermal_mass}, we find 
\begin{equation}
\label{eq:mb_saddle_pt}
	\tilX_{2,\sfsp} \equiv \mb^2 = \alpha_m \, T,
\end{equation}
where $\alpha_m$ is a constant depending on $\alpha$ and $\lamphi$. It can be determined by self-consistently solving Eq.~(\ref{eq:mb2_sub-eval2}). 
In contrast to clean case, in which the thermal mass term is proportional to $T \ln T$ \cite{thermal_mass_Bellac,thermal_mass_Torroba}, 
our leading result is not modified by the logarithm. 
The bosonic propagator at saddle-point level is then
\begin{equation}
\label{eq:DR_saddle_pt}
	D^R_{\sfsp}(\Omega,\textbf{q})
	=
	- \frac{1}{2}
	\frac{1}{
	\textbf{q}^2 + \mb^2- i\alpha \Omega
	}.
\end{equation}
Physically, $\mb^2$ represents the thermal mass induced by the quartic bosonic interaction. 
The Yukawa interaction mediated by the quantum-critical bosons is therefore screened 
at any finite temperature. The thermal mass $\mb^2$ serves as the phase relaxation rate for the bosons, 
crucial for describing the dynamics in the quantum relaxational regime \cite{quantum_relax_Sachdev_PRB_94,quantum_relax_Sachdev_PRB_99,NFL_Subir_book_CUP_11}. 
The proportionality of the relaxation rate to $T$ is not surprising, since it is the only relevant energy scale.

Compared to the 
zero-temperature
bosonic self-energy in the clean limit \cite{NFL_SU_N_disO_Raghu_PRL_20,NFL_SU_N_Raghu_PRL_19}, 
\begin{equation}
	\Pi^R_{\sfclean}(\Omega,\textbf{q})
	=
	M_D^2 
	\left(\frac{-i\Omega}{q}\right),
\end{equation}
where the Landau damping scale $M_D^2 = \frac{1}{N} \frac{g^2 \nu_0}{v_F}$, the singularity $\sim \Omega/q$ at $q \rightarrow 0$ 
is neutralized, rendering the bosons diffusive in Eq.~(\ref{eq:DR_saddle_pt}) \cite{MFL_Galitski_PRB_05}. 
In fact, this form of the bosonic propagator is quite generic and not unexpected---it
is the simplest possible form allowed by symmetry in the quantum-relaxational regime, 
when non-analyticities are smeared out by disorder. 

\begin{figure}[b!]
\centering
\includegraphics[width=0.35\textwidth]{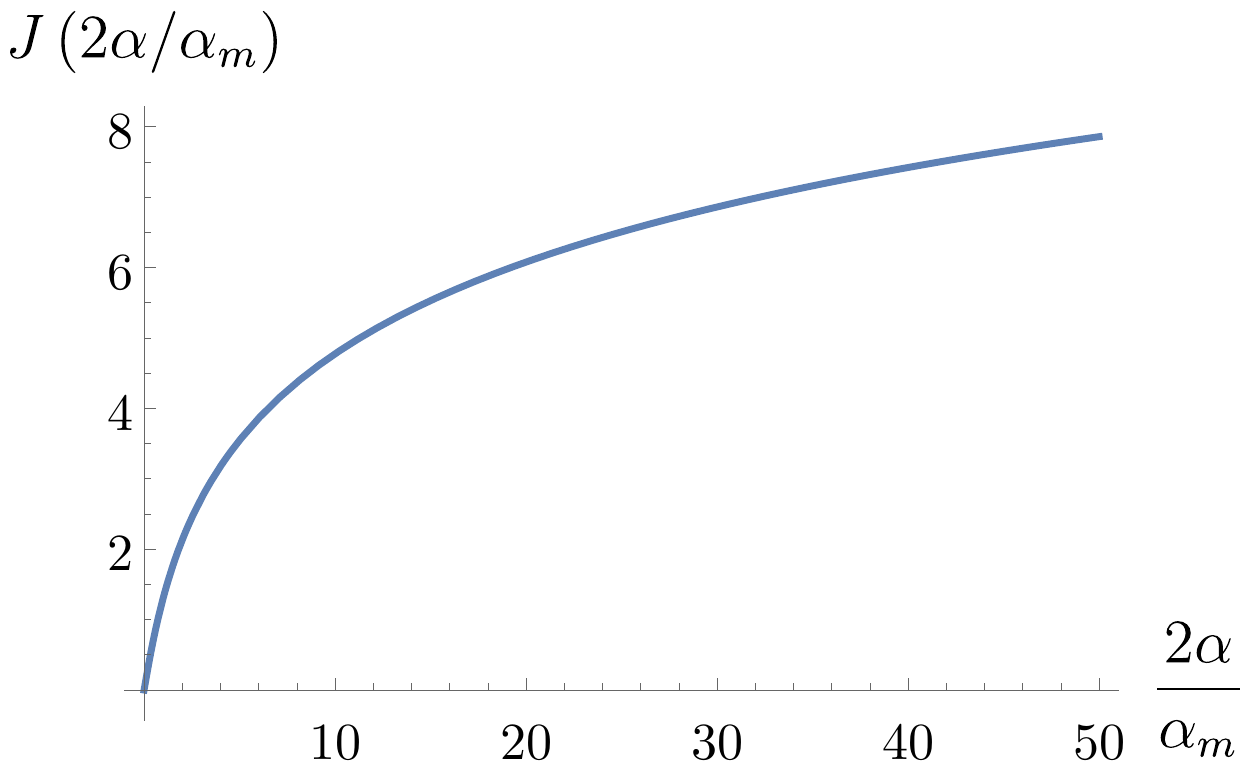}
\caption{
Plot of the dimensionless coefficient $J$ that determines the
temperature dependence of the MFL fermion self-energy in Eq.~(\ref{eq:Sig_sp_final}).
The integral defining $J$ is given by Eq.~(\ref{eq:MFL_J}); 
the argument $2 \alpha / \alpha_m$ is the ratio of the inverse-diffusion constant
[Eq.~(\ref{eq:PiR_saddle_pt})] and the temperature coefficient of the thermal mass [Eq.~(\ref{eq:mb_saddle_pt})]
for the quantum-relaxational bosons. 
}
\label{fig:plot_J}
\end{figure}

To evaluate the fermionic self-energy due to the Yukawa interaction, we feed Eq.~(\ref{eq:DR_saddle_pt}) back to Eq.~(\ref{eq:Sig_R_formal}). 
By performing the momentum integral up to a cutoff $q_{\sfmax}$ below which Eq.~(\ref{eq:PiR_saddle_pt}) is justified, we obtain
\begin{widetext}
\begin{equation}\label{MFL-IntRep}
	\Sigma^R_{\sfsp}(\omega)
	=
	\frac{g^2}{(2\pi)^2 \gammael}
	\int_{-\infty}^{\infty} d\Omega
	\left\{
		\dfrac{i}{2}
		\tan^{-1}
		\left(\dfrac{\alpha \Omega}{\mb^2}\right)
		\left[
			\tanh\left(\dfrac{\Omega + \omega}{2T}\right)
			-
			\coth\left(\dfrac{\Omega}{2T}\right)
		\right]
		-
		\dfrac{1}{4}
		\ln
		\left[
		\dfrac{q_{\sfmax}^4}{
		(\alpha \Omega)^2
		+
		\mb^4
		}
		\right]
		\tanh\left( \dfrac{\Omega + \omega}{2T} \right)
	\right\}.
\end{equation}
\end{widetext}
Using the expression for $\mb^2$ in Eq. (\ref{eq:mb_saddle_pt}), we find, up to logarithmic accuracy, that
\begin{equation}
\label{eq:Sig_sp_final}
\begin{gathered}
	\Sigma^R_{\sfsp}(\omega)
	=
	-
	\gbar^2
	\left[
		\omega 
		\ln 
		\left(\frac{\omega_c}{x}\right)
		+
		i
		\frac{\pi}{2}
		x
	\right],
\\
	x \equiv \max(|\omega|,J \, T),
\end{gathered}
\end{equation}
which is the expression shown at the beginning [Eq.~(\ref{eq:Sig_MFL_intro})]. 
Here
$\gbar^2 = g^2/(4\pi^2 \gammael)$ 
is the dimensionless square of the reduced Yukawa coupling 
and $\omega_c > 0$ is a cutoff below which the MFL form holds.  
The dimensionless temperature coefficient $J$ is determined by the thermal mass coefficient;
it is expressed as the integral
\begin{equation}
\label{eq:MFL_J}
	J(A)
	=
	\frac{4}{\pi}
	\int_{0}^{\infty}dy
	\;
	\tan^{-1}
	\left(A \, y\right)
	\left(
	\coth y
	-
	\tanh y
	\right),
\end{equation}
where $A \equiv 2 \alpha / \alpha_m$, with 
$\alpha$ and $\alpha_m$ respectively defined via Eqs.~(\ref{eq:PiR_saddle_pt}) and (\ref{eq:mb_saddle_pt}). 
A plot of $J$ as a function of $2\alpha/\alpha_m$ is shown in Fig. \ref{fig:plot_J}; 
it has the asymptotic behaviors 
\begin{align}\label{JAsym}
	J(A \rightarrow 0) \simeq \pi A / 2, 
	\qquad
	J(A \rightarrow \infty) \simeq 2 \ln(A). 
\end{align}

The form of $\Sigma^R_{\sfsp}$ in Eq.~(\ref{eq:Sig_sp_final})
is reminiscent of the fermionic self-energy in the MFL phenomenology developed for the cuprates \cite{MFL_Varma_CuO_PRL_89}. 
Our calculations suggest the combined effect of disorder and interaction with critical collective modes serve as a possible origin for the 
MFL behavior. 
The expression for $\Sigma^R_{\sfsp}$ is in sharp contrast with the clean case, in which the fermionic self-energy acquires a 
NFL form \cite{NFL_SU_N_disO_Raghu_PRL_20,NFL_SU_N_Raghu_PRL_19} as
\begin{equation}
	\Sigma^R_{\sfclean}(\omega)
	=
	-
	\frac{g^2}{2\pi \sqrt{3} v_F}
	\frac{1}{
	M_D^{2/3}
	}
	(-i\omega)^{2/3}.
\end{equation}
The MFL form in Eq.~(\ref{eq:Sig_sp_final}) behaves in a less singular manner; this a physical consequence
of disorder broadening. 

More importantly, the temperature-dependence of the MFL self-energy in Eq.~(\ref{eq:Sig_sp_final})
is only stabilized for a nonzero bosonic thermal mass $\alpha_m > 0$ [Eq.~(\ref{eq:mb_saddle_pt})]. 
Eq.~(\ref{JAsym}) shows that the temperature coefficient $J$ diverges logarithmically 
in the limit $\alpha_m \rightarrow 0$.

\section{Non-linear sigma model (``MFL-FNLsM'')}
\label{sec:NLsM}

We now consider fluctuations around the saddle points obtained in Sec.~\ref{sec:saddle_pt_eqn},
which leads to an effective $\hq$-matrix field theory \cite{NLsM5_Finkelshtein_83}
that we dub the ``marginal-Fermi liquid Finkel'stein nonlinear sigma model'' (MFL-FNLsM). 
Specifically,  we only allow $\hG$, $\hSig$ and $\hq$ to fluctuate,
while freezing the bosonic fields $\hD$, $\hat{\Pi}$, and $\tilX$ at their saddle-point values. 
We do this since our primary interest here is in fermion charge transport; in this paper, 
we do not consider the conduction of the conserved SU($N$) ``spin.'' 
We also note that fluctuations in the bosonic sector are expected to be suppressed by $1/N$,
and that the saddle-point boson propagator [Eq.~(\ref{eq:DR_saddle_pt})]
is already damped by the thermal mass in the quantum relaxational regime.  

As usual in a matrix sigma model for quantum (de)localization \cite{NLsM5_Finkelshtein_83,disO_review_Mirlin_RMP_08,disO_review_Kirkpatrick_94,Keldysh_conv4_Kamenev_AdvPhy_09,NLsM2_Matt_Yun_Ann_17}, 
we ignore the massive fluctuations of $\hq$ and focus on the Goldstone modes, which can be parametrized as
\begin{equation}\label{qNLsM}
	\hq 
	=
	\hU^{\dagger}
	\hqsp
	\hU,
\end{equation}
where $\hU$ is a unitary rotation matrix acting upon the composite space of 
frequencies $\otimes$ $\{$retarded, advanced$\}$ $\otimes$ SU($N$) flavors; 
the saddle-point $\hqsp$ is given by Eq.~(\ref{qSPDef}). 
The massless modes correspond to diffusons, quantum two-particle diffusion modes
in the particle-hole channel. 

In this work, we focus on the diffusive regime in which $\gammael \gg \omega, v_F k$,
where $k$ is a fermion rendered from the Fermi surface.  
To derive an effective action involving only $\hq$, we integrate out the fluctuation $\delta \hG$, 
followed by $\delta \hSig$. 
We then perform a gradient expansion to obtain the FNLsM, following the standard procedures 
\cite{NLsM1_Ludwig_PRB_99,NLsM2_Matt_Yun_Ann_17,NLsM3_Matt_PRB_08,NLsM4_Matt_PRB_06,NLsM5_Finkelshtein_83,NLsM6_Kamenev_CUP_11,NLsM7_Kane_Stone_AoP_81,NLsM8_Keldysh_SC_PRB_20}. 
Leaving the technical details for the derivation in Appendix~\ref{app:derivation_FNLsM}, we 
state the resulting action below: 
\begin{eqnarray}
\label{eq:Z_eff}
	Z[V]
	&=&
	\int 
	\calD \hq 
	\;
	\e^{-S[\hq]},
\\
\label{eq:S_eff_q}
	S[\hq]
	&=&
	S_D
	+
	S_{\sfintI}
	+	
	S_{\sfintII}
	+
	S_{qV}
	+
	S_V,
\end{eqnarray}
where the $\hq$ matrix is subjected to the constraints 
\begin{equation}
	\hq^2 = \hat{1}, 
\qquad
	\Tr [\hq] = 0,
\qquad
	\hq = \hq^{\dagger},
\end{equation}
which follow from Eq.~(\ref{qNLsM}). 
The matrix $\hat{q} \rightarrow q_{i j; \omega \omega'}^{a b}$ with
frequency $\{\omega,\omega'\}$, 
SU($N$) flavor $\{i,j\}$, 
and
Keldysh $\{a,b\}$ indices displayed. 

The first term in the action is 
\begin{widetext}
\begin{equation}
\label{eq:S_D_q}
	S_D
	=
	\frac{1}{2\lambda}
	\int\limits_{\textbf{x}}
	\Tr\left[
	\Nabla \hq
	\cdot
	\Nabla \hq
	\right]
	+
	2ih
	\int\limits_{\textbf{x}}
	\Tr\left[
	(
	\hat{\omega}
	+ i\eta \htau^3
	-\hSig_{\sfsp}
	)
	\hq
	\right],
\end{equation}
where
\begin{equation}
\label{eq:def_lambda_h}
	\frac{1}{\lambda}  = D h, 
\quad
	h = \frac{\pi \nu_0}{2},
\quad
	D = \frac{v_F^2}{4\gammael}.
\end{equation}
Here $D$ is the semiclassical diffusion constant.
In Eq.~(\ref{eq:S_D_q}) and what follows, matrix traces $\Tr(\cdots)$ run over all indices not explicitly 
displayed. 
We focus on the weak-disorder limit in which $\gammael \ll \mu$ ($\mu$ is the chemical potential)
so that $\lambda$, 
proportional to the inverse conductance, can be treated as a small parameter. 
The only difference from the usual unitary-class sigma model in Eq.~(\ref{eq:S_D_q}) is 
the MFL fermion self-energy 
$\hSig_{\sfsp} = \diag\left\{\Sigma_{\sfsp}^R,\Sigma_{\sfsp}^A\right\}_\tau$ [Eq.~(\ref{eq:Sig_sp_final})]. 

Meanwhile, interparticle interactions mediated by the quantum-relaxational bosons via the Yukawa coupling 
give rise to the terms 
\begin{align}
\label{eq:S_intI_q}
	S_{\sfintI}
	=&\,
	\Gamma_1	
	\int\limits_{1-4,\vex{k}}
	\delta_{1+3,2+4}
	\,
	\Tr
	\left[
		\hq_{ij;1,2}(-\vex{k})
		\,
		\hgamma^{s}_{2,1}
	\right]
	D^{ss'}_{\sfsp,2-1}(\vex{k})
	\,
	\Tr\left[
		\hgamma^{s'}_{4,3}
		\,
		\hq_{ji;3,4}(\vex{k})
	\right],
\\
\label{eq:S_intII_q}
	S_{\sfintII}
	=&\,
	- 
	\Gamma_2
	\int\limits_{1-4,\vex{k}}
	\delta_{1+3,2+4}
	\,
	\Tr
	\bigg[
		\hq_{ii;1,2}(-\vex{k}) 
		\,
		\hgamma^s_{2,3}
		\,
		\hq_{jj;3,4}(\vex{k})
		\,
		\hgamma^{s'}_{4,1}
	\bigg]
		\,
		\int\limits_{\vex{k'}}
		D^{ss'}_{\sfsp,1-4}(\vex{k'}),
\end{align}
$\phantom{0}$\\
\end{widetext}
where the coupling constants
\begin{equation}
\label{eq:def_Gamma_1_and_2}
	\Gamma_1 \equiv i \frac{2h^2 g^2}{N},
\qquad
	\Gamma_2 \equiv i \frac{h g^2}{\gammael N}.
\end{equation}
Here, we defined the shorthands $\int_{1-4} = \int_{1,2,3,4}$, 
$\left\lbrace 1,2,3,4 \right\rbrace \equiv \left\lbrace \omega_1,\omega_2,\omega_3,\omega_4 \right\rbrace$ 
and the energy-conserving Dirac-$\delta$ function $\delta_{1+3,2+4} = 2\pi \delta(\omega_1 + \omega_3 -\omega_2 - \omega_4)$. 
The numerical subscripts thus stand for the corresponding frequency labels. 
In addition, $s,s' \in\{\sfcl,\sfq\}$ are boson Keldysh indices, while the traces run over the fermion Keldysh space $a \in \{R,A\}$.

\begin{figure}[t!]
	\centering
	{\includegraphics[width=6cm]{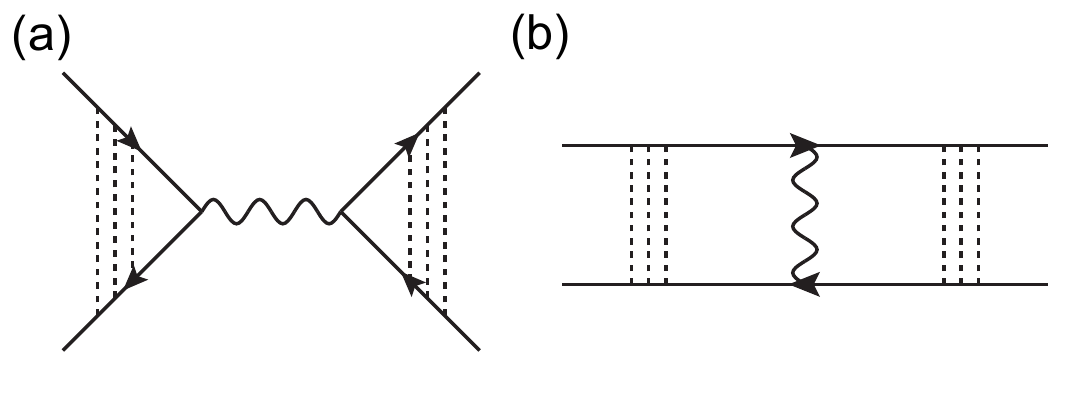} }%
	\caption{
	Diagrammatic representation for the matrix-boson mediated interactions terms
	(a) $S_{\sfintI}$ and (b) $S_{\sfintII}$, in the action of the  
	MFL-FNLsM in Eqs.~(\ref{eq:S_eff_q}), (\ref{eq:S_intI_q}), and  (\ref{eq:S_intII_q}).
	Solid lines and wavy lines correspond respectively to $\hG_{\sfsp}$ and $\hD_{\sfsp}$. 
	The dotted-line ladder represents lowest-order quantum (diffuson) fluctuations of the
	constrained matrix field $\hq$. 
	}  	
	\label{fig:Sint_I_II}
\end{figure}

The interactions are depicted diagrammatically in Fig.~\ref{fig:Sint_I_II}. 
The first one [$S_{\sfintI}$, Fig.~\ref{fig:Sint_I_II}(a)] is the usual interaction term appearing in the FNLsM. 
The second one [$S_{\sfintII}$, Fig.~\ref{fig:Sint_I_II}(b)], which involves a vertical interaction line, is not considered in the original paper by Finkel'stein \cite{NLsM5_Finkelshtein_83}. 
However, in the case of the MFL-FNLsM, this term is essential for fixing the Ward identity for physical response functions. 
In the context of disordered $d$-wave superconductors, interaction terms similar to $S_{\sfintII}$ were also taken into account in the derivation of 
the sigma model \cite{SintII1_Altland_PRL_03,SintII2_Altland_IJMP_04}.

Finally, the terms in the action [Eq.~(\ref{eq:S_eff_q})] 
involving the source field are
\begin{equation}
\label{eq:S_qV_q}
	S_{qV}
	=
	-
	2 i h 
	\int\limits_{t,\textbf{x}}
	\Tr
	\left\lbrace  
	\hq_{t,t} (\textbf{x})
	\left[
		V^{\sfcl}(t,\textbf{x}) \hgamma^{\sfcl}
		+
		V^{\sfq}(t,\textbf{x}) \hgamma^{\sfq}
	\right]
	\right\rbrace 
\end{equation}
and
\begin{equation}
\label{eq:S_V}
	S_V
	=
	-
	2 i N \nu_0  
	\int\limits_{t,\textbf{x}}
	V^{\sfq}(t,\textbf{x})
	V^{\sfcl}(t,\textbf{x}).
\vspace{2pt}
\end{equation}
The second term arises from the integration involving products of retarded-retarded and advanced-advanced fermionic Green's functions
in the second-order gradient expansion \cite{NLsM2_Matt_Yun_Ann_17,NLsM6_Kamenev_CUP_11}. 
It gives the static-compressibility component of the density response function.\\

\subsection{$\pi$-$\sigma$ parametrization}

We employ the ``$\pi$-$\sigma$'' parametrization for the $\hq$ matrix
\begin{equation}
\label{eq:q=q0_q1_q2}
\begin{aligned}[b]
\!\!\!\!
	\hat{q}
	=
	\begin{bmatrix}
	\sqrt{\hat{1} - \hW \hWdag}
	& 
	\hW
	\\
	\hWdag
	&
	-\sqrt{\hat{1} - \hWdag \hW}
	\end{bmatrix}_{\tau}
	\equiv&\,
	\hat{q}^{(0)}
	+
	\hat{q}^{(1)}
	\\
	&\,
	+
	\hat{q}^{(2)}
	+
	\ldots
\!\!
\end{aligned}
\end{equation}
where
\begin{eqnarray}
	\label{eq:pi_sigma_0}
	\hq^{(0)}
	&=&
	\begin{bmatrix}
	\hat{1} & 0
	\\
	0 & -\hat{1}
	\end{bmatrix}_{\tau},
\\
	\label{eq:pi_sigma_1}
	\hq^{(1)}
	&=&
	\begin{bmatrix}
	0 & \hW
	\\
	\hWdag & 0
	\end{bmatrix}_{\tau},
\\
	\label{eq:pi_sigma_2}
	\hq^{(2)}
	&=&
	-
	\frac{1}{2}
	\begin{bmatrix}
	\hW \hWdag & 0
	\\
	0 & - \hWdag \hW
	\end{bmatrix}_{\tau},
\\
	\label{eq:pi_sigma_3}
	\hq^{(3)}
	&=&
	0,
\\
	\label{eq:pi_sigma_4}
	\hq^{(4)}
	&=&
	-
	\frac{1}{8}
	\begin{bmatrix}
	(\hW \hWdag)^2 & 0
	\\
	0 & - (\hWdag\hW)^2
	\end{bmatrix}_{\tau},
\end{eqnarray}
and 
$\hW \rightarrow W_{ij;\omega_1,\omega_2}$ is a matrix field containing flavor indices $i,j$ and frequency labels $\omega_1,\omega_2$.     
In the following, we expand the action in Eq.~(\ref{eq:S_eff_q}) up to quartic order in $\hW$. 
After the expansion, we will rescale 
$\hW \rightarrow \sqrt{\lambda} \hW$, 
$\hWdag \rightarrow \sqrt{\lambda} \hWdag$ 
to facilitate the perturbative calculation 
in $\lambda$.

\begin{widetext}

\subsubsection{Diffusion term}

By plugging Eqs.~(\ref{eq:q=q0_q1_q2})--(\ref{eq:pi_sigma_4}) into $S_D$ [Eq.~(\ref{eq:S_D_q})], 
we have
\begin{equation}
\label{eq:S_D_2}
	S_D^{(2)}
	[W,\Wdag]
	=
	\int\limits_{1,2,\textbf{q}}
	\left[
	\textbf{q}^2
	-
	ih \lambda
	(
	\omega_2 -\hSig^R_2 
	-
	\omega_{1}  
	+
	\hSig^A_1 
	)
	\right]
	\Tr\left[
	\hWdag_{12} (\textbf{q})
	\hW_{21} (\textbf{q})
	\right]
\end{equation}
and
\begin{equation}
	S_D^{(4)}[W,\Wdag]
	=
	\lambda
	\int\limits_{1-4,\textbf{q}_i}
	\Box_{1,2,3,4}^{\textbf{q}_1,\textbf{q}_2,\textbf{q}_3,\textbf{q}_4}
	\delta_{\textbf{q}_1 + \textbf{q}_3,\textbf{q}_2 + \textbf{q}_4}
	\,
	\Tr
	\left[
	\hWdag_{12}(\textbf{q}_1)
	\hW_{23}(\textbf{q}_2)
	\hWdag_{34}(\textbf{q}_3)
	\hW_{41}(\textbf{q}_4)
	\right].
\end{equation}
Here, $\int_{\textbf{q}_i} \equiv \int_{\textbf{q}_1,\textbf{q}_2,\textbf{q}_3,\textbf{q}_4}$,
the trace runs over the Keldysh space and we defined
\begin{equation}
\label{eq:S_D4_box}
	\Box_{1,2,3,4}^{\textbf{q}_1,\textbf{q}_2,\textbf{q}_3,\textbf{q}_4}
	=
	\dfrac{1}{2^3}
	\begin{bmatrix}
		-2(\textbf{q}_1 \cdot \textbf{q}_3 + \textbf{q}_2 \cdot \textbf{q}_4)
		+
		(\textbf{q}_1 + \textbf{q}_3)
		\cdot
		(\textbf{q}_2 + \textbf{q}_4)
	\\
	-
	ih\lambda
	(\omega_2 + \omega_4 - \omega_1 - \omega_3)
	+
	ih\lambda
	\left(
	\Sigma^R_{\sfsp,2} + \Sigma^R_{\sfsp,4} - \Sigma^A_{\sfsp,1} - \Sigma^A_{\sfsp,3}
	\right)
\end{bmatrix}.
\end{equation}

\subsubsection{Interaction terms}


By substituting Eqs. (\ref{eq:pi_sigma_0})--(\ref{eq:pi_sigma_4}) into $S_{\sfintI}$ 
[Eq.~(\ref{eq:S_intI_q})], we obtain the corresponding quadratic, cubic, and quartic interaction terms, which are respectively given by
\begin{equation}
\label{eq:q1_Gamma_q1}
\begin{aligned}
&S_{\sfintI}^{(2)}
=
-
2\lambda \Gamma_1
\int\limits_{1-4,\textbf{q}}
(F_1 - F_2)
\Wdag_{ij;1,2}(\textbf{q}) 
W_{ji;3,4}(\textbf{q})
D^{R}_{\sfsp,2-1}(\textbf{q})
\delta_{1+3,2+4}
,
\end{aligned}
\end{equation}
\begin{equation}
\label{eq:q1_Gamma_q2_sum}
S_{\sfintI}^{(3)}
=
-
\lambda^{3/2}
\Gamma_1
\int\limits_{1-5,\textbf{p},\textbf{q}}
\begin{Bmatrix}
\Wdag_{ij;1,2}(\textbf{q})
\hW_{jk;3,5}(\textbf{p} + \textbf{q})
\Wdag_{ki;5,4}(\textbf{p})
D^R_{\sfsp,2-1}(\textbf{\textbf{q}})
\left[
(F_1 - F_2)
\left(
\dfrac{1}{F_{3-4}}
+
F_4
\right)
\right]
\\
+
\Wdag_{ij;1,2}(\textbf{q})
\hWdag_{jk;3,5}(\textbf{p})
W_{ki;5,4}(\textbf{p} + \textbf{q})
D^R_{\sfsp,2-1}(\textbf{q})
\left[
(F_1 - F_2)
\left(
F_3 - \dfrac{1}{F_{3-4}}
\right)
\right]
\\
-
W_{ij;1,2}(\textbf{q})
\left[
W_{jk;3,5}(\textbf{p})
\Wdag_{ki;5,4}(\textbf{p} + \textbf{q})
-
\Wdag_{jk,3,5}(\textbf{p} + \textbf{q})
W_{ki;5,4}(\textbf{p})
\right]
D^A_{\sfsp,2-1}(-\textbf{q})
\end{Bmatrix}
\delta_{1+3,2+4}
,
\end{equation}
and
\begin{equation}
\label{eq:q2_Gamma_q2}
S_{\sfintI}^{(4)}
=
\frac{\lambda^2}{2}
\Gamma_1
\int\limits_{1-6,\textbf{k},\textbf{p},\textbf{q}}
\begin{bmatrix}
W_{ik;1,5}(\textbf{p})
\Wdag_{kj;5,2}(\textbf{p} + \textbf{q})
W_{jl;3,6}(\textbf{k} )
\Wdag_{li;6,4}(\textbf{k} -  \textbf{q})
\left(
\dfrac{1}{F_{3-4}} + F_4
\right)
\\
+
W_{ik;1,5}(\textbf{p})
\Wdag_{kj;5,2}(\textbf{p} + \textbf{q})
\Wdag_{jl;3,6}(\textbf{k} - \textbf{q})
W_{li;6,4}(\textbf{k})
\left(
-\dfrac{1}{F_{3-4}} + F_3
\right)
\\
+
\Wdag_{ik;1,5}(\textbf{p} + \textbf{q})
W_{kj;5,2}(\textbf{p})
W_{jl;3,6}(\textbf{k})
\Wdag_{li;6,4}(\textbf{k} - \textbf{q})
\left(
-\dfrac{1}{F_{3-4}} - F_4
\right)
\\
+
\Wdag_{ik;1,5}(\textbf{p} + \textbf{q})
W_{kj;5,2}(\textbf{p})
\Wdag_{jl;3,6}(\textbf{k} - \textbf{q})
W_{li;6,4}(\textbf{k})
\left(
\dfrac{1}{F_{3-4}} - F_3
\right)
\end{bmatrix}
D^R_{\sfsp,2-1}(\textbf{q})
\delta_{1+3,2+4},
\end{equation}
where $F_{\omega} \equiv F(\omega)$.


Repeating the above procedures for $S_{\sfintII}$ [Eq. (\ref{eq:S_intII_q})], we obtain the type II interaction terms
\begin{equation}
S_{\sfintII}^{(2)}
=
-  2\lambda \Gamma_2
\int\limits_{1-4,\textbf{k},\textbf{q}}
\Wdag_{ii;1,2}(\textbf{q})
W_{jj;3,4}(\textbf{q})
\bigg[
D^K_{\sfsp,1-4}(\textbf{k})
-
F_1 D^R_{\sfsp,1-4}(\textbf{k})
+
F_{2} D^A_{\sfsp,1-4}(\textbf{k})
\bigg]
\delta_{1+3,2+4},
\end{equation} 
and
\begin{equation}
\begin{aligned}
&S_{\sfintII}^{(3)}
=
\lambda^{3/2}
\Gamma_2
\int\limits_{1-4,\textbf{k},\textbf{p},\textbf{q}}
\delta_{1+3,2+4}
\\
&\times
\begin{Bmatrix}
-W_{jk;3,5}(\textbf{p}) \Wdag_{kj;5,4}(\textbf{p} + \textbf{q})
W_{ii;1,2}(\textbf{q})
D^A_{\sfsp,1-4}(\textbf{k})
\\
+
\Wdag_{jk;3,5}(\textbf{p} + \textbf{q}) W_{kj;5,4}(\textbf{p})
W_{ii;1,2}(\textbf{q})
D^R_{\sfsp,1-4}(\textbf{k})
\\
+
W_{jk;3,5} (\textbf{p} + \textbf{q})
\Wdag_{kj;5,4}(\textbf{p})
\Wdag_{ii;1,2}(\textbf{q})
\left[
(F_1 - F_4) F_{2} D^A_{\sfsp,1-4}(\textbf{k})
-
(1 - F_1 F_4) D^R_{\sfsp,1-4}(\textbf{k})
+
(F_1 - F_4) D^K_{\sfsp,1-4}(\textbf{k})
\right]
\\
-
\Wdag_{jk;3,5}(\textbf{p})
W_{kj;5,4}(\textbf{p} + \textbf{q})
\Wdag_{ii;1,2}(\textbf{q})
\left[
-(
1 - F_{2} F_{3}
)
D^A_{\sfsp,1-4}(\textbf{k})
-
F_1(F_{3} - F_{2})
D^R_{\sfsp,1-4}(\textbf{k})
+
(F_{3} - F_{2})
D^K_{\sfsp,1-4}(\textbf{k})
\right]
\end{Bmatrix}
.
\end{aligned}
\end{equation}

The contribution of the quartic terms from $S_{\sfintII}$ to the density correlation function is subleading and thus omitted here.

\subsubsection{Source field term}

Similarly, the $S_{qV}$ [Eq.~(\ref{eq:S_qV_q})] becomes 
\begin{equation}
\label{eq:S_qV}
\begin{aligned}
S_{qV}
=
-2ih 
\int\limits_{1-3,\textbf{p},\textbf{q}}
\begin{Bmatrix}
\sqrt{\lambda}
V^{\sfcl}_{2-1}(\textbf{q}) (F_1 - F_2) \Wdag_{ii;1,2}(\textbf{q})
\delta_{\textbf{p},\textbf{q}}
\delta_{2,3}
\\
+
\sqrt{\lambda}
\left[
-W_{ii;1,2}(\textbf{q})V^{\sfq}_{2-1}(-\textbf{q})
\delta_{\textbf{p},\textbf{q}} 
- 
(1 - F_1 F_2) 
\Wdag_{ii;1,2}(\textbf{q})
V^{\sfq}_{2-1}(\textbf{q})
\delta_{\textbf{p},\textbf{q}} 
\right]
\delta_{2,3}
\\
-
\frac{\lambda}{2}
\left[
V^{\sfcl}_{2-1}(\textbf{q})
W_{i,j;1,3} (\textbf{p}) \Wdag_{ji;3,2}(\textbf{p} + \textbf{q})
-
V^{\sfcl}_{2-1}(\textbf{q})
\Wdag_{ij;1,3}(\textbf{p} + \textbf{q})
W_{ji;3,2}(\textbf{p})
\right]
\\
-
\frac{\lambda}{2}
\left[
V^{\sfq}_{2-1}(\textbf{q})
W_{ij;1,3}(\textbf{p}) \Wdag_{ji;3,2} (\textbf{q} + \textbf{k})
F_2
+
V^{\sfq}_{2-1}(\textbf{q})
\Wdag_{ij;1,3}(\textbf{p}+\textbf{q})
W_{ji;3,2}(\textbf{p})
F_1
\right]
\end{Bmatrix}.
\end{aligned}
\end{equation}

\end{widetext}

\subsection{Feynman rules}

\subsubsection{Bare propagator}

\newpage
From Eq.~(\ref{eq:S_D_2}), we can read off the bare diffuson propagator to be
\begin{equation}
\label{eq:free_W_propagator}
\begin{aligned}
&
\Delta^R_{1,2}(\textbf{q})
\delta_{1,4}
\delta_{2,3}
\\
&\equiv
\left\langle 
W_{ij;1,2}
(\textbf{q})
W_{ji;3,4}^{\dagger}
(\textbf{q})
\right\rangle_0
\\
&=
\frac{
\delta_{1,4}
\delta_{2,3}
}{
\textbf{q}^2
-
ih\lambda(\omega_1 - \omega_2)
-
ih\lambda
[
\Sigma^A_{\sfsp}(\omega_2)
-
\Sigma^R_{\sfsp}(\omega_1)
]
}.
\end{aligned}
\end{equation}
This is in contrast with the diffuson propagator in the Fermi liquid 
case \cite{Keldysh_conv4_Kamenev_AdvPhy_09,NLsM1_Ludwig_PRB_99,NLsM2_Matt_Yun_Ann_17,NLsM3_Matt_PRB_08,NLsM5_Finkelshtein_83,NLsM6_Kamenev_CUP_11}, 
\begin{equation}
\label{eq:diffson_FL}
\Delta^R_{\sfFL;1,2}(\textbf{q})
=
\Delta^R_{\sfFL;1-2}(\textbf{q})
=
\frac{1}{
\textbf{q}^2 - ih \lambda (\omega_1 - \omega_2)
},
\end{equation}
which explicitly conserves particles. 
Our diffuson in Eq.~(\ref{eq:free_W_propagator}) is now anomalous due to the appearance of the 
MFL fermionic self-energy $\Sigma^{R/A}_{\sfsp}(\omega)$. 

The anomalous diffuson propagator is represented diagrammatically in Fig.~\ref{fig:vertices_S_D}(a) by two black
solid lines with arrows pointing in the opposite directions. 
The frequency indices of the matrix fields $\hW$ and $\hWdag$ are labeled by the numbers. 
The flavor indices are implicit. 
Along a solid line, the frequency label and flavor index remain unchanged. 
The momentum flowing through the diffuson is labeled by an arrow in the middle of the two black solid lines. 
The momentum arrow points inwards (outwards) for the field $\hW$ ($\hWdag$).

\subsubsection{Interaction vertices}

\begin{figure}[t!]
	\centering
	{\includegraphics[width=7cm]{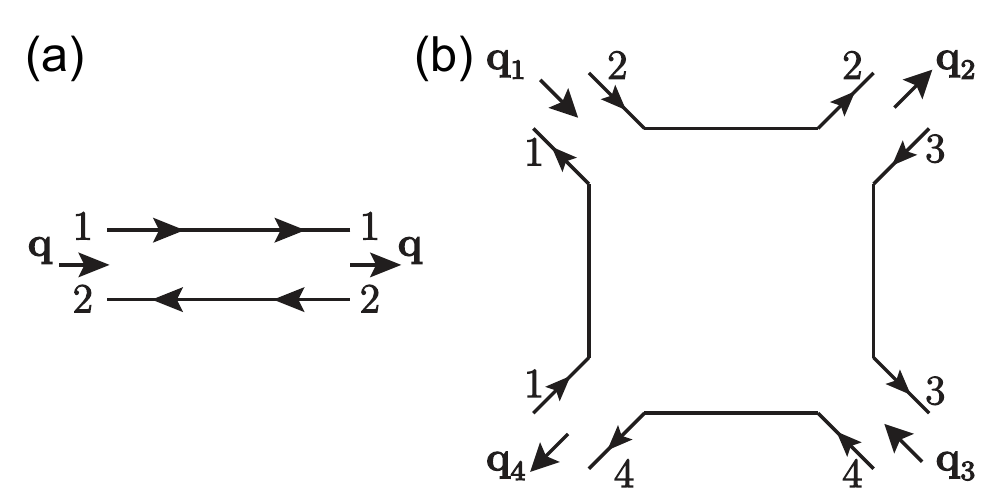} }%
	\caption{
	(a) The bare propagator and (b) quartic vertex generated from $S_D$. 
	}  	
	\label{fig:vertices_S_D}
\end{figure}

\begin{figure}
	\centering
	{\includegraphics[width=8cm]{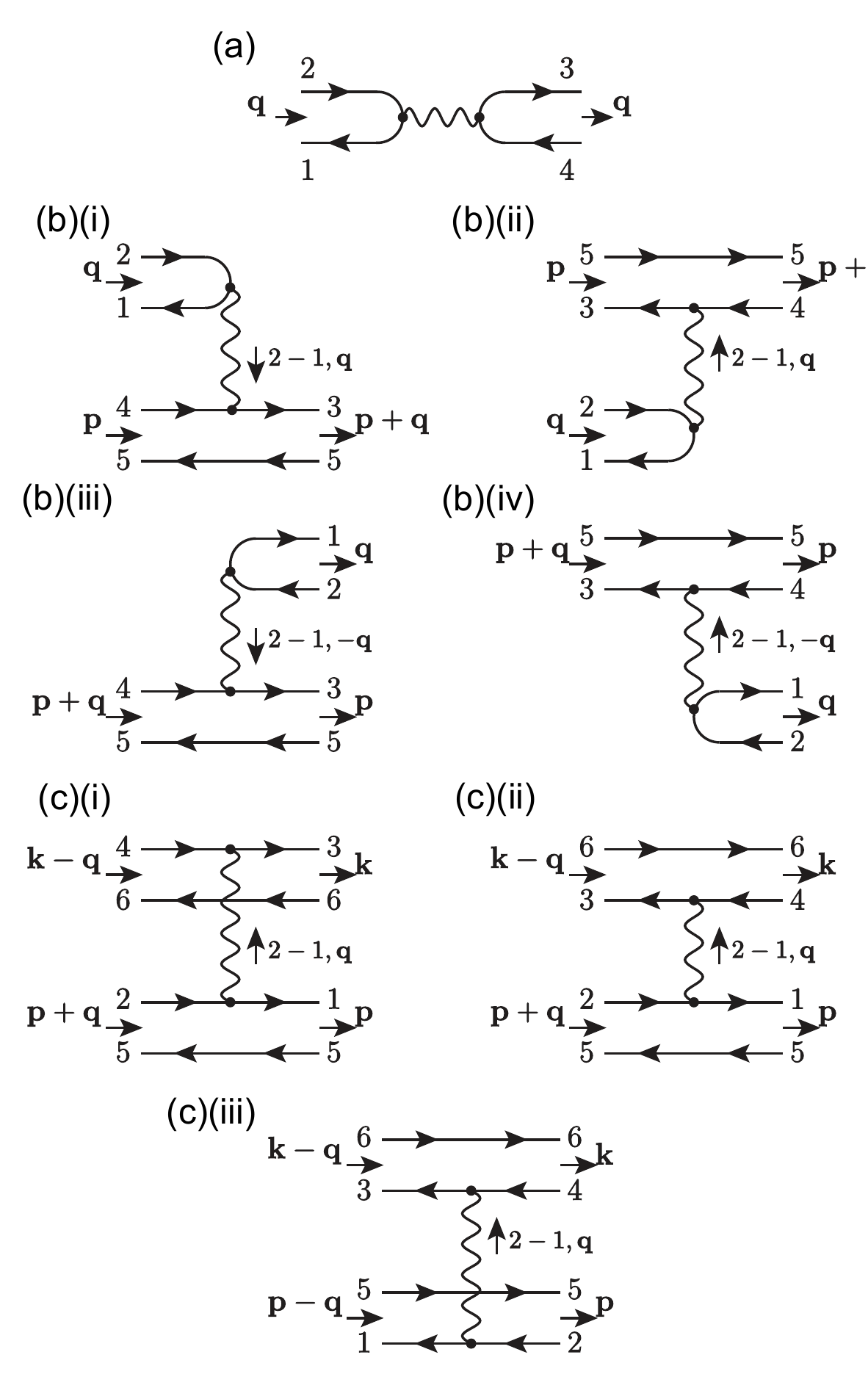} }%
	\caption{
	The set of vertices due to $S_{\sfintI}$. 
	Diagram (a) represents the quadratic vertex, (b) represents the cubic interaction vertices, and (c) represents the quartic interaction vertices. 
	}  	
	\label{fig:vertices_SintI}
\end{figure}

\begin{figure}
	\centering
	{\includegraphics[width=8cm]{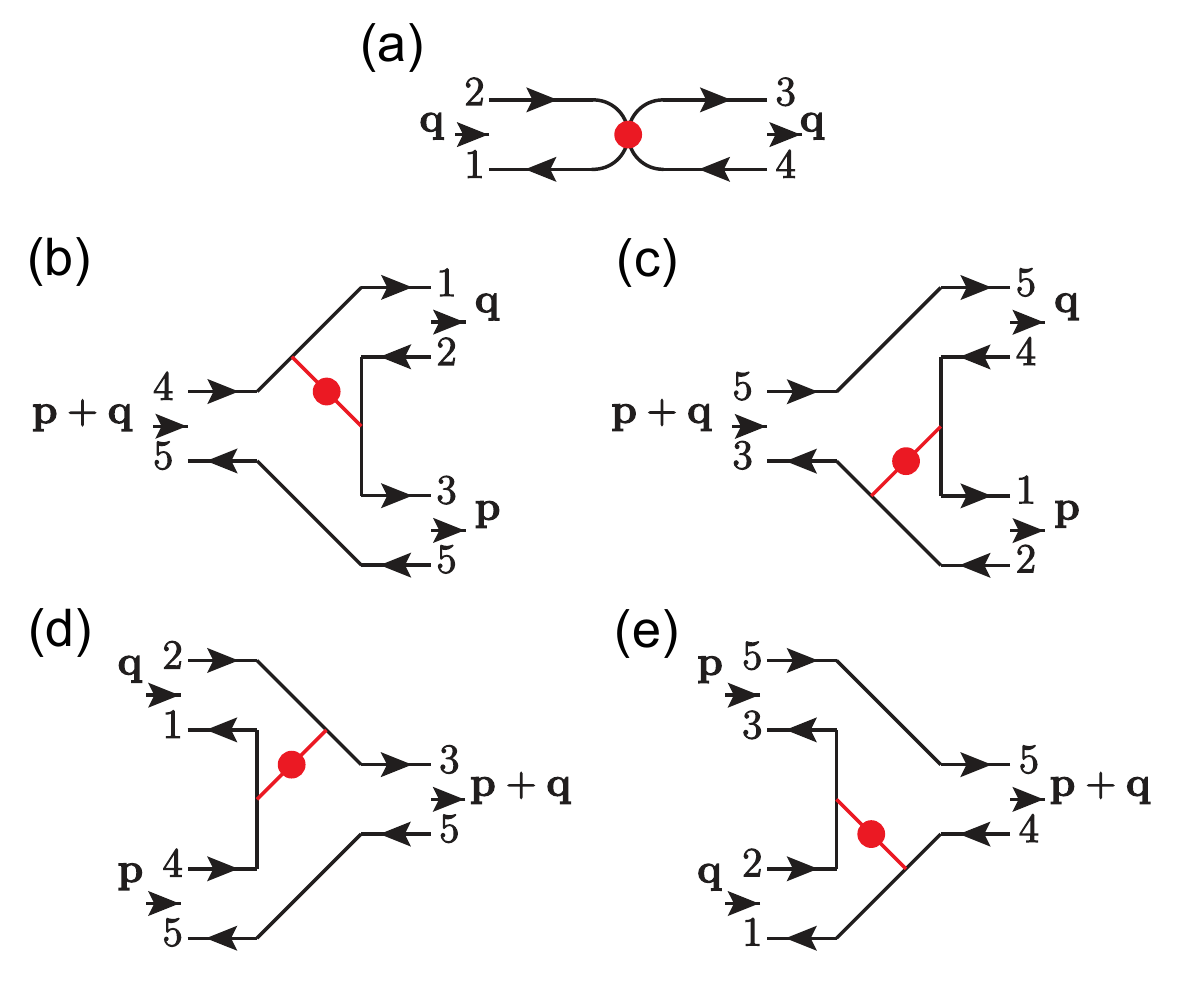} }%
	\caption{
	The set of vertices due to $S_{\sfintII}$. Diagram (a) represents the quadratic vertex and (b)--(e) represent the cubic vertices. 
	}  	
	\label{fig:vertices_SintII}
\end{figure}

\begin{figure}[t!]
	\centering
	{\includegraphics[width=8cm]{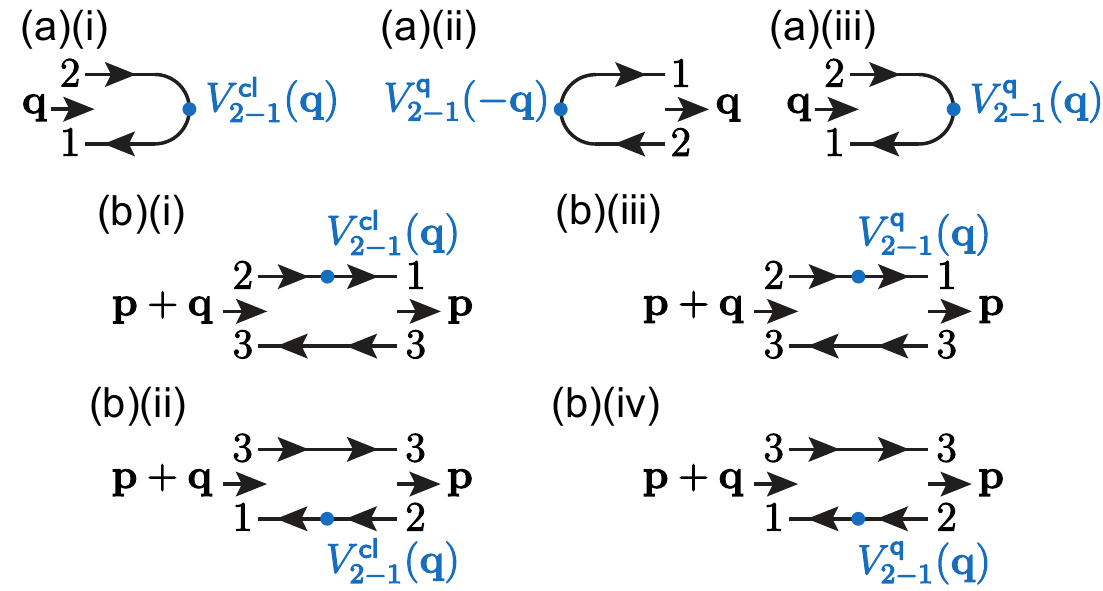} }%
	\caption{
	The set of vertices due to $S_{qV}$. Diagram (a) shows the coupling between $V^{\sfcl/\sfq}$ and $\hq^{(1)}$ while (b) shows the coupling between $V^{\sfcl/\sfq}$ and $\hq^{(2)}$. 
	}  	
	\label{fig:vertices_qV}
\end{figure}

Fig.~\ref{fig:vertices_S_D}(b) and Figs.~\ref{fig:vertices_SintI}--\ref{fig:vertices_qV} 
illustrate the interaction vertices. In particular, the amplitudes of the terms 
linear 
in $\hW$ are
\begin{equation}
\text{Fig. 
\ref{fig:vertices_qV}(a)(i)
}
=
2ih \sqrt{\lambda}
(F_1 - F_2)
V^{\sfcl}_{2-1}(\textbf{q})
,
\end{equation}
\begin{equation}
\text{Fig. 
\ref{fig:vertices_qV}(a)(ii)
}
=
-2ih \sqrt{\lambda}
V^{\sfq}_{2-1}(-\textbf{q})
,
\end{equation}
\begin{equation}
\text{Fig. 
\ref{fig:vertices_qV}(a)(iii)
}
=
-2 ih \sqrt{\lambda}
(1 - F_1 F_2)
V^{\sfq}_{2-1}(\textbf{q})
.
\end{equation}
%
Those 
quadratic 
in $\hW$ are
\begin{equation}
\text{Fig. 
\ref{fig:vertices_SintI}(a)
}
=
2\lambda \Gamma_1
(
F_1 - F_2
)
D^R_{\sfsp,2-1}(\textbf{q})
\delta_{1+3,2+4}
,
\end{equation}
\begin{equation}\label{sintii_quad}
\begin{aligned}
&\text{Fig. 
\ref{fig:vertices_SintII}(a)
}
=
2\lambda \Gamma_2 \delta_{1+3,2+4}
\\
&\times
\int\limits_{\textbf{k}}
\bigg[
D^K_{\sfsp,1-4}(\textbf{k})
-
F_1 D^R_{\sfsp,1-4}(\textbf{k})
+
F_{2} D^A_{\sfsp,1-4}(\textbf{k})
\bigg]
,
\end{aligned}
\end{equation}
\begin{eqnarray}
\text{Fig. 
\ref{fig:vertices_qV}(b)(i)
}
&=&
-ih\lambda
V^{\sfcl}_{2-1}(\textbf{q})
,
\\
\text{Fig.
\ref{fig:vertices_qV}(b)(ii)
}
&=&
ih\lambda
V^{\sfcl}_{2-1}(\textbf{q})
,
\\
\text{Fig. 
\ref{fig:vertices_qV}(b)(iii)
}
&=&
-ih\lambda
V^{\sfq}_{2-1}(\textbf{q})
F_2
,
\\
\text{Fig.
\ref{fig:vertices_qV}(b)(iv)
}
&=&
-ih\lambda
V^{\sfq}_{2-1}(\textbf{q})
F_1
.
\end{eqnarray}
The 
cubic
ones are 
\begin{equation}
\begin{aligned}
&\text{Fig. \ref{fig:vertices_SintI}(b)(i)}
\\
&=
\lambda^{3/2} \Gamma_1
\delta_{1+3,2+4}
D^R_{\sfsp,2-1}(\textbf{q})
(F_1 - F_2)
\left(
\frac{1}{F_{3-4}}
+
F_4
\right)
,
\end{aligned}
\end{equation}
\begin{equation}
\begin{aligned}
&
\text{Fig. \ref{fig:vertices_SintI}(b)(ii)}
\\
&=
\lambda^{3/2} \Gamma_1
\delta_{1+3,2+4}
D^R_{\sfsp,2-1}(\textbf{q})
(F_1 - F_2)
\left(
F_3
-
\frac{1}{F_{3-4}}
\right)
,
\end{aligned}
\end{equation}
\begin{equation}
\text{Fig. \ref{fig:vertices_SintI}(b)(iii)}
=
-
\lambda^{3/2} \Gamma_1
\delta_{1+3,2+4}
D^A_{\sfsp,2-1}(-\textbf{q})
,
\end{equation}
\begin{equation}
\text{Fig. \ref{fig:vertices_SintI}(b)(iv)}
=
\lambda^{3/2} \Gamma_1
\delta_{1+3,2+4}
D^A_{\sfsp,2-1}(-\textbf{q})
,
\end{equation}
\begin{equation}
\text{Fig. \ref{fig:vertices_SintII}(b)} 
=
\lambda^{3/2} \Gamma_2
\delta_{1+3,2+4}
\int\limits_{\textbf{k}}
D^A_{\sfsp,1-4}(\textbf{k})
,
\end{equation}
\begin{equation}
\text{Fig. \ref{fig:vertices_SintII}(c)}
=
-
\lambda^{3/2} \Gamma_2
\delta_{1+3,2+4}
\int\limits_{\textbf{k}}
D^R_{\sfsp,1-4}(\textbf{k})
,
\end{equation}
\begin{equation}
\begin{aligned}
\text{Fig. \ref{fig:vertices_SintII}(d)} 
&=
-
\lambda^{3/2} \Gamma_2
\delta_{1+3,2+4}
\\
&\times
\int\limits_{\textbf{k}}
\begin{bmatrix}
(F_1 - F_4) F_2 D^A_{\sfsp,1-4}(\textbf{k})
\\
-
(1 - F_1 F_4) D^R_{\sfsp,1-4}(\textbf{k})
\\
+
(F_1 - F_4) D^K_{\sfsp,1-4}(\textbf{k})
\end{bmatrix}
,
\end{aligned}
\end{equation}
\begin{equation}
\begin{aligned}
\text{Fig. \ref{fig:vertices_SintII}(e)} 
&=
-
\lambda^{3/2} \Gamma_2
\delta_{1+3,2+4}
\\
&\times 
\int\limits_{\textbf{k}}
\begin{bmatrix}
(1 - F_2 F_3)D^A_{\sfsp,1-4}(\textbf{k})
\\
+
F_1 ( F_3 - F_2) D^R_{\sfsp,1-4}(\textbf{k})
\\
-
(F_{3} - F_{2})
D^K_{\sfsp,1-4}(\textbf{k})
\end{bmatrix}
.
\end{aligned}
\end{equation}

The 
quartic 
ones are 
\begin{equation}
\label{eq:S_D_4_vertice_amp}
\text{
Fig. 
\ref{fig:vertices_S_D}(b)
}
=
-2\lambda 
\;
\Box_{1,2,3,4}^{\textbf{q}_1,\textbf{q}_2,\textbf{q}_3,\textbf{q}_4}
\;
\delta_{\textbf{q}_1 + \textbf{q}_3,\textbf{q}_2 + \textbf{q}_4},
\end{equation}
\begin{equation}
\begin{aligned}
&
\text{
Fig. 
\ref{fig:vertices_SintI}(c)(i)
}
\\
&=
-\frac{\lambda^2}{2}
\Gamma_1
\delta_{1+3,2+4}
D^R_{\sfsp,2-1}(\textbf{q})
\left(
\dfrac{1}{F_{3-4}} + F_4
\right),
\end{aligned}
\end{equation}

\begin{equation}
\begin{aligned}
&
\text{
Fig. 
\ref{fig:vertices_SintI}(c)(ii)
}
\\
&=
-\frac{\lambda^2}{2}
\Gamma_1
\delta_{1+3,2+4}
\begin{bmatrix}
D^R_{\sfsp,2-1}(\textbf{q})
\left(
-\dfrac{1}{F_{3-4}} + F_3
\right)
\\\\
-
D^R_{\sfsp,4-3}(\textbf{q})
\left(
\dfrac{1}{F_{1-2}}
+
F_2
\right)
\end{bmatrix}
,
\end{aligned}
\end{equation}

\begin{equation}
\begin{aligned}
&
\text{
Fig. 
\ref{fig:vertices_SintI}(c)(iii)
}
\\
&=
-\frac{\lambda^2}{2}
\Gamma_1
\delta_{1+3,2+4}
D^R_{\sfsp,2-1}(\textbf{q})
\left(
\dfrac{1}{F_{3-4}} - F_3
\right),
\end{aligned}
\end{equation}
where $\Box_{1,2,3,4}^{\textbf{q}_1,\textbf{q}_2,\textbf{q}_3,\textbf{q}_4}$ is defined in Eq.~(\ref{eq:S_D4_box}). 
In Eq.~(\ref{eq:S_D_4_vertice_amp}), we included a symmetry factor of $2$. 
We also made use of the fact that $D^R_{\sfsp,\Omega}(\textbf{q}) = D^R_{\sfsp,\Omega}(-\textbf{q})$.

\section{Semiclassical result for the density response function}
\label{sec:semi_classic}

We now evaluate the density linear response function using the NLsM derived above. 
Under the Keldysh response theory, the retarded density-density correlation function is defined as
\begin{equation}
\label{eq:def_density_resp}
	\pi^R(\Omega,\textbf{q})
	=
	-
	\frac{1}{2i}
	\frac{
	\delta^2 Z[V]
	}{
	\delta V^{\sfcl}_{\Omega}(\textbf{q})
	\delta V^{\sfq}_{-\Omega}(-\textbf{q})
	}
	\bigg|_{V^{\sfcl} = V^{\sfq} = 0}.
\end{equation}
Using the partition function in Eq.~(\ref{eq:Z_eff}), we have
\begin{equation}\label{eq:dens_resp_exp}
	\pi^R(\Omega,\textbf{q})
	=
	-N\nu_0
	+
	\pi^R_{\sfdyn}(\Omega,\textbf{q}),
\end{equation}
where the first term is the static contribution and the second one is the dynamical contribution given by
\begin{equation}\label{eq:dens_resp_exp_dyn}
\begin{aligned}
\pi^R_{\sfdyn}(\Omega,\textbf{q})
&=
i
\frac{
(2h)^2
}{2}
\int\limits_{1,2}
\left(
F_1 - F_{1+\Omega}
\right)
\\
&\times
\left\langle 
W_{ii;2+\Omega,2}(\textbf{q})
W_{jj;1,1 + \Omega}^{\dagger}(\textbf{q})
\right\rangle.
\end{aligned}
\end{equation}
The appearance of $N$ in the static term is due to $N$ flavors of fermions.

In this section, we consider the semiclassical contribution to the density response function to leading order of $1/N$, 
i.e.\ we only perform the calculation at the Gaussian level. If we ignore interaction corrections and just make use of the bare 
diffuson propagator in Eq.~(\ref{eq:free_W_propagator}), the dynamical density response function is
\begin{equation}
	\pi^{R(0)}_{\sfdyn}(\Omega,\textbf{q})
	\simeq
	N \nu_0
	\frac{
	-i h \lambda \Omega
	}{
	\textbf{q}^2 - ih\lambda \Omega - ih\lambda r_{\Omega}
	},
\end{equation}
where we can approximate to logarithmic accuracy that
\begin{equation}
\Delta^R_{2+\Omega,2}(\textbf{q})
\simeq
\frac{1}{
\textbf{q}^2
-
ih \lambda \Omega
-
ih \lambda r_{\Omega}
}
\equiv
\Delta^R_{\Omega}(\textbf{q}),
\end{equation} 
with
\begin{equation}
\label{eq:r_Omega}
r_{\Omega}
=
\gbar^2 \Omega \ln \frac{2\omega_c}{|\Omega|}.
\end{equation}

\begin{figure}
	\centering
	{\includegraphics[width=7cm]{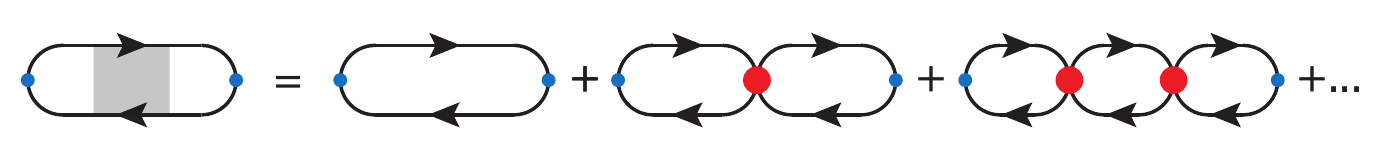} }%
	\caption{
	Diagrammatic representation for $ \pi_{\sfdyn}^{R}(\Omega,\textbf{q}) = \sum_{n = 0}^{\infty} \pi_{\sfdyn}^{R(n)}(\Omega,\textbf{q})$ [Eq.~(\ref{eq:pi_R_dyn_sum})].
	The $n$th term on the right hand side is denoted by $\pi_{\sfdyn}^{R(n)}$, which contains $n$ red dot(s) defined in Fig.~\ref{fig:vertices_SintII}(a). 
	The blue dots are scalar potential vertices defined in Fig.~\ref{fig:vertices_qV}(a). 
	}  	
	\label{fig:den_corr_semiclassic_ladder_sum}
\end{figure}

Due to the presence of the fermionic self-energy in the free propagator [Eq.~(\ref{eq:free_W_propagator})], 
the full density response function in Eq.~(\ref{eq:dens_resp_exp})
does not go to zero at $\textbf{q} \rightarrow 0$, violating particle conservation. 
To rectify this, one must take $S_{\sfintII}^{(2)}$ into account and consider the series of diagrams shown in 
Fig.~\ref{fig:den_corr_semiclassic_ladder_sum}. 
We note that a similar set of diagrams have to be summed in order to satisfy particle conservation in the context of 
clean non-Fermi and marginal Fermi liquids \cite{Ward1_Chubukov_PRB_05,Ward2_PALee_PRB_88,Ward3_PALee_PRB_86,Ward4_Varma_PRB_09}. 

Denote the $n$th term in the series as $\pi^{R(n)}_{\sfdyn}(\Omega,\textbf{q})$, 
which contains $n$ interaction vertices due to $S_{\sfintII}^{(2)}$. The dynamical part of the density response function is then 
\begin{equation}
\label{eq:pi_R_dyn_sum}
	\pi^R_{\sfdyn}(\Omega,\textbf{q})
	=
	\sum_{n = 0}^{\infty}
	\pi^{R(n)}_{\sfdyn}(\Omega,\textbf{q}).
\end{equation}
The $n=1$ term (second term on the RHS of Fig.~\ref{fig:den_corr_semiclassic_ladder_sum}) is 
\begin{equation}
\begin{aligned}
&
	\pi^{R(1)}_{\sfdyn}(\Omega,\textbf{q})
	\\
	&=
	2i N^2 \lambda
	h^2
	\left(
	2\lambda \Gamma_2
	\right)
\\
	&\times
	\int\limits_{1,2,\textbf{k}}
	\Delta_{1 + \Omega,1}^R(\textbf{q})
	\Delta_{2 + \Omega,2}^R(\textbf{q})
	(
	F_1 - F_{1 + \Omega}
	)
\\
	&\times
	\left[
	D^K_{\sfsp,2-1}(\textbf{k} )
	+
	F_{2+ \Omega}
	D^A_{\sfsp,2-1}(\textbf{k} )
	-
	F_2
	D^R_{\sfsp,2-1}(\textbf{k} )
	\right]
	.
\end{aligned}
\end{equation}
Performing frequency integrals and focusing on the $T = 0$ limit, we have
\begin{equation}
\begin{aligned}
	&
	\pi^{R(1)}_{\sfdyn}(\Omega,\textbf{q})
\\
	&
	\simeq
	2iN^2
	\lambda
	h^2
	\left(
	2\lambda \Gamma_2
	\right)
	\left[
	\Delta^R_{\Omega}(\textbf{q})
	\right]^2
\\
	&\times
	\int\limits_1 
	(
	F_1 - F_{\Omega + 1}
	)
	\int_0^{q_{\sfmax}^2}
	\frac{dx}{4\pi}
	\left(-
	\frac{1}{2\pi}
	\frac{2}{
	i\alpha
	}
	\right)
	\ln
	\left(
	\frac{
	x
	}{
	x - i\alpha \Omega/2 
	}
	\right)
\\
	&\simeq
	2iN \lambda^2
	h^2
	\left[
		\Delta^R_{\Omega}(\textbf{q})
	\right]^2
	\int\limits_1 
	(
	F_1 - F_{1 + \Omega }
	)
	(-ih r_{\Omega})
	,
\end{aligned}
\end{equation}
For $n > 1$, the diagrams can be evaluated in a similar fashion, resulting in the following geometric series
\begin{equation}
\label{eq:shaded_box}
\begin{aligned}
	&\pi^R_{\sfdyn}(\Omega,\textbf{q})
\\
	&=
	2iN \lambda
	h^2
	\int\limits_{1}
	(
	F_1 - F_{1 + \Omega}
	)
	\Delta_{\Omega}^R(\textbf{q})
	\left[
		1
		-
		ih \lambda r_{\Omega}
		\Delta_{\Omega}^R(\textbf{q})
		+
		\ldots
	\right]
\\
	&=
	2iN \lambda
	h^2
	\int\limits_{1}
	(
	F_1 - F_{1 + \Omega}
	)
	\Delta_{\sfFL,\Omega}^R(\textbf{q})
\\
&=
	\nu_0 N
	\frac{-ih \lambda \Omega}{
	\textbf{q}^2
	-
	ih \lambda \Omega
}
.
\end{aligned}
\end{equation}
where $\Delta_{\sfFL,\Omega}$ is defined in Eq.~(\ref{eq:diffson_FL}).
The dynamical term cancels the static one at $\textbf{q} = 0$ in Eq.~(\ref{eq:dens_resp_exp}), 
as required by the Ward identity. \emph{Despite the strongly dissipative MFL self-energy,
the semiclassical density-density response is purely diffusive, with the 
usual semiclassical diffusion constant defined via Eq.~(\ref{eq:def_lambda_h}).}

The conductivity can be obtained via the continuity relation 
\begin{equation}
	\sigma^R
	=
	\lim_{\textbf{q} \rightarrow 0}
	e^2
	\frac{i\Omega}{\textbf{q}^2}
	\pi^R(\Omega,\textbf{q})
	=
	e^2 ND \nu_0.
\end{equation}
We thus recover the semiclassical Drude conductivity at zero temperature.

In Appendix~\ref{app:no_lin_T}, we extend the above calculation
to finite temperature, and show that the temperature-dependent
contributions to the semiclassical resistivity from the MFL
self-energy and bosonic vertex corrections cancel exactly. 
This leads to $\rho(t) = \rho_{\sfDrude}$ independent of temperature, 
consistent with the results obtained in Ref.~\cite{SYK_Patel_linearT_arxiv_22}.
Linear-$T$ temperature dependence arises in our model only via the AA 
correction.

\section{Quantum interaction corrections to the density response function}
\label{sec:interaction_AA_corr}

We now examine the interaction correction due to quantum interference
to the density response function, analogous to the AA correction 
in disordered metals \cite{AA1_Altshuler_JETP_79,AA2_Altshuler_PRL_80,AA3_Aleiner_PRB_01,disO_review_PALee_85,AA5_N_flav_Finkel_Sci_05,AA6_N_flavor_Mirlin_PRL_10}. 
In the following, we first discuss the vertex correction to the $q^{(2)}$--$V$ vertices shown in Fig.~\ref{fig:vertices_qV}(b). 
We then study the effect of dynamical screening on the interaction. Finally, we present the Feynman diagrams responsible 
for the correction of density response, by taking into account both vertex corrections and the effect of dynamical screening.

\subsection{Vertex corrections to the $q^{(2)}$--$V$ coupling}
\label{sec:vertex_corr_q2V}

We now consider vertex corrections to the $q^{(2)}$--$V$ (matrix field--source) coupling, 
due to interactions. 
The bare vertices are captured by the last two lines of $S_{qV}$ in Eq.~(\ref{eq:S_qV}). 
Suppose the modified vertices are described by the following action
\begin{widetext}
\begin{equation}
\begin{aligned}
\label{eq:S_{q2V}_eff}
S_{q^{(2)}V}^{\mathsf{eff}}
=
ih
\lambda
\int\limits_{1-3,\Omega,\textbf{p},\textbf{q}}
\begin{Bmatrix}
V^{\sfcl}_{\Omega}(\textbf{q})
\begin{bmatrix}
(1 + \delta\Gamma_{V^{\sfcl}}^{(1)} )
W_{i,j;1,2} (\textbf{p}) \Wdag_{ji;2,1+\Omega}(\textbf{p} + \textbf{q})
\\
-
(1 + \delta \Gamma_{V^{\sfcl}}^{(2)})
\Wdag_{ij;1,2}(\textbf{p} + \textbf{q})
W_{ji;2,1+\Omega}(\textbf{p})
\end{bmatrix}
\\
+
V^{\sfq}_{-\Omega}(-\textbf{q})
\begin{bmatrix}
(1 + \delta \Gamma_{V^{\sfq}}^{(1)})
W_{ij;1+\Omega,2}(\textbf{p} + \textbf{q}) \Wdag_{ji;2,1} (\textbf{p})
F_{1}
\\
+
(1 + \delta \Gamma_{V^{\sfq}}^{(2)})
\Wdag_{ij;1+\Omega,2}(\textbf{p})
W_{ji;2,1}(\textbf{p} + \textbf{q})
F_{1+\Omega}
\end{bmatrix}
\end{Bmatrix},
\end{aligned}
\end{equation}
\end{widetext}
where $\delta \Gamma_{V^{\sfcl/\sfq}}^{(i)}$ are vertex corrections that we are going to evaluate below. 
To facilitate the calculation of the density correlation function in the next subsection, we explicitly define the frequency passing through $V^{\sfcl/\sfq}$ as $\pm \Omega$. 

\begin{figure}[b!]
	\centering
	{\includegraphics[width=8.5cm]{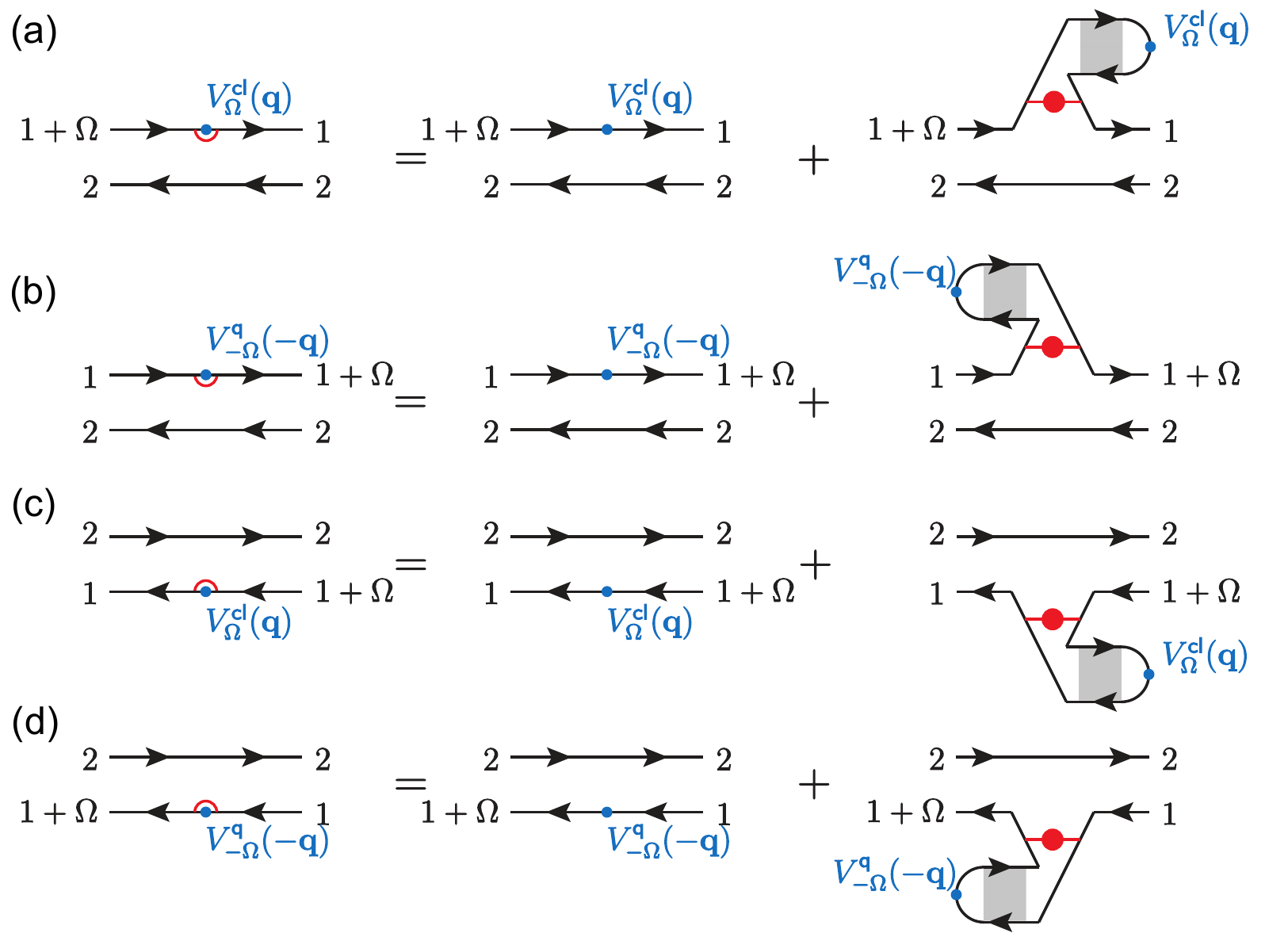} }%
	\caption{
	The effective coupling between $q^{(2)}$ and $V$ [Eq.~(\ref{eq:S_{q2V}_eff})].
	The blue dot represents external scalar potential $V^{\sfcl/\sfq}$. 
	The dressed vertex is indicated by an additional red semicircle. 
	In each subfigure, the first term on the right hand side is the bare coupling and the subsequent diagram is the vertex correction for 
	(a) $\delta \Gamma_{V^{\sfcl}}^{(1)}$, 
	(b) $\delta \Gamma_{V^{\sfq}}^{(1)}$, 
	(c) $\delta \Gamma_{V^{\sfcl}}^{(2)}$, and 
	(d) $\delta \Gamma_{V^{\sfq}}^{(2)}$. 
	The shaded box denotes the vertex correction demonstrated in Fig.~\ref{fig:den_corr_semiclassic_ladder_sum}.
	}  	
	\label{fig:vertex_corr_V_by_Sint2}
\end{figure}

We only consider the corrections originated from $S_{\sfintII}$, since they are not suppressed by $1/N$. 
The corresponding Feynman diagrams are shown in Fig.~\ref{fig:vertex_corr_V_by_Sint2}. 
In each panel, the first term on the right-hand-side is the bare coupling term [see Fig.~\ref{fig:vertices_qV}(b)] 
and the subsequent term is the corresponding vertex correction $\delta \Gamma^{(i)}_{V^{\sfcl/\sfq}}$. 
The shaded box in the diagrams denotes the vertex correction demonstrated in Fig.~\ref{fig:den_corr_semiclassic_ladder_sum}.

\subsubsection{Vertex correction $\delta \Gamma^{(i)}_{V^{\sfcl}}$}

The diagrams responsible for $\delta \Gamma^{(i)}_{V^{\sfcl}}$  are shown on the right-hand-side of Fig.~\ref{fig:vertex_corr_V_by_Sint2}(a). They are constructed using the vertices shown in Figs.~\ref{fig:vertices_SintII}(a) and (b). 
Using the corresponding Feynman rules, we find the leading order correction to be
\begin{equation}
\begin{aligned}
&\quad
\delta \Gamma_{V^{\sfcl}}^{(1)}
\\
&=
-
2N \Gamma_2 \lambda
\int\limits_{\varepsilon,\textbf{k}}
D^A_{\sfsp,\varepsilon-1}(\textbf{k})
(
F_{\varepsilon} - F_{\varepsilon+\Omega}
)
\Delta_{\sfFL,\varepsilon+\Omega,\varepsilon}^R(\textbf{q})
\\
&=
-
2N \Gamma_2 \lambda
\int\limits_{\varepsilon}
\frac{-1}{8\pi}
\ln \frac{\omega_c}{i  (\varepsilon-\omega_1)}
(
F_{\varepsilon} - F_{\varepsilon+\Omega}
)
\Delta_{\sfFL,\Omega}^R(\textbf{q})
\\
&\simeq
-ih \lambda \gbar^2 \Omega
\ln \frac{\omega_c}{| \Omega/2+\omega_1|}
\Delta_{\sfFL,\Omega}^R(\textbf{q})
+
\ldots,
\end{aligned}
\end{equation}
where $\ldots$ are subleading terms. 
Here, we have made use of the result in Eq.~(\ref{eq:shaded_box}).   
{The modified coupling is thus
\begin{equation}
\begin{aligned}
1
+
\delta \Gamma_{V^{\sfcl}}^{(1)}
&\simeq
\left(
\textbf{q}^2
-ih\lambda \Omega
-ih \lambda \gbar^2 \Omega
\ln \frac{\omega_c}{|\omega_1|}
\right)
\Delta_{\sfFL,\Omega}^R(\textbf{q})
.
\end{aligned}
\end{equation}

The correction $\delta \Gamma_{V^{\sfcl}}^{(2)}$ can be computed in a similar fashion by considering the Feynman diagrams shown in Fig.~\ref{fig:vertex_corr_V_by_Sint2} (b), which involve  the vertices shown in Figs.~\ref{fig:vertices_SintII}(a) and (c). 
Following the lines of calculations as above, we find that
\begin{equation}
\delta \Gamma_{V^{\sfcl}}^{(2)}
=
\delta \Gamma_{V^{\sfcl}}^{(1)},
\end{equation}
as expected.

\subsubsection{Vertex correction $\delta \Gamma^{(i)}_{V^{\sfq}}$}

Next, we consider the correction $\delta \Gamma_{V^{\sfq}}^{(1)}$, which is diagrammatically described in Figs.~\ref{fig:vertex_corr_V_by_Sint2}(b). 
The diagrams are constructed using the vertices shown in Figs.~\ref{fig:vertices_SintII}(a) and (d). 
By restricting our attention to terms with the same thermal factor ($F_{1}$) as the bare coupling, we find 
\begin{equation}
\begin{aligned}
&
\delta \Gamma_{V^{\sfq}}^{(1)}
=
-2N \Gamma_2
\lambda
\int\limits_{\varepsilon,\textbf{k}}
\Delta_{\sfFL,{\varepsilon}+\Omega,\varepsilon}^R(\textbf{q})
\\
&\times
\bigg[
-
F_{ \varepsilon + \Omega}
D^A_{\sfsp,\varepsilon-1}(\textbf{k})
+
F_{\varepsilon}
D^R_{\sfsp,\varepsilon-1}(\textbf{k})
\bigg]
\\
&=
-2N\Gamma_2\lambda
\int\limits_{\varepsilon}
\frac{-1}{8\pi}
\ln \frac{\omega_c}{|\varepsilon - \omega_1|}
(
F_{\vareps}
-
F_{\vareps + \Omega}
)
\Delta^R_{\sfFL,\Omega}(\textbf{q})
+
\ldots
\\
&\simeq
-ih \lambda \gbar^2 \Omega
\ln \frac{\omega_c}{| \Omega/2 + \omega_1|}
\Delta_{\sfFL,\Omega}^R(\textbf{q}).
\end{aligned}
\end{equation}
Hence, the modified coupling is again
\begin{equation}
1
+
\delta \Gamma_{V^{\sfq}}^{(1)}
\simeq
\left(
\textbf{q}^2
-ih\lambda \Omega
-ih \lambda \gbar^2 \Omega
\ln \frac{\omega_c}{|\omega_1|}
\right)
\Delta_{\sfFL,\Omega}^R(\textbf{q})
.
\end{equation}

Lastly, for $\delta \Gamma_{V^{\sfq}}^{(2)}$, we have to consider a similar set of diagrams shown in 
Fig.~\ref{fig:vertex_corr_V_by_Sint2}(d) which are constructed with the vertices in Figs.~\ref{fig:vertices_SintII}(a) and (e). 
This time, we focus on terms with thermal factor $F_{1+\Omega}$, as in the bare coupling. 
Evaluating the diagrams explicitly, we find 
\begin{equation}
\delta \Gamma_{V^{\sfq}}^{(2)}
=
\delta \Gamma_{V^{\sfq}}^{(1)}, 
\end{equation}
which is again what we expect.

\begin{figure}[t!]
	\centering
	{\includegraphics[width=6cm]{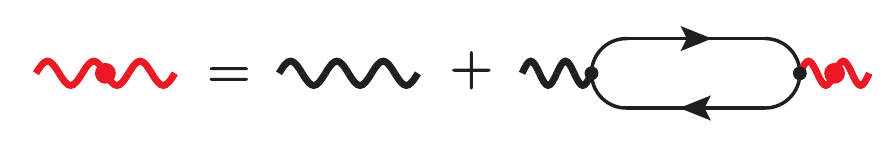} }%
	\caption{
	Diagrammatic representation of the dynamically screened bosonic propagator $D^R_{\sfscr}$ under random phase approximation (RPA). 
	The black wavy line (red wavy line with a dot) represents the interaction without (with) dynamical screening.  
	}  	
	\label{fig:dyn_screen_boson}
\end{figure}

\begin{figure}[b!]
	\centering
	{\includegraphics[width=8cm]{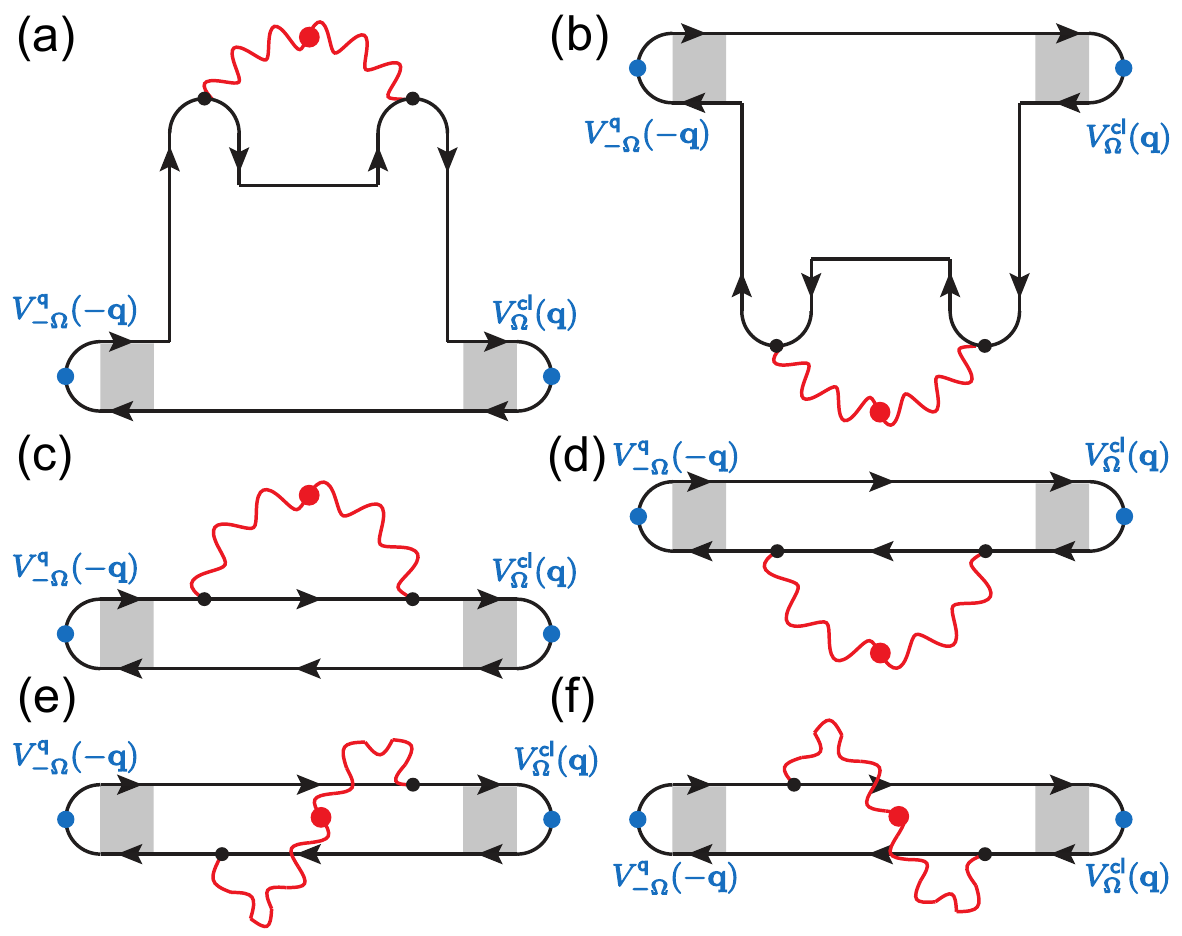} }%
	\caption{
	The set of type-A diagrams contributing to the interaction correction of the density response function to leading order of $1/N$ and $\lambda$.
	The blue dots are vertices between the diffuson and scalar potential. 
	The red wavy line is the dynamically screened interaction [Fig.~\ref{fig:dyn_screen_boson}]. 
	The gray box represents the vertex correction shown in Fig.~\ref{fig:den_corr_semiclassic_ladder_sum}. 
	}  	
	\label{fig:AA_Hikami_Fock}
\end{figure}

\subsection{Dynamical screening}

As in the case of disordered Fermi liquid, dynamical screening is crucial for evaluating the interaction correction 
to the density response and conductivity \cite{AA1_Altshuler_JETP_79,AA2_Altshuler_PRL_80,NLsM1_Ludwig_PRB_99,NLsM2_Matt_Yun_Ann_17,NLsM6_Kamenev_CUP_11},
particularly in the temperature $T \rightarrow 0$ limit.  
In the following, we treat dynamical screening under the random phase approximation (RPA). 
 The dynamically screened bosonic propagator is represented by a red wavy line in Fig.~\ref{fig:dyn_screen_boson}.
This effectively modifies the bosonic propagator in the quartic vertices in Fig.~\ref{fig:vertices_SintI}(c).  
Such vertex corrections can alternatively be obtained by joining the cubic and quadratic vertices in Figs.~\ref{fig:vertices_SintI}(a)--(b).
Since the bosons are quantum critical, mass terms in the self-energy can be fine-tuned to zero at $T = 0$. 
We therefore throw away the constant term $\sim \nu_0$, which is generated from terms $\sim G^R G^R + G^A G^A$ during the gradient expansion
\cite{NLsM2_Matt_Yun_Ann_17,NLsM6_Kamenev_CUP_11}, in the polarization bubble [as already discussed at the saddle-point level, 
below Eq.~(\ref{eq:PiR_saddle_pt})].

Explicitly, the retarded bosonic self-energy due to dynamical screening at $T = 0$ is
\begin{equation}
\label{eq:Pi_scr}
\begin{aligned}
\Pi_{\sfscr}^R(\Omega,\textbf{q})
&=
\Gamma_1
\lambda
\int\limits_{1}
\Delta^R_{1+\Omega,1}(\textbf{q})
(F_{1} - F_{1 + \Omega})
\\
&
\simeq
-
\frac{2i}{\pi N}
\frac{
h^2  g^2 \lambda 
\Omega
}{
\textbf{q}^2 - ih \lambda \Omega  - ih \lambda r_{\Omega}
}
.
\end{aligned}
\end{equation}
The retarded component of the screened bosonic propagator under RPA is then
\begin{equation}
\label{eq:DR_dyn_scn}
\begin{aligned}
D^R_{\sfscr}(\Omega,\textbf{q})
&=
[(D^R_{\sfsp})^{-1}(\Omega,\textbf{q})
-
\Pi_{\sfscr}^R(\Omega,\textbf{q})]^{-1}
\\
&=
-
\frac{
1
}{
2(\textbf{q}^2 + \mb^2 - i\alpha \Omega) + \Pi^R_{\sfscr}(\Omega,\textbf{q})
}
.
\end{aligned}
\end{equation}

\subsection{Feynman diagrams for the interaction correction}

We divide the leading order (in $\lambda$ and $1/N$) interaction correction  to the retarded density response into two types, A and B, 
\begin{equation}
\delta \pi^R(\Omega,\textbf{q})
=
\delta \pi^R_{\text{type A}}(\Omega,\textbf{q})
+
\delta \pi^R_{\text{type B}}(\Omega,\textbf{q})
,
\end{equation}
which are diagrammatically depicted in Figs.~\ref{fig:AA_Hikami_Fock}--\ref{fig:AA_donut}. 
The type-A diagrams [Fig.~\ref{fig:AA_Hikami_Fock}] involve contributions from the self-energy correction of the diffuson. 
Meanwhile, the type-B diagrams [Fig.~\ref{fig:AA_WFR}] are analogous to the wavefunction renormalization correction in the context of 
the disordered FL \cite{NLsM2_Matt_Yun_Ann_17}. 
They are obtained by considering the vertices in Figs.~\ref{fig:vertices_qV}(a)--(b). 

In both the type A and B diagrams, vertex corrections and the effect of dynamical screening have been taken into account. 
In particular, the red wavy line represents the dynamically screened interaction in Fig.~\ref{fig:dyn_screen_boson}. 
On the other hand, the shaded diffuson denotes the vertex correction shown in Fig.~\ref{fig:den_corr_semiclassic_ladder_sum}, 
while the scalar potential vertices with a red semicircle are the vertex corrections considered in Fig.~\ref{fig:vertex_corr_V_by_Sint2}.  
Note, however, that two of the diffusons in Figs.~\ref{fig:AA_Hikami_Fock}(a)--(b) and the diffusons in the middle of the interaction line in 
Figs.~\ref{fig:AA_Hikami_Fock}(c)--(f) and \ref{fig:AA_WFR} are not shaded since those corrections are suppressed by $1/N$.

\begin{figure}[b!]
	\centering
	{\includegraphics[width=8.5cm]{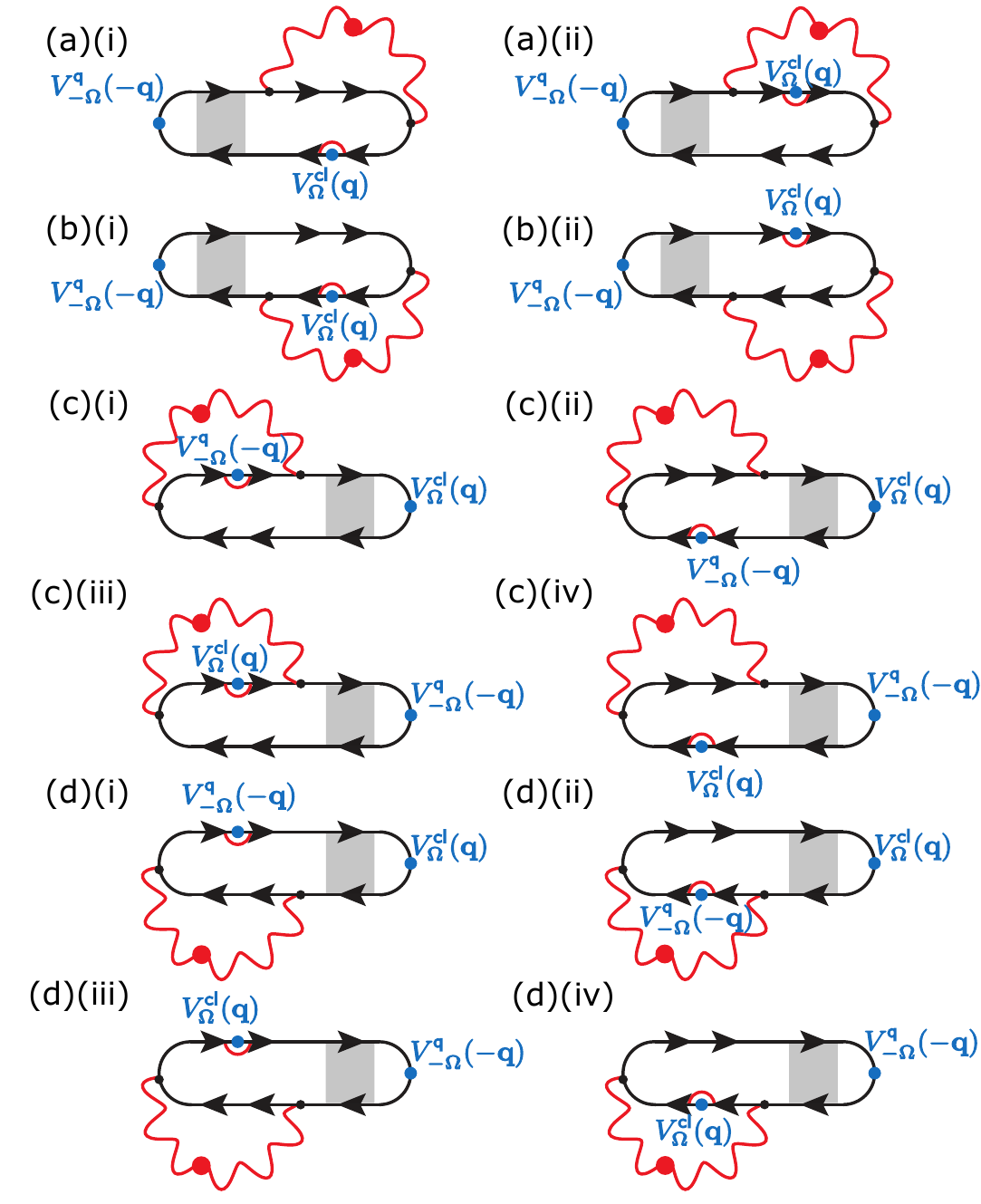} }%
	\caption{
	The set of type-B diagrams contributing to the interaction correction of the density response function to leading order of $1/N$ and $\lambda$.
	The convention here is the same as Fig.~\ref{fig:AA_Hikami_Fock}. 
	The red semicircle denotes the vertex corrections shown in Fig.~\ref{fig:vertex_corr_V_by_Sint2}. 
	}  	
	\label{fig:AA_WFR}
\end{figure}

\begin{figure}[b!]
	\centering
	{\includegraphics[width=9cm]{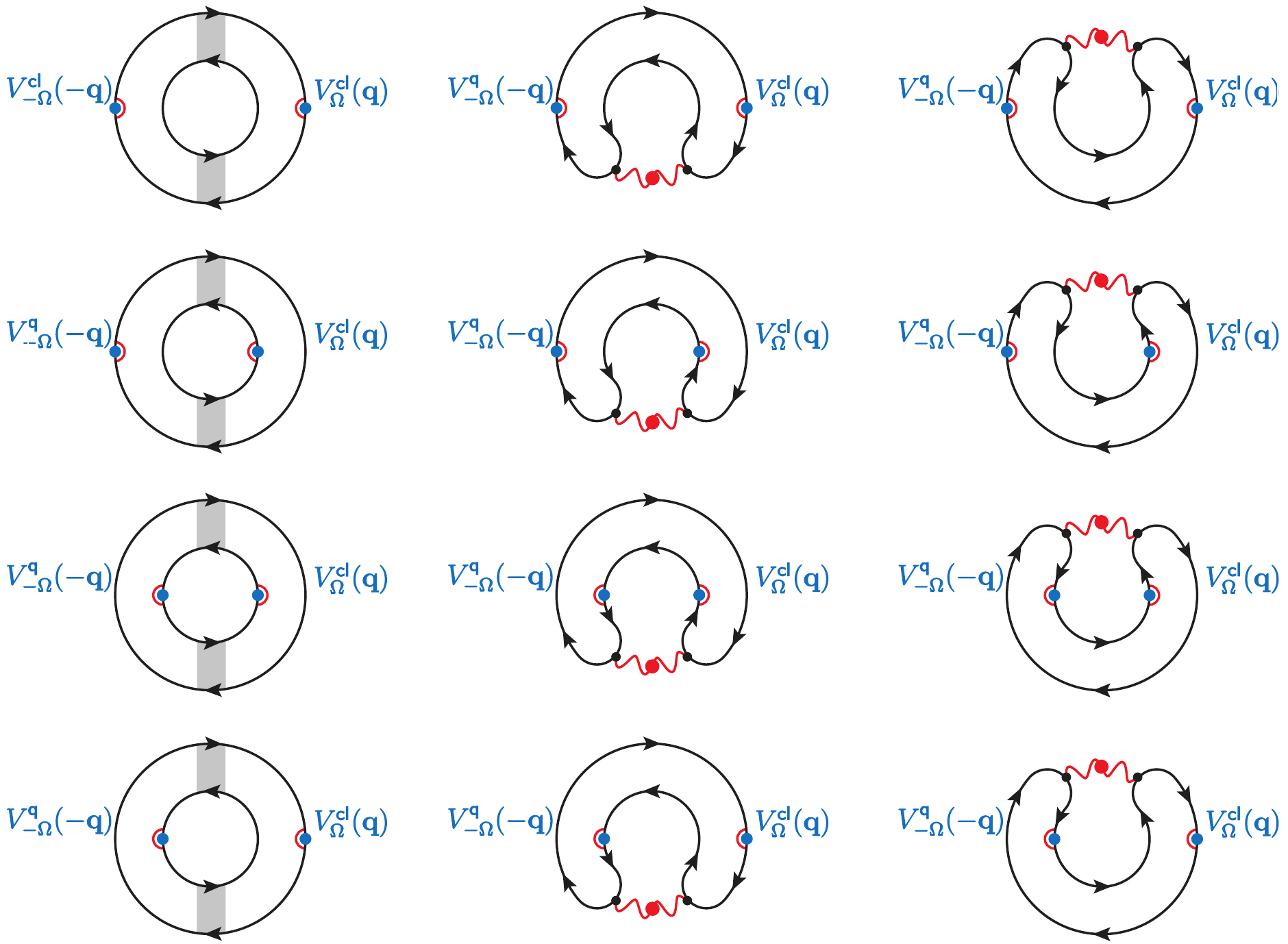} }%
	\caption{
	The set of type-C diagrams for the interaction correction of the density response function to leading order of $1/N$ and $\lambda$. 
	They are summed to zero.
	}  	
	\label{fig:AA_donut}
\end{figure}

Finally, there is an additional set of diagrams [Fig.~\ref{fig:AA_donut}] formed with the purely quadratic vertices in Fig.~\ref{fig:vertices_qV}(b). 
However, they are summed to zero and thus we do not discuss them further. These ``donut'' diagrams are important 
in non-standard universality classes \cite{NLsM2_Matt_Yun_Ann_17}.

\clearpage

\begin{widetext}

\subsubsection{Type-A diagrams}

After summing the diagrams in Fig.~\ref{fig:AA_Hikami_Fock} and using the definition in Eq.~(\ref{eq:def_density_resp}), we find
\begin{equation}
\begin{aligned}
\label{eq:AA_typeA}
	&\delta \pi^R_{\text{type A}}(\Omega,\textbf{q})
	=
	- \frac{(2h)^2 \lambda}{2i}
	\int\limits_{\omega}
	\left(
	F_{\omega}
	-
	F_{\omega + \Omega}
	\right)
	[\Delta_{\sfFL;\Omega}^R(\textbf{q})]^2
	\begin{Bmatrix}
	\nsum_{t=a,b,c,d}
	[\Sigma_{W;i,j;j,i}^{\ref{fig:AA_Hikami_Fock}(t)}(\textbf{q})]_{\omega+\Omega,\omega;\omega,\omega+\Omega}
	\delta_{ij}
	\\\\
	+
	\nsum_{t=e,f}
	[\Sigma_{W;i,i;j,j}^{\ref{fig:AA_Hikami_Fock}(t)}(\textbf{q})]_{\omega+\Omega+\xi,\omega+\xi;\omega,\omega+\Omega}
	\end{Bmatrix}
	,
\end{aligned}
\end{equation}
where 
$ \Delta_{\sfFL;\Omega}^R(\textbf{q})$ is given by Eq.~(\ref{eq:diffson_FL}) and 
$\Sigma_W$ represents the self-energy of the matrix field $\hW$. 
The subscripts in $\Sigma_W$ indicate the flavor indices and the external frequencies. 
Meanwhile, the superscript in it indicates the correspondence to the diagrams in Fig.~\ref{fig:AA_Hikami_Fock}. 
In particular, the frequency-diagonal self-energies are given by
\begin{equation}
\begin{aligned}
	[\Sigma_{W;i,j;j,i}^{\ref{fig:AA_Hikami_Fock}(a)}(\textbf{q})]_{\omega+\Omega,\omega;\omega,\omega+\Omega}
	&=
	-\frac{\lambda^2}{2} N\Gamma_1
	\int\limits_{\xi,\textbf{k}}
	(F_{\omega + \Omega - \xi} - F_{\omega + \Omega})
	D^R_{\sfscr,\xi}(\textbf{k})
	\begin{Bmatrix}
	[\Delta^R_{\omega+\Omega,\omega}(\textbf{q})]^{-1}
	[\Delta^R_{\omega+\Omega,\omega+\Omega-\xi}(\textbf{k})]^2
	\\
	+
	\Delta^R_{\omega+\Omega,\omega+\Omega-\xi}(\textbf{k})
	\end{Bmatrix},
\end{aligned}
\end{equation}
\begin{equation}
\begin{aligned}
	[\Sigma_{W;i,j;j,i}^{\ref{fig:AA_Hikami_Fock}(b)}(\textbf{q})]_{\omega+\Omega,\omega;\omega,\omega+\Omega}
	&=
	-\frac{\lambda^2}{2} N  \Gamma_1
	\int\limits_{\xi,\textbf{k}}
	(F_{\omega} - F_{\omega + \xi})
	D^R_{\sfscr,\xi}(\textbf{k})
	\begin{Bmatrix}
	[\Delta^R_{\omega+\Omega,\omega}(\textbf{q})]^{-1}
	[\Delta^R_{\omega+\xi,\omega}(\textbf{k})]^2
	\\
	+
	\Delta^R_{\omega+\xi,\omega}(\textbf{k})
	\end{Bmatrix}
	,
\end{aligned}
\end{equation}
\begin{equation}
\begin{aligned}
	[\Sigma_{W;i,j;j,i}^{\ref{fig:AA_Hikami_Fock}(c)}(\textbf{q})]_{\omega+\Omega,\omega;\omega,\omega+\Omega}
	&=
	- \frac{\lambda^2}{2} N \Gamma_1
	\int\limits_{\xi,\textbf{k}}
	\Delta^R_{\omega + \Omega + \xi,\omega}(\textbf{k}+\textbf{q})
	\begin{bmatrix}
	D^R_{\sfscr,\xi}(\textbf{k})
	\left(
	\dfrac{1}{F_{\xi}}
	+
	F_{\omega + \Omega}
	\right)
	\\	
	+
	D^A_{\sfscr,\xi}(\textbf{k})
	\left(
	-\dfrac{1}{F_{\xi}}
	+
	F_{\omega + \Omega+\xi}
	\right)
	\end{bmatrix}
	,
\end{aligned}
\end{equation}
and
\begin{equation}
\begin{aligned}
	[\Sigma_{W;i,j;j,i}^{\ref{fig:AA_Hikami_Fock}(d)}(\textbf{q})]_{\omega+\Omega,\omega;\omega,\omega+\Omega}
	&=
	- \frac{\lambda^2}{2} N \Gamma_1
	\int\limits_{\xi,\textbf{k}}
	\Delta^R_{\omega + \Omega, \omega - \xi}(\textbf{k}+\textbf{q})
	\begin{bmatrix}
	D^R_{\sfscr,\xi}(\textbf{k})
	\left(
	\dfrac{1}{F_{\xi}}
	-
	F_{\omega}
	\right)
	\\
	+
	D^A_{\sfscr,\xi}(\textbf{k})
	\left(
	-\dfrac{1}{F_{\xi}}
	+
	F_{\xi - \omega}
	\right)
\end{bmatrix}
.
\end{aligned}
\end{equation}
On the other hand, the frequency-off-diagonal self-energies are
\begin{equation}
\begin{aligned}
	[\Sigma_{W;i,i;j,j}^{\ref{fig:AA_Hikami_Fock}(e)}(\textbf{q})]_{\omega+\Omega+\xi,\omega+\xi;\omega,\omega+\Omega}
	&=
	- \frac{\lambda^2}{2} \Gamma_1
	\int\limits_{\xi,\textbf{k}}
	\Delta^R_{\omega + \Omega,\omega - \xi}(\textbf{k}+\textbf{q})
	\begin{bmatrix}
	D^R_{\sfscr,\xi}(\textbf{k})
	\left(
	-\dfrac{1}{F_{\xi}}
	-
	F_{\omega + \Omega - \xi}
	\right)
	\\
	+
	D^A_{\sfscr,\xi}(\textbf{k})
	\left(
	\dfrac{1}{F_{\xi}}
	-
	F_{\xi-\omega}
	\right)
	\end{bmatrix}
\end{aligned}
\end{equation}
and
\begin{equation}
\begin{aligned}
	[\Sigma_{W;i,i;j,j}^{\ref{fig:AA_Hikami_Fock}(f)}(\textbf{q})]_{\omega+\Omega+\xi,\omega+\xi;\omega,\omega+\Omega}
	&=
	- \frac{\lambda^2}{2} \Gamma_1
	\int\limits_{\xi,\textbf{k}}
	\Delta^R_{\omega + \Omega + \xi,\omega}(\textbf{k}+\textbf{q})
	\begin{bmatrix}
	D^R_{\sfscr,\xi}(\textbf{k})
	\left(
	-\dfrac{1}{F_{\xi}}
	+
	F_{\omega + \xi}
	\right)
	\\
	+
	D^A_{\sfscr,\xi}(\textbf{k})
	\left(
	\dfrac{1}{F_{\xi}}
	-
	F_{\omega + \Omega  + \xi}
	\right)
	\end{bmatrix}
.
\end{aligned}
\end{equation}
Here, the momentum ($\textbf{k}$) integrals are performed over the whole space and the frequency ($\xi$) integrals 
are bounded in the ultraviolet by $\Lambda = \gammael$. We will perform integrals to logarithmic accuracy in the 
ultraviolet cutoff $\Lambda$ by expanding in the external frequency $\Omega$ and momentum $\textbf{q}$.

\subsubsection{Type-B diagrams}

The evaluation of the type-B diagrams in Fig.~\ref{fig:AA_WFR} is straightforward yet tedious. 
We leave the detailed expressions of the diagrams to Appendix \ref{app:typeB_diag}. 
In the large-$N$ limit, the total contribution of the type-B diagrams to the density response can be approximated as
\begin{equation}
\begin{aligned}
\label{eq:AA_typeB}
	\delta \pi^R_{\text{type B}}(\Omega,\textbf{q})
	&=
	- \frac{C}{2i}
	\int\limits_{\xi,\omega,\textbf{k}}
	\Delta_{\sfFL;\Omega}^R(\textbf{q})
	[\Delta_{\xi,0}^R(\textbf{k})]^2
	D_{\sfscr,\xi}^R(\textbf{k})
	\begin{bmatrix}
	\left(
	F_{\omega + \xi}
	+
	F_{\omega + \Omega}
	-
	F_{\omega}
	-
	F_{\omega + \Omega - \xi}
	\right)
	\left(
	F_{\omega}
	-
	F_{\omega + \Omega}
	\right)
	\Phi_1
	\\
	+
	\left(
	-
	F_{\omega + \xi - \Omega}
	+
	F_{\omega + \xi}
	+
	F_{\omega + \Omega}
	-
	F_{\omega}
	\right)
	\left(
	F_{\omega}
	-
	F_{\omega + \xi}
	\right)
	\Phi_2
	\end{bmatrix}
\end{aligned}
\end{equation}
where
$ C 
= 
-2h^2 \lambda^3 \Gamma_1 N^2
=
- 4i N
h^4
 g^2 \lambda^3
$,
\begin{equation}
\Phi_1(\Omega,\xi,\textbf{q})
=
\left(
\textbf{q}^2 - ih \lambda \Omega - ih\lambda \gbar^2\Omega \ln \frac{\omega_c}{|\xi|}
\right)
\Delta^R_{\sfFL;\Omega}(\textbf{q})
,
\end{equation}
and
\begin{equation}
\Phi_2(\Omega,\textbf{q})
=
\left(
\textbf{q}^2 - ih \lambda \Omega - ih\lambda \gbar^2\Omega \ln \frac{\omega_c}{|\Omega/2|}
\right)
\Delta^R_{\sfFL;\Omega}(\textbf{q})
.
\end{equation}
The factor $\Phi_i$ is due to the vertex corrections discussed in Sec.~\ref{sec:vertex_corr_q2V}.
The first line in the big square parenthesis is originated from the diagrams in Figs.~\ref{fig:AA_WFR}(a)--(b), 
while the second line in it is originated from the diagrams in Figs.~\ref{fig:AA_WFR}(c)--(d). 
Here, we only retain terms in leading order of external frequency $\Omega$ and momentum $\textbf{q}$.

\subsubsection{Total contribution}

After summing the expressions in Eqs.~(\ref{eq:AA_typeA}) and (\ref{eq:AA_typeB}) in leading order of $\textbf{q}$ and $\Omega$, we have
\begin{equation}
\begin{aligned}
\label{eq:sum_AA}
	\delta \pi^R(\Omega,\textbf{q})
	=
	\frac{4C}{2i} 
	\textbf{q}^2
	\,
	[\Delta_{\sfFL;\Omega}^R(\textbf{q})]^2
	\int\limits_{\xi,\textbf{k}}
	(
	-
	B_{\xi - \Omega}
	+
	B_{\xi}
	)
	\,
	\textbf{k}^2 
	\,
	[\Delta^R_{\xi,0}(\textbf{k})]^3
	D^R_{\sfscr,\xi}(\textbf{k})
	,
\end{aligned}
\end{equation}
where 
\begin{equation}
B_{\omega}
=
\frac{\omega}{\pi}
\coth \frac{\omega}{2T}
=
\int\limits_{\omega'}
(
1
-
F_{\omega + \omega'}
F_{\omega'}
)
.
\end{equation}
Note that $\delta \pi^R(\Omega,\textbf{q}\rightarrow 0 ) = 0$, as required by the Ward identity. 
This is further supported with our numerics in Appendix \ref{app:numerical_Ward}. 
This shows that the vertex corrections we considered are essential for obtaining the correct density response function, 
due to the anomalous MFL self-energy term in the bare diffuson, which arises from the interplay between disorder and interactions 
with the quantum-critical collective modes. Without such corrections, particle conservation would be violated.

The interaction correction to conductivity can be found using the continuity relation
\begin{equation}
\label{eq:AA_density_to_cond}
	\sigma_{\sfAA}(\Omega)
	=
	\lim_{\textbf{q} \rightarrow 0}
	e^2\frac{i\Omega}{\textbf{q}^2}
	\delta \pi^R(\Omega,\textbf{q}).
	\end{equation} 
In the dc limit $\Omega \rightarrow 0 $, we find 
\begin{equation}
\label{eq:AA_delta_sig_formal}
\begin{aligned}
&
	\sigma_{\sfAA,\sfdc}
	=
	2ie^2 g^2 \lambda (2h)^2  N
	\int\limits_{\xi,\textbf{k}}
	\frac{\partial}{\partial \xi}
	\left(
	\frac{\xi}{\pi}
	\coth \frac{\xi}{2T}
	\right)
	\textbf{k}^2 
	[\Delta^R_{\xi,0}(\textbf{k})]^3
	D^R_{\sfscr,\xi}(\textbf{k}).
\end{aligned}
\end{equation}
This is the main result of this section. 
In the FL limit, Eq.~(\ref{eq:AA_delta_sig_formal}) reduces to the well-known formula for 
the AA correction when the anomalous diffuson propagator is replaced by the normal one
\cite{AA1_Altshuler_JETP_79,AA2_Altshuler_PRL_80,AA3_Aleiner_PRB_01,NLsM6_Kamenev_CUP_11}.

\subsection{Evaluation of the AA correction \label{Sec:AAEval}}
\subsubsection{Finite temperature, dc limit}

Here we compute the AA correction in Eq.~(\ref{eq:AA_delta_sig_formal}).
At temperatures $T \gtrsim g^2/N$, we can ignore the effects of dynamical screening in the boson propagator
$D^R_{\sfscr,\xi}(\textbf{k})$. 
Then Eq.~(\ref{eq:AA_delta_sig_formal}) takes the form 
\begin{align}
\label{eq:AA-Eval1}
	\sigma_{\sfAA,\sfdc}
	=&\,
	- 
	\frac{N i \lambda (2 e h g)^2}{\pi}
	\int\limits_{\xi}
	\tilde{F}(\xi)
	\int_0^\infty
	\frac{d x}{4 \pi}
	\frac{x}{\left\{x - i h \lambda\left[\xi - \Sigma^R_{\sfsp}(\xi)\right]\right\}^3\left\{x - i \alpha \xi + \mb^2\right\}},
\end{align}
where we have changed variables to $x \equiv k^2$, 
$\Sigma^R_{\sfsp}(\xi)$ is the retarded MFL fermion self-energy at finite temperature [Eqs.~(\ref{eq:Sig_MFL_intro}) and (\ref{eq:Sig_sp_final})],
and the thermal factor is
\begin{align}\label{tFDef}
	\tilde{F}(\xi)
	=
	\frac{\partial}{\partial \xi}
	\left[
	\xi
	\coth \left(\frac{\xi}{2T}\right)
	\right]
	\simeq
	\tanh\left(\frac{\xi}{3 T}\right).
\end{align}
Performing the squared-momentum $x$-integral in Eq.~(\ref{eq:AA-Eval1}), 
we can cast the AA correction as follows:
\begin{align}
\label{eq:AA-Eval2}
	\sigma_{\sfAA,\sfdc}
	=&\,
	\frac{N e^2}{2 \pi^2}
	\left(\frac{g^2}{T}\right)
	\mathcal{G}_{AA},
	\qquad
	T \gtrsim g^2/N,
\end{align}
where
\begin{align}\label{GAADef}
	\mathcal{G}_{AA}
	=
	\left(\frac{h}{\pi \alpha_m}\right)
	\mathcal{I}_1\left[D \alpha_m,D \alpha,\gbar^2,\log\left(\frac{\omega_c}{T}\right)\right],
\end{align}
and the full expression for the kernel $\mathcal{I}_1$ is given by Eq.~(\ref{AA--Ikernel}) in Appendix \ref{app:AA_integral}.
This kernel depends on dimensionless combinations of 
the semiclassical diffusion constant $D$ [Eq.~(\ref{eq:def_lambda_h})],
the boson inverse-diffusion constant $\alpha$ [Eqs.~(\ref{eq:PiR_saddle_pt}) and (\ref{eq:DR_saddle_pt})],
and 
the boson thermal mass coefficient $\alpha_m$ [Eq.~(\ref{eq:mb_saddle_pt})]. 
Through the MFL self-energy [Eq.~(\ref{eq:Sig_sp_final})], it also depends
separately on $\gbar^2$ and the temperature (via the logarithm). 
Eq.~(\ref{eq:AA-Eval2}) gives the second term in the square brackets 
in Eq.~(\ref{eq:resist_dc}) in the Introduction.

A simple limiting case takes $\alpha = 0$ (static, massive boson) 
and $\gbar^2 = 0$ (the Fermi liquid case, ignoring the MFL self-energy). 
Then 
\begin{align}\label{AA--FLKernel}
	\mathcal{I}_1\left(D \alpha_m\right)
	=
	\int_0^\infty
	d y 
	\,
	\tanh\left(\frac{y D \alpha_m}{3}\right)
	\left[
		\frac{(1 - 3 y^2)(y^2 + 1 - \pi y) + 2 y(y^3 - 3 y)\ln(y)}{y(y^2+1)^3}
	\right].
\end{align}
This kernel is independent of $N$ and temperature $T$. The integration is finite,
with a result that is logarithmic in $D \alpha_m$ for values of this parameter much larger
than one. This is the usual logarithm associated to AA corrections in 2D \cite{AA4_Altshuler_Review_85,AAG99}. 
Here there is no explicit dependence on a UV cutoff, because we have retained the 
formally irrelevant $k^2$ momentum-dependence of the boson propagator [Eq.~(\ref{eq:DR_saddle_pt}) with $\alpha = 0$]. 
\end{widetext}

In the general case with nonzero $\alpha$, retaining the MFL self-energy, the kernel $\mathcal{I}_1$
exhibits weak temperature dependence.
In Figs.~\ref{fig:GAA_z_plot1}--\ref{fig:GAA_T_plot} we plot $\mathcal{G}_{AA} =  \mathcal{I}_1$ (with $h/\pi \alpha_m \rightarrow 1$),
comparing the result to the static, Fermi liquid case in Eq.~(\ref{AA--FLKernel}).

\begin{figure}
	\centering
	{\includegraphics[width=0.4\textwidth]{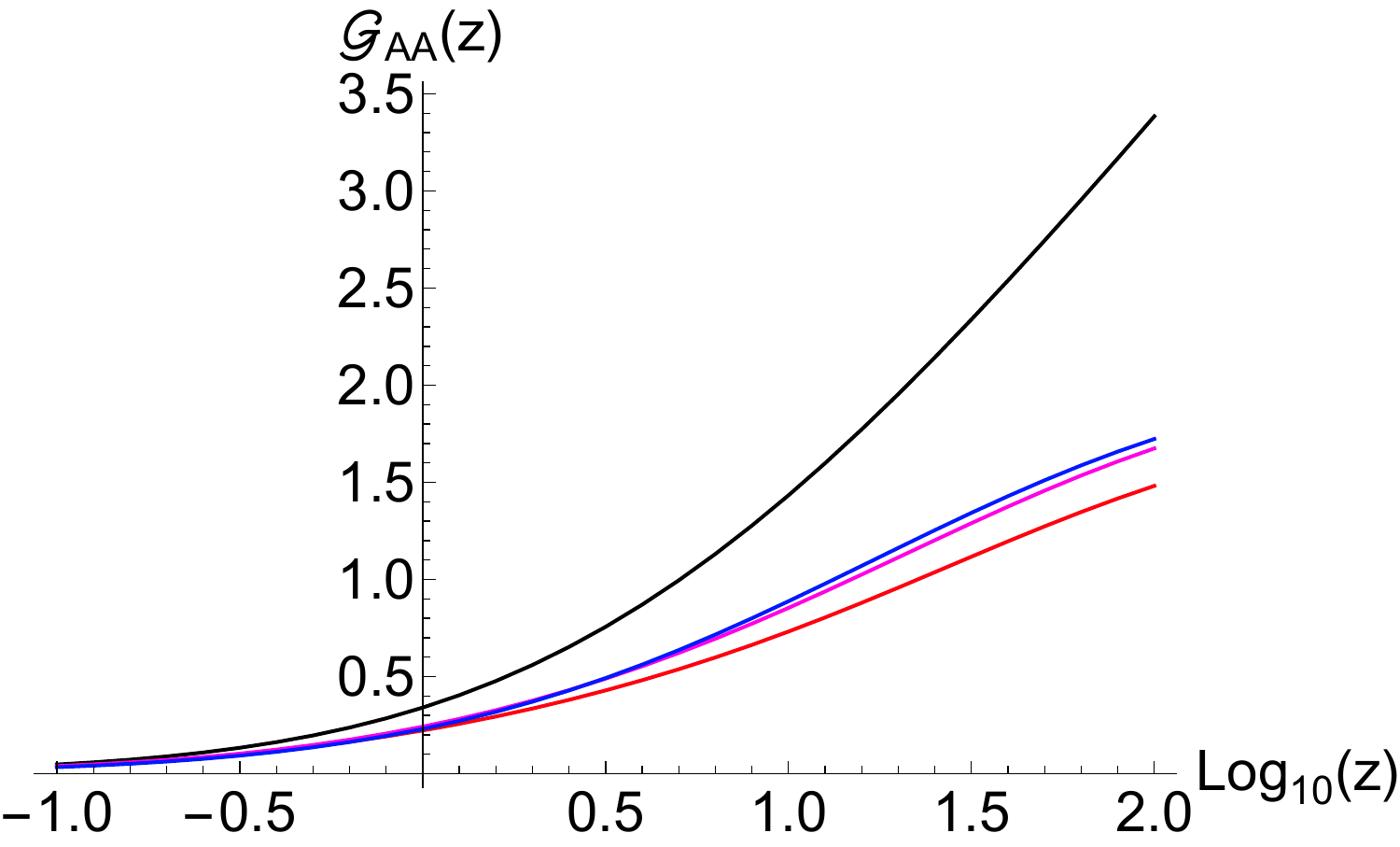} }%
	\caption{
	Linear-log plot of the coefficient $\mathcal{G}_{AA}$ that determines the AA correction [Eq.~(\ref{eq:AA-Eval2}), (\ref{GAADef}) and (\ref{AA--Ikernel})],
	as a function of the variable $z \equiv D \alpha_m \equiv f \, D \alpha$, for fixed $f$ and $\gbar^2$ 
	and variable $T/\omega_c$. Here $f = 0.1$ and $\gbar^2 = 0.5$. 
	The curves have $T/\omega_c = \{0.25,0.5,0.75\}$ (blue to red, top to bottom). 
	For comparison, the black curve is the $\alpha = 0$, $\gbar^2 = 0$ (static boson, Fermi-liquid electron) case [Eq.~(\ref{AA--FLKernel})].
	We set the prefactor $h/\pi \alpha_m = 1$ in Eq.~(\ref{GAADef}).
	}  	
	\label{fig:GAA_z_plot1}
\end{figure}

\begin{figure}
	\centering
	{\includegraphics[width=0.4\textwidth]{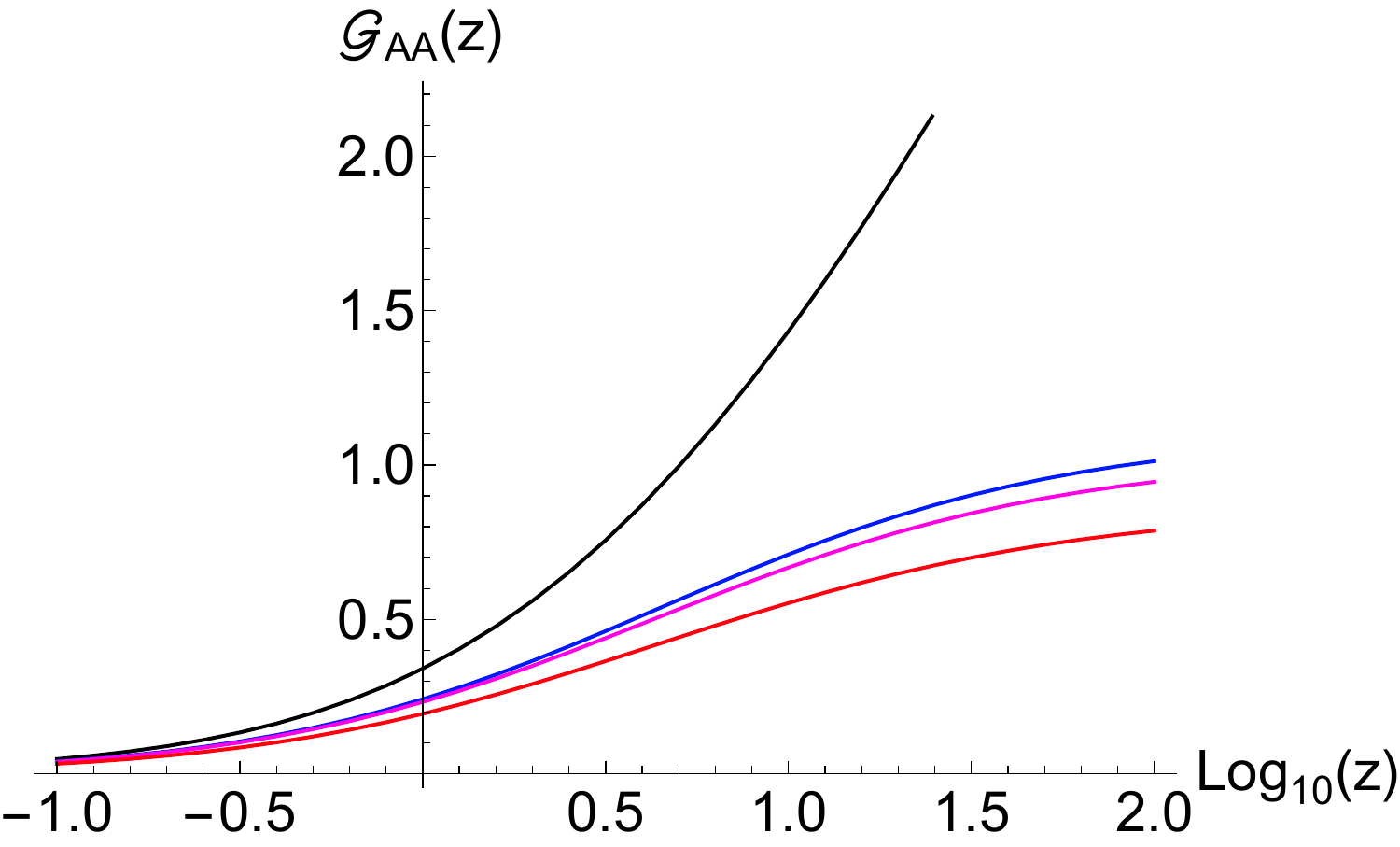} }%
	\caption{
	Same as Fig.~\ref{fig:GAA_z_plot1}, but for $f = 0.5$ and $\gbar^2 = 0.5$. 	
	The curves have $T/\omega_c = \{0.25,0.5,0.75\}$ (blue to red, top to bottom). 
	For comparison, the black curve is the $\alpha = 0$, $\gbar^2 = 0$ (static boson, Fermi-liquid electron) case [Eq.~(\ref{AA--FLKernel})].
	}  	
	\label{fig:GAA_z_plot2}
\end{figure}

\begin{figure}
	\centering
	{\includegraphics[width=0.4\textwidth]{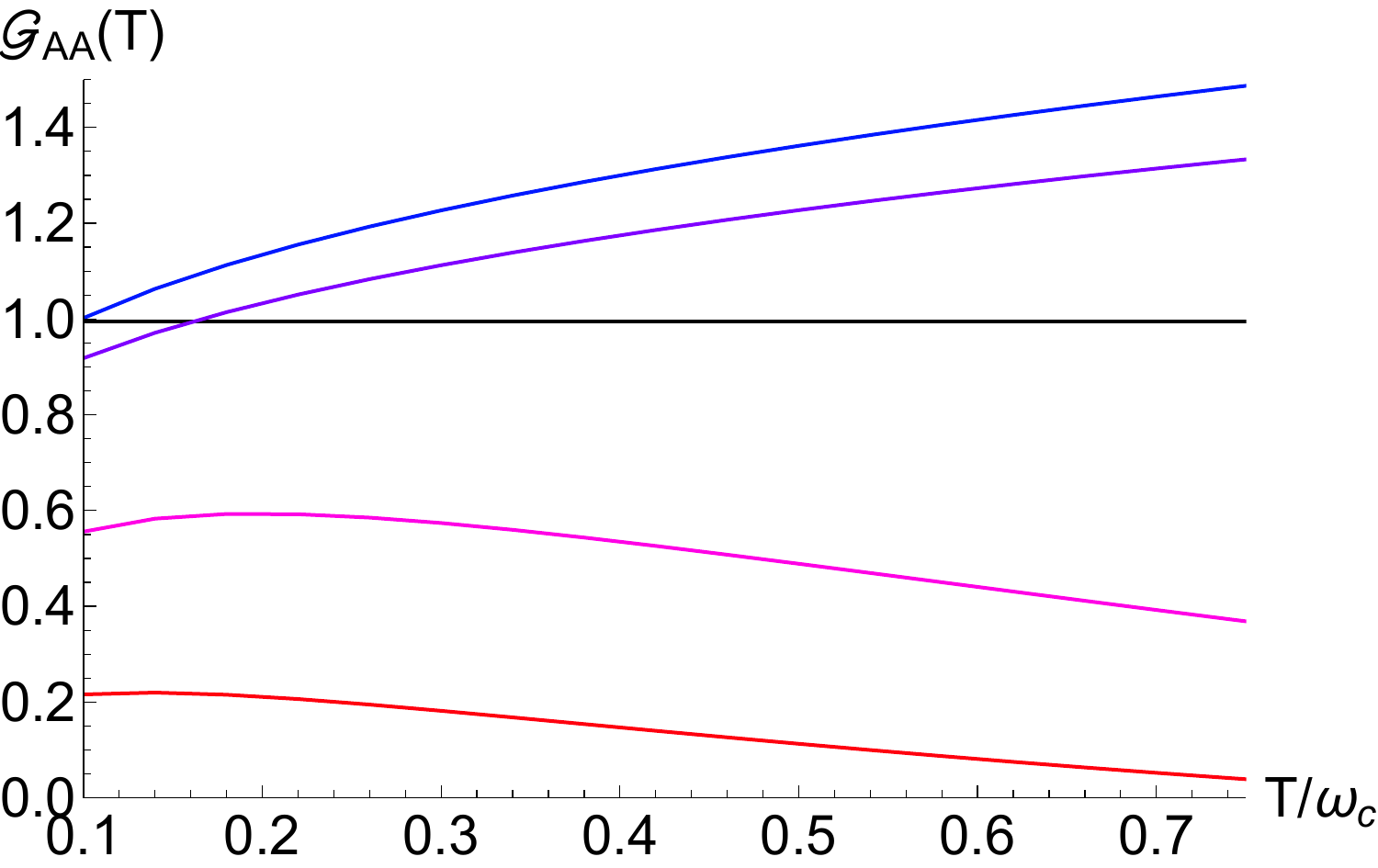} }%
	\caption{
	Plot of the coefficient $\mathcal{G}_{AA}$ that determines the AA correction [Eq.~(\ref{eq:AA-Eval2}), (\ref{GAADef}) and (\ref{AA--Ikernel})],
	as a function of the temperature $T$, for fixed $D \alpha_m$ and variable $D\alpha$, $\gbar^2$. 
	All curves have $D \alpha_m = 5$. 
	The horizontal black curve is the $\alpha = 0$, $\gbar^2 = 0$ (static boson, Fermi-liquid electron) case [Eq.~(\ref{AA--FLKernel})].
	The general results have $\{D \alpha, \gbar^2\} = \{\{0.01,0.1\},\{0.1,0.1\},\{0.1,0.5\},\{0.5,1\}\}$ (blue to red, top to bottom). 
	We set the prefactor $h/\pi \alpha_m = 1$ in Eq.~(\ref{GAADef}).
	}  	
	\label{fig:GAA_T_plot}
\end{figure}

So far we have ignored the effects of dynamical screening in the bosonic propagator. 
The latter becomes important at very low temperatures, or to determine the frequency-dependent ac conductivity 
at zero temperature. 
The bosonic propagator 
in Eq.~(\ref{eq:DR_dyn_scn}) can be approximated as 
\begin{equation}
	D_{\sfscr}^R(\xi,\textbf{k})
	\simeq
	-
	\frac{1}{2}
	\frac{
	\textbf{k}^2   - i \fraka
	}{
	\textbf{k}^4
	+
	(
	\mb^2
	-
	i\fraka
	)
	\textbf{k}^2
	-
	i \frakb
	},
\end{equation}
where we have defined
\begin{eqnarray}
	\fraka(\xi)
	&=&
	h \lambda  \xi 
	\left(
	1 + \gbar^2\ln \frac{\omega_c}{|\xi|}
	\right),
\\
	\frakb(\xi)
	&=&
	b \xi
	,
	\quad
	b = 
	\frac{
	h^2 g^2 \lambda
	}{
	\pi N
	}. 
\end{eqnarray}
The AA conductivity correction can be written as
\begin{equation}
\label{eq:sigma_AA1}
\begin{aligned}
	\sigma_{\sfAA,\sfdc}
	&=
	\frac{2
	e^2  \lambda (2h)^2  g^2 N}{8\pi^3}
	\int_{0}^{\infty}
	d\xi
	\,
	\tilde{F}(\xi)
	\,
	\calM(\fraka ,\frakb,\mb^2), 
\end{aligned}
\end{equation}
where the kernel
\begin{equation}
\label{eq:AA_kernel_M}
\begin{aligned}
&
	\calM(\fraka,\frakb,\mb^2)
	\\
	&=
	\im
	\int_0^{\infty} dx
	\;
	\frac{x}{
	\left(
	x - i\fraka 
	\right)^2
	[
	x^2
	+
	(
	\mb^2
	-
	i\fraka
	) 
	x
	-
	i \frakb
	]
	}.
\end{aligned}
\end{equation}
The $x$-integral in $\calM(\fraka,\frakb,\mb^2)$ can be done exactly, 
with the detailed expression given in Appendix \ref{app:AA_integral}. 
To make further analytical progress, we approximate the thermal factor as 
\begin{equation}
	\tilde{F}(\xi)
	\simeq
	\frac{\xi}{3T}
	\theta(3T - \xi)
	+
	\theta(\xi - 3T),
\quad
\xi > 0,
\end{equation}
where $\theta(x)$ is the Heaviside step function. 
We 
assume a large cutoff $\omega_c$ such that we can approximate
\begin{equation}
\label{eq:AA_approx_fraka}
	\fraka(\xi)	
	\simeq
	h \lambda \xi  
	\left(
	1
	+
	\gbar^2
	\ln 
	\frac{\omega_c}{T}
	\right)
	\equiv
	a \xi.
\end{equation}
With above approximations, the frequency integral can be performed exactly and we obtain
\begin{equation}
\label{eq:sigma_AA2}
	\sigma_{\sfAA,\sfdc}(T)
	\simeq
	\frac{2
	e^2  \lambda (2h)^2  g^2 N}{8\pi^3}
	\calI_2(a,b,\mb^2,T),
\end{equation}
where the full expression of kernel $\calI_2$ is given by Eq.~(\ref{eqApp:kernel_I}) in Appendix \ref{app:AA_integral}.  

The low-temperature behavior ($T \ll \gammael$) of the kernel $\calI_2$ is governed by the relative strength between the thermal mass 
$\mb^2$ and the dynamical screening coupling $ \sim g^2/N$.  
Specifically, we have 
\begin{equation}
\label{eq:AA_kernelI}
	\calI_2(a,b,\mb^2,T)
	\simeq
	\begin{cases}
	\dfrac{\pi}{24 a^2 T}
	,
	\quad
	& T \gg g^2/N
	,
	\\\\
	\dfrac{1}{4b}
	\ln^2\left(\dfrac{a^2 T}{b}\right)
	,
	&
	T \ll g^2/N 
	.
\end{cases}
\end{equation}
Hence, at low temperatures the AA correction acquires the following simple asymptotic form 
\begin{equation}
\label{eq:sigma_AA3}
	\sigma_{\sfAA,\sfdc}
	\simeq
	\dfrac{
	N^2 e^2
	}{
	4\pi^2
	}
	\ln^2 
	\left(
	\dfrac{N\lambda}{4\pi}
	\gbar^2 t
	\right),
	\qquad
	T \ll g^2/N,
\end{equation}
where $t \equiv T/\gammael$, as in Eq.~(\ref{eq:resist_dc}). 

In the $T \gtrsim g^2/N$ regime, the quantum relaxational thermal mass is the major screening mechanism for the interaction,
leading to the AA correction quoted in Eqs.~(\ref{eq:resist_dc}) and (\ref{eq:AA-Eval2}). 
On the other hand, for $T \ll g^2/N$, thermal screening is no longer effective and dynamical screening becomes important. 
In both cases, the AA correction is positive, indicating its antilocalizing nature. 

Our low-temperature result in Eq.~(\ref{eq:sigma_AA3})
should be compared with the AA conductance correction in a disordered Fermi liquid subjected to SU($N$) ferromagnetic spin-spin interactions
\begin{equation}\label{eq:AA_FLN}
	\sigma^{\sfFL}_{\sfAA,\sfdc}(T)
	\propto
	\frac{N^2 e^2}{2 \pi^2}
	\ln\left(\frac{\gammael}{T}\right),
\end{equation}
valid to leading order in $N$. Eq.~(\ref{eq:AA_FLN}) is proportional to $N^2$ due to the exchange carrier-carrier scattering 
among fermions with different flavors; in this case, the AA correction diverges as log instead of squared-log 
\cite{AA5_N_flav_Finkel_Sci_05,disO_review_Kirkpatrick_94}.

\subsubsection{Zero temperature limit with finite external frequency}

At $T = 0$ with finite external frequency $\Omega$, we evaluate the AA correction by making use of Eqs.~(\ref{eq:sum_AA}) and (\ref{eq:AA_density_to_cond}). 
For small $\Omega$, we have 
\begin{equation}
\label{eq:sigma_AA1_zeroT}
\begin{aligned}
&\sigma_{\sfAA}(\Omega)
\simeq
\frac{2
e^2  \lambda (2h)^2  g^2}{8\pi^3}
\\
&\times
\int_{0}^{\infty}d\xi
\left[
\frac{2\xi}{\Omega}
\theta( \Omega/2  - \xi)
+
\theta( \xi - \Omega/2  )
\right]
\calM_0(\fraka ,\frakb)
\end{aligned}
\end{equation}
where $\theta(x)$ is the Heaviside step function and the kernel 
\begin{equation}
\calM_0(\fraka ,\frakb)
=
\calM(\fraka,\frakb,0)
=
\frac{1}{
4\frakb
}
R
\left(
\frac{\fraka^2}{\frakb}
\right)
,
\end{equation}
with 
\begin{equation}
\begin{aligned}
R(x)
&=
-\pi x
- 2 \ln(e^2 x)
\\
&+
2\im 
\left[
\frac{x - 3i}{\sqrt{1- 4i/x}}
\ln
\frac{
1 - \sqrt{1- 4i/x}
}{
1 + \sqrt{1- 4i/x}
}
\right].
\end{aligned}
\end{equation}
By making the same approximation as in Eq.~(\ref{eq:AA_approx_fraka}), we can perform the remaining frequency integral exactly, resulting in
\begin{equation}
\sigma_{\sfAA}(\Omega)
\simeq
\frac{2
e^2  \lambda (2h)^2  g^2 N}{8\pi^3}
\frac{1}{4}
\calR(a,b,\Omega),
\end{equation}
where 
\begin{equation}
\calR(a,b,\Omega)
=
\calR_1(a,b,\Omega)
+
\calR_2(a,b,\Omega)
\end{equation}
with
\begin{equation}
\label{eq:IntJ1}
\begin{aligned}
&\calR_1(a,b,\Omega)
=
\frac{2}{a^2 \Omega}
\int_0^{x_0}
dx
\,
R
\left(
x
\right)
\\
&=
-\frac{x_0}{a^2 \Omega}
\left[
4 + \pi x_0 + 4 \ln x_0
-
2x_0
\im 
\left(
\varphi
\ln
\frac{
1
-
\varphi
}{
1
+
\varphi
}
\right)
\right]
\end{aligned}
\end{equation}
and
\begin{equation}
\label{eq:IntR2}
\begin{aligned}
&\calR_2(a,b,\Omega)
=
\frac{1}{b}
\int_{x_0}^{\infty}
dx
\,
\frac{1}{x}
R
\left(
x
\right)
\\
&=
\frac{1}{b}
\begin{Bmatrix}
4 - \frac{\pi^2}{4}
+
\pi x_0
+
4\ln x_0
+
\ln^2 x_0
\\
+
4 x_0
\im
\left[
\varphi
\tanh^{-1} \varphi
\right]
\\
-
4 \re 
(\tanh^{-1} \varphi)^2
\end{Bmatrix}
.
\end{aligned}
\end{equation}
In the above expressions, we defined 
\begin{equation}
\varphi = \sqrt{
\frac{x_0 - 4i}{x_0}
}
,
\qquad
x_0 = \frac{a^2 \Omega}{2b}
.
\end{equation}
In the small $\Omega$ limit, we find 
\begin{equation}
\sigma_{\sfAA}(\Omega)
\sim 
N^2 \ln^2 \Omega.
\end{equation}

\section{Discussion and outlook}
\label{sec:discuss_conclu}

In this paper, we have investigated the combined effects of disorder and quantum-critical interactions 
on the dc electric transport by incorporating both the semiclassical contribution and the Altshuler-Aronov (AA) quantum correction. 
Through self-consistently solving the saddle-point equations in the large-$N$ limit at finite temperature, we uncovered the 
quantum relaxational nature of the bosons and a MFL self-energy for the fermions, signaling a breakdown of the quasiparticle picture.  
Even though the concept of quasiparticle is no longer well-defined, we argued that the hydrodynamic modes corresponding to the 
diffusive motion of electrons can remain quantum coherent. 
We demonstrated that the quantum interference among these hydrodynamic modes results in an antilocalizing AA correction, which depends on 
temperature $T$ as (i) $N/T$ 
when thermal screening for the interaction dominates, and (ii) $N^2 \ln^2 T$ when dynamical screening for the interaction becomes important
at very low temperatures. 
We found that the AA correction 
can give rise to the characteristic linear-$T$ resistivity expected for a strange metal. 
As temperature approaches zero, the resistivity is driven to zero by the singularity in the AA correction.      

We have also elucidated how the Ward identity for particle conservation can be satisfied by introducing a new interaction term in the 
MFL-FNLsM, which also includes the MFL fermion self-energy. By incorporating various vertex corrections, we showed in leading order of 
disorder and $1/N$ that our density response function fulfills the Ward identity at both the classical and quantum levels at $T = 0$. 
The verification is extended to finite $T$ at the semiclassical level in Appendix~\ref{app:no_lin_T}, 
while we further confirm the Ward identity for the quantum correction via numerics in Appendix~\ref{app:numerical_Ward}.

We comment on a few differences between the present work and some of the existing ones. 
First, instead of starting with the clean NFL fixed point \cite{NFL_SU_N_disO_Raghu_PRL_20}, 
the self-energies for the fermions and bosons are dictated by a set of finite-temperature saddle-point equations. 
These capture the effects of disorder, Yukawa interactions, and thermal screening. 
While a similar AA formula analogous to Eq.~(\ref{eq:AA_delta_sig_formal}) was employed in Ref.~\cite{Maslov_PRL_05}, the effects of the disorder smearing on the bosonic self-energy and thermal screening were not considered. 
Our work presents for the first time the effects of quantum-relaxational bosons on the AA corrections. 
Second, the randomness of our model enters through the impurity potential instead of the Yukawa coupling \cite{SYK_Patel_PRB_21,SYK_Patel_linearT_arxiv_22}. 
Our model is thus manifestly SU($N$) invariant for a fixed realization of disorder. 
As a consequence, the SU($N$) flavor polarization forms a slow hydrodynamic mode that survives on time and length scales larger than those set 
by the impurity scattering. These slow flavor degrees of freedom are responsible for the AA correction in 
Eq.~(\ref{eq:resist_dc}).

Our results are potentially applicable to physical systems with fermions coupled to magnetic fluctuations near a the QCP, 
although $N$ is practically $2$ in reality. 
For instance, the Ising nematic transition (ferromagnetic QCP) and linear-$T$ resistivity have indeed been observed in iron-based 
superconductors \cite{NFL_FeHTSC_Fisher_Science_16,NFL_FeHTSC_Shibauchi_AnnuRevCMP_14} 
(some heavy fermion compounds \cite{NFL_heavy_fermion_Lohneysen_RMP_07,NFL_heavy_fermion_Brando_Science_13}). 
Recently, linear-$T$ resistivity has also been reported in a non-superconducting iron-pnictide 
Ba(Fe$_{1/3}$Co$_{1/3}$Ni$_{1/3}$)$_{2}$As$_{2}$, which possesses a ferromagnetic QCP at zero temperature and magnetic 
field \cite{NFL_JP_BaFeCoNiAs_PhysComm_16}.  
Intriguingly, at zero field, the linear-$T$ resistivity demonstrates a downturn at low temperature, which is in 
qualitative agreement with the AA quantum correction discussed in this work, Eq.~(\ref{eq:resist_dc}). 
Our calculations are also potentially
relevant for Si metal-oxide-semiconductor field-effect transistors, in which formation of spin droplets was reported in the 
insulating phase \cite{disO_spin_droplet_MOSFETs_Reznikov_PRL_12,disO_spin_droplet_MOSFETs_Pudalov_PRB_16}.
In particular, linear-$T$ resistivity was reported in the low-temperature regime for samples with low carrier densities \cite{disO_spin_droplet_MOSFETs_Pudalov_PRB_16}. 
We note that linear-$T$ resistivity has been observed down to the lowest temperature available in some experiments, 
possibly indicating that thermal screening plays the crucial role until even lower temperatures are accessed.

We close by mentioning a couple of interesting avenues and open questions that warrant further investigations. 
(i) 
It is desirable to carry out an RG analysis of our MFL-FNLsM and clarify the stability of the interaction term. 
Specifically, a key question is whether there is \emph{multifractal enhancement} for the interaction \cite{Feigelman07,Feigelman10,Burmistrov12,Foster12,Foster14,Mayoh15}. 
Based on our quantum correction in Eq.~(\ref{eq:sigma_AA3}), we expect to have double-logarithmic terms in the RG equations. 
While double-logarithmic divergences are not common in conventional field theories \cite{DoubleLogRG_Peskin}, they have been reported in the 
studies of zero-bias anomaly in  disordered metals  \cite{NLsM5_Finkelshtein_83,AA1_Altshuler_JETP_79} and bilayer graphene \cite{DoubleLogRG_Nandkishore_PRB_10},
and can arise due to subtle variations in the infrared structure of the quantum-loop propagators. 
 A systematic RG analysis for the MFL-FNLsM, incorporating higher-loop contributions, might be conveniently performed with the background field method \cite{NLsM3_Matt_PRB_08}. 
(ii) 
It would be interesting to extend our MFL-FNLsM to other symmetry classes. 
In particular, the current work assumes weakly broken time-reversal symmetry, 
and neglects the contribution of Cooperons which give rise to the weak-localization correction 
\cite{disO_review_PALee_85,NLsM6_Kamenev_CUP_11,NLsM2_Matt_Yun_Ann_17,disO_Liao_Matt_dephasing_PRL_18,Dephasing_AAK_PhysC_82,Dephasing_Seth_PRB_20}.   
A crucial question is how would these corrections affect the electric transport and (especially) the BCS pairing 
instability 
\cite{NFL_Raghu_BCS_PRB_15,SYK_Chowdhury_Berg_AnnPhy_20,SYK_YBKim_arxiv_21,NFL_Mandal_BCS_PRB_16,SYK_e_ph_SC_Schmalian_PRB_19,SYK_quantum_dot_YWang_PRL_20,NFL_SC_Chubukov_I_PRB_20,NFL_SC_Chubukov_II_PRB_20,NFL_SC_Chubukov_III_PRB_20,NFL_SC_Chubukov_IV_PRB_21,NFL_SC_Chubukov_V_PRB_21,NFL_SC_Chubukov_VI_PRB_21}. 
(iii) 
Various mesoscopic effects, such as level statistics and zero-bias anomaly \cite{NLsM6_Kamenev_CUP_11,NLsM9_Burmistrov_PRB_18}, can in principle be derived using the MFL-FNLsM. 
(iv)
While the concept of a quasiparticle is not well-defined in our system due to the MFL self-energy, the kinetic equation governing the distribution functions can still be 
formally applied \cite{NLsM6_Kamenev_CUP_11,NEQM_PALee_PRB_07,NEQM_PALee_PRB_95}. 
Determining the nonequilibrium distribution allows one to go beyond linear response theory and study phenomena such as shot noise 
\cite{NLsM6_Kamenev_CUP_11,NEQM_Yuval_PRB_01} and transport due to out-of-equilibrium bosons \cite{NEQM_Levchenko_AnnOfPhys_20}.\\

\section{Acknowledgment}

We thank Alexander Altland, Alex Levchenko, Patrick Lee, Subir Sachdev, and especially Aavishkar A.\ Patel for useful discussions. 
This work was supported by the Welch Foundation Grant No.~C-1809 (T.C.W.\ and M.S.F.), 
and by the Simons Foundation ``Ultra-Quantum Matter'' Research Collaboration (Y.L.).

\appendix

\begin{widetext}

\section{Evaluation of the thermal mass $\mb^2$}
\label{app:thermal_mass}

In the following, we derive the thermal mass result shown in Eq.~(\ref{eq:mb_saddle_pt}). 
Consider the saddle-point equation
\begin{equation}
	\mb^2
	=
	2i\lamphi
	\int\limits_{\Omega,\textbf{q}}
	D^K_{\sfsp}(\Omega,\textbf{q})
	=
	2 \lamphi
	\int\limits_{\Omega,\textbf{q}}
	\frac{
	\alpha \Omega
	}{
	(\textbf{q}^2 + \mb^2)^2 + (\alpha \Omega)^2
	}
	\coth\left(\frac{\Omega}{2T}\right).
\end{equation}
Since the model is tuned to a QCP, we have to subtract off the zero-temperature contribution such that
\begin{align}\label{eq:mb2_sub}
	\mb^2
	\rightarrow&\,
	\frac{\lamphi}{2 \pi^2}
	\int_{0}^{\infty} 
	d \Omega
	\int_0^{\infty} 
	d x
	\left[
		\frac{
		\alpha \Omega
		}{
		(x + \mb^2)^2 + (\alpha \Omega)^2
		}
		\coth\left(\frac{\Omega}{2T}\right)
	-
		\frac{
		\alpha \Omega
		}{
		x^2 + (\alpha \Omega)^2
		}
	\right]
\nonumber\\
	=&\,
	\frac{\lamphi}{2 \pi^2}
	\int_{0}^{\omega_c} 
	d \Omega
	\left\{
		\left[
			\frac{\pi}{2}
			-
			\tan^{-1}
			\left(
			\frac{\mb^2}{\alpha \Omega}
			\right)
		\right]
		\coth\left(\frac{\Omega}{2T}\right)
		-
		\frac{\pi}{2}
	\right\},
\end{align}
which ensures that $\mb^2(T = 0) = 0$. 
On the second line in Eq.~(\ref{eq:mb2_sub}), we have restricted the frequency
integral below the UV cutoff $\omega_c$; this integral is still logarithmically 
divergent. 
We isolate the divergence by writing  
\begin{align}\label{eq:mb2_sub-eval1}
	\mb^2
	=&\,
	\frac{\lamphi}{2 \pi^2}
\left\{
	(2 T)
	\int_{0}^{\infty} 
	d y
	\left[
		\coth(y)
		-
		1
	\right]
	\left[
		\frac{\pi}{2}
		-
		\tan^{-1}
		\left(
		\frac{\mb^2}{2 \alpha T y}
		\right)
	\right]	
	-
	\int_{0}^{\omega_c} 
	d \Omega
	\,
	\tan^{-1}
	\left(
	\frac{\mb^2}{\alpha \Omega}
	\right)
\right\}
\nonumber\\
	\simeq&\,
	\frac{\lamphi \, T}{2 \pi^2}
\left[
	I\left(\frac{\alpha_m}{2 \alpha}\right)
	-
	\left(\frac{\alpha_m}{\alpha}\right)
	\ln\left(\frac{e \alpha \omega_c}{\alpha_m T}\right)
\right],
\end{align}
valid to logarithmic accuracy, and where
\begin{align}
	I(z)
	\equiv
	2
	\int_{0}^{\infty} 
	d y
	\left[
		\coth(y)
		-
		1
	\right]
	\left[
		\frac{\pi}{2}
		-
		\tan^{-1}
		\left(
		\frac{z}{y}
		\right)
	\right].
\end{align}		
In Eq.~(\ref{eq:mb2_sub-eval1}), we have parametrized $\mb^2 \equiv \alpha_m \, T$.
The function $I(z)$ has the asymptotic behaviors
\begin{align}
	I(z \rightarrow 0) \simeq -\pi \ln(2 z), 
\qquad
	I(z \rightarrow \infty) \simeq \frac{\pi^2}{6 z}.
\end{align}
Therefore for $(\alpha_m/\alpha) \lesssim 1$, we can neglect the second term
in the square brackets on the right-hand side of Eq.~(\ref{eq:mb2_sub-eval1}), and approximate 
\begin{align}\label{eq:mb2_sub-eval2}
	\alpha_m 
	\simeq&\,
	\frac{\lamphi}{2 \pi^2}
	\,
	I\left(\frac{\alpha_m}{2 \alpha}\right).
\end{align}
The $T$-independent coefficient $\alpha_m$ can be obtained by solving Eq.~(\ref{eq:mb2_sub-eval2}).

\section{Derivation of the MFL-FNLsM}
\label{app:derivation_FNLsM}

In this appendix, we provide some technical details for the derivation of the FNLsM from Eq.~(\ref{eq:S_GDPiSig}). 
The FNLsM is derived by considering fluctuations around the saddle points discussed in Sec.~\ref{sec:saddle_pt_eqn}. 
By dropping the fluctuations of $\{\hD,\hat{\Pi},\tilX_1,\tilX_2\}$ and keeping only fluctuations in the fermionic sector, we write 
\begin{equation}
\label{eqApp:_fluc}
	\hG = \hG_{\sfsp} \otimes \hIdSUN+ \delta \hG,
	\qquad
	\hSig = \hSig_{\sfsp} \otimes\hIdSUN + \delta \hSig,
	\qquad
	\hq = \hq_{\sfsp} + \delta \hq,
\end{equation}
where the saddle points $\hG_{\sfsp}$, $\hSig_{\sfsp}$ and $\hqsp$ can be found in Sec.~\ref{sec:saddle_pt_eqn}. 
Here, $\delta \hG \rightarrow \delta G_{ij}^{ab}(x,x')$ and $\delta \hSig \rightarrow \delta \Sigma_{ij}^{ab}(x,x')$ are bilocal matrix 
fields containing flavor indices $i,j$, Keldysh indices $a,b$ and spacetime  $x = (t,\textbf{x})$. 
Meanwhile, $\delta \hq \rightarrow \delta q_{ij;t,t'}^{ab} (\textbf{x})$ is a matrix field that is bilocal in time but local in space. 
Plugging Eq.~(\ref{eqApp:_fluc}) into Eq.~(\ref{eq:S_GDPiSig}), we obtain the action for the fluctuating matrix fields 
\begin{equation}
\label{eqApp:delta_S}
\begin{aligned}
	\delta S
	=&\,
	-
	\Tr \ln[
	\hat{1}
	-
	\hG_{\sfsp} \, \delta \hSig
	+
	i \gammael 
	\hG_{\sfsp} \, \delta \hq
	-
	\hG_{\sfsp} \, \hat{\cal V } \otimes \hIdSUN
	]
\\
&\,
	+
	\int\limits_{x,x'}
	\delta \Sigma_{ij}^{ab}(x,x')
	\delta G_{ji}^{ba}(x',x)
	+
	i\frac{g^2}{2N}
	\int\limits_{x,x'}
	\delta G_{ii}^{b'a} (x',x)
	\gamma_{ab}^s
	D^{ss'}_{\sfsp}(x,x')
	\gamma_{a'b'}^{s'}
	\delta G_{jj}^{ba'}(x,x')
\\
&\,
	+
	\frac{\pi \nu_0 \gammael }{2}
	\int\limits_{\textbf{x}}
	\Tr
	\left[
	\hat{q}(\textbf{x})^2
	\right]
	,
\end{aligned}
\end{equation}
where we dropped unimportant constant terms and omitted terms linear in $\delta \hSig$ and $\delta \hG$ that
will be canceled with the corresponding terms in the $\Tr \ln[\ldots]$, owing to the saddle point conditions discussed in 
Sec.~\ref{sec:saddle_pt_eqn}. 
Here, $\int_x = \int_{t,\textbf{x}} = \int dt \, d^2 \textbf{x}$.

By expanding the $\Tr \ln[\ldots]$ in $\delta S$ [Eq.~(\ref{eqApp:delta_S})] up to quartic order, we have
\begin{equation}
\label{eqApp:delta_S_Trln}
\begin{aligned}
&
-
\Tr \ln[
\hat{1}
-
\hG_{\sfsp} \delta \hSig
+
i \gammael 
\hG_{\sfsp} \delta \hq
-
\hG_{\sfsp} \hat{\cal V } \otimes \hIdSUN
]
\\
&=
\frac{1}{2} 
\Tr [
-
\hG_{\sfsp} \delta \hSig
+
i \gammael 
\hG_{\sfsp} \delta \hq
-
\hG_{\sfsp} \hat{\cal V } \otimes \hIdSUN
]^2
-
\frac{1}{3}
\Tr [
-
\hG_{\sfsp} \delta \hSig
+
i \gammael 
\hG_{\sfsp} \delta \hq
-
\hG_{\sfsp} \hat{\cal V } \otimes\hIdSUN
]^3
\\
&+
\frac{1}{4}
\Tr [
-
\hG_{\sfsp} \delta \hSig
+
i \gammael 
\hG_{\sfsp} \delta \hq
-
\hG_{\sfsp} \hat{\cal V }  \otimes\hIdSUN
]^4
+
\ldots
\\
&=
-
\frac{\gammael^2}{2}
\Tr
\left[
\hG_{\sfsp} \delta \hq
\,
\hG_{\sfsp} \delta \hq
\right]
-
i\gammael 
\Tr
\left[
\hG_{\sfsp} \delta \hq
\,
\hG_{\sfsp} \hat{\cal V }
\right]
+
\frac{N}{2}
\Tr
\left[
\hG_{\sfsp} \hat{\cal V }
\,
\hG_{\sfsp} \hat{\cal V }
\right]
\\
&
\qquad
+
\frac{1}{2}
\Tr
\left[
\hG_{\sfsp} \delta \hSig
\,
\hG_{\sfsp} \delta \hSig
\right]
-
i \gammael
\Tr
\left[
\hG_{\sfsp} \delta \hSig
\,
\hG_{\sfsp} \delta \hq
\right]
-
\frac{\gammael^2}{2}
\Tr
\left[
\hG_{\sfsp} \delta \hSig
\,
\hG_{\sfsp} \delta \hq
\,
\hG_{\sfsp} \delta \hSig
\,
\hG_{\sfsp} \delta \hq
\right]
+
\ldots
\end{aligned}
\end{equation}
In the second line, we again dropped the linear terms due to the saddle point conditions. 
In the last step, we neglected the cubic terms which generate interaction terms that are less important. 
As we are going to show below, the terms in the first line give rise to 
$S_D$ [Eq.~(\ref{eq:S_D_q})], $S_{qV}$ [Eq.~(\ref{eq:S_qV_q}] and $S_{V}$ [Eq.~(\ref{eq:S_V})], 
respectively, 
while those in the second line give rise to $S_{\sfintI}$ [Eq.~(\ref{eq:S_intI_q})] and $S_{\sfintII}$ [Eq.~(\ref{eq:S_intII_q})].

\subsection{Derivation for $S_D$, $S_{qV}$ and $S_V$}

The diffusive part of the FNLsM, $S_D$ in Eq.~(\ref{eq:S_D_q}), can be derived following the standard gradient expansion. 
The Keldysh-space-diagonal components of $\delta \hq$ correspond to the massive fluctuations around the saddle point, 
while the off-diagonal components correspond to massless Goldstone modes that are important for the quantum transport. 
By ignoring the diagonal components, we can write 
\begin{equation}
\label{eqApp:GqGq}
\begin{aligned}
&
-
\frac{\gammael^2}{2}
\Tr
\left[
\hG_{\sfsp} \delta \hq
\,
\hG_{\sfsp} \delta \hq
\right]
\\
&=
-
\frac{\gammael^2}{2}
\int\limits_{\omega_1,\omega_2,\textbf{q}}
\delta q_{\omega_1,\omega_2;ij}^{ab}(-\textbf{q})
\left[
\int\limits_{\textbf{p}}
G^{aa}_{\sfsp}(\omega_1,\textbf{p})
G^{bb}_{\sfsp}(\omega_2,\textbf{p} + \textbf{q})
\right]
\delta 
q_{\omega_2,\omega_1;ji}^{ba}(\textbf{q})
\\
&\simeq
\frac{1}{2}
\int\limits_{\omega_1,\omega_2,\textbf{q}}
\delta q_{\omega_1,\omega_2;ij}^{ab}(-\textbf{q})
\left\lbrace 
-
\pi \nu_0 \gammael 
+
\frac{1}{\lambda}
\textbf{q}^2
-
ih 
\left\lbrace 
(\omega_2 - \Sigma^{bb}_{\sfsp,\omega_2}) [\htau^3]^{bb} +  
(\omega_1 - \Sigma^{aa}_{\sfsp,\omega_1})[\htau^3]^{aa}
\right\rbrace 
\right\rbrace 
\delta 
q_{\omega_2,\omega_1;ji}^{ba}(\textbf{q}),
\end{aligned}
\end{equation}
where $\lambda$ and $h$ are given in Eq.~(\ref{eq:def_lambda_h}). 
The first term exactly cancels with the last term in Eq.~(\ref{eqApp:delta_S}) when the latter is expanded to second order in $\delta \hq$, 
indicating the off-diagonal fluctuations are indeed soft modes of the system. By further making use of the constraint 
$\hq^2 = (\hqsp + \delta \hq)^2 = \hat{1}$
and going back to real space, we have 
\begin{equation}
\begin{aligned}
S_D
=
\frac{1}{2\lambda}
\int\limits_{\textbf{x}}
\Tr\left[
\nabla \hq
\cdot
\nabla \hq
\right]
+
2ih
\int\limits_{\textbf{x}}
\Tr\left[
(
\hat{\omega}
+ i\eta \htau^3
-\hSig_{\sfsp}
)
\hq
\right]
.
\end{aligned}
\end{equation}

On the other hand, the coupling term between $\delta \hq$ and the source field $V$ is
\begin{equation}
\begin{aligned}
-
i\gammael 
\Tr
\left[
\hG_{\sfsp} \delta \hq
\,
\hG_{\sfsp} \hat{\cal V }
\right]
&=
-
i\gammael 
\int\limits_{\omega_1,\omega_2\textbf{q}}
\delta q_{ii;\omega_1,\omega_2}^{ab}(\textbf{q})
\left[
\int\limits_{\textbf{p}}
G_{\sfsp}^{aa}(\omega_1,\textbf{p} + \textbf{q}) 
G_{\sfsp}^{bb}(\omega_2,\textbf{p}) 
\right]
{\cal V }^{ba}_{\omega_2,\omega_1}(-\textbf{q})
\\
&\simeq
-i \pi \nu_0
\int\limits_{\omega_1,\omega_2,\textbf{q}}
\delta q_{ii;\omega_1,\omega_2}^{ab}(\textbf{q})
{\cal V }^{ba}_{\omega_2,\omega_1}(-\textbf{q})
.
\end{aligned}
\end{equation}
In real space, this can be written as 
\begin{equation}
S_{qV}
=
-
2i h 
\int\limits_{t,\textbf{x}}
\Tr\left[ 
\hq_{t,t} (\textbf{x})
\left(
V^{\sfcl}(t,\textbf{x}) \hgamma^{\sfcl}
+
V^{\sfq}(t,\textbf{x}) \hgamma^{\sfq}
\right)
\right],
\end{equation}
which is Eq.~(\ref{eq:S_qV_q}). 

Lastly, for the source field term, one has to be careful when handling the integral involving retarded-retarded and advanced-advanced 
combinations of Green's function components:
\begin{equation}
\begin{aligned}
S_V
&=
\frac{N}{2}
\Tr
\left[
\hG_{\sfsp} \hat{\cal V }
\,
\hG_{\sfsp} \hat{\cal V }
\right]
\\
&=
\frac{N}{2}
\int\limits_{\omega,\Omega,\textbf{p},\textbf{q}}
G_{\sfsp}^{aa}(\omega + \Omega,\textbf{p}+\textbf{q})
{\cal V}^{ab}_{\omega + \Omega,\omega}(\textbf{q})
G_{\sfsp}^{bb}(\omega,\textbf{p}) 
{\cal V }^{ba}_{\omega,\omega + \Omega}(-\textbf{q})
\\
&\simeq
N
\int\limits_{\Omega,\textbf{q}}
\left\lbrace 
\int\limits_{\omega,\textbf{p}}
F_{\omega}
\left[
G_{\sfsp}^{R}(\omega,\textbf{p})
G_{\sfsp}^{R}(\omega,\textbf{p})
-
G_{\sfsp}^{A}(\omega,\textbf{p})
G_{\sfsp}^{A}(\omega,\textbf{p})
\right]
\right\rbrace 
V^{\sfcl}(\Omega,\textbf{q})
V^{\sfq}(-\Omega,-\textbf{q})
\\
&\simeq
-2i N\nu_0 
\int\limits_{t,\textbf{x}}
V^{\sfcl}(t,\textbf{x})
V^{\sfq}(t,\textbf{x}),
\end{aligned}
\end{equation}
which is Eq.~(\ref{eq:S_V}).

\begin{figure}[b!]
	\centering
	{\includegraphics[width=6cm]{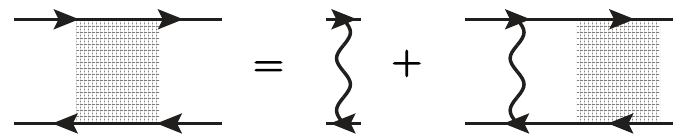} }%
	\caption{
	Diagrammatic depiction of the propagator of $\delta \hSig$. 
	Contraction of a pair of $\delta \hSig$ effectively inserts the interaction ladder. 
	}  	
	\label{fig:delta_Sig_propagator}
\end{figure}

\subsection{Derivation for $S_{\sfintI}$ and $S_{\sfintII}$}
\label{sec:app_delta_Sig_contraction}

We now show that $S_{\sfintI}$ and $S_{\sfintII}$ can be derived from the last two terms of Eq.~(\ref{eqApp:delta_S_Trln}). 
To obtain the interaction terms $S_{\sfintI}$ and $S_{\sfintII}$, we have to first derive the contraction rule for $\delta \hSig$. 
To do so,  we combine the second line of Eq.~(\ref{eqApp:delta_S}) and the first term in the 
last line of Eq.~(\ref{eqApp:delta_S_Trln}) such that
\begin{equation}
\begin{aligned}
\delta S_{\delta \Sigma}
&=
\frac{1}{2}
\Tr
\left[
\left(
\hG_{\sfsp} \delta \hSig
\right)^2
\right]
+
\int\limits_{x,x'}
\delta \Sigma_{ij}^{ab}(x,x')
\delta G_{ji}^{ba}(x',x)
+
\frac{1}{2}
\int\limits_{x,x'}
\delta G_{ij}^{ab}(x,x')
{\cal T}_{ji;lk}^{ba;dc} (x,x')
\delta G_{kl}^{cd}(x',x),
\end{aligned}
\end{equation}
where we defined the kernel
\begin{equation}
{\cal T}_{ji;lk}^{ba;dc} (x,x')
=
i\frac{g^2}{N}
\gamma_{bc}^s
D^{ss'}_{\sfsp}(x,x')
\gamma_{da}^{s'}
\delta_{ij} \delta_{kl}.
\end{equation}
Integrating out $\delta G$, we have
\begin{equation}
\begin{aligned}
\delta S_{\delta \Sigma}
&=
\frac{1}{2}
\Tr
\left[
\hG_{\sfsp} \delta \hSig_{ij}
\hG_{\sfsp} \delta \hSig_{ji}
\right]
-
\int\limits_{x_1,\ldots,x_4}
\delta \Sigma_{ij}^{ab}(x_1,x_2)
[{\cal T}^{-1}]_{ji;lk}^{ba;dc} (x_1,x_2)
\delta_{x_4,x_1} \delta_{x_2,x_3}
\delta \Sigma_{kl}^{cd}(x_3,x_4)
\\
&=
\frac{1}{2}
\int\limits_{x_1,\ldots,x_4}
\delta \Sigma_{ij}^{ab}(x_1,x_2)
\begin{Bmatrix}
-
[{\cal T}^{-1}]_{ji;lk}^{ba;dc} (x_1,x_2)
\delta_{x_4,x_1} \delta_{x_2,x_3}
+
G^{bc}_{\sfsp}(x_2,x_3)
G^{da}_{\sfsp}(x_4,x_1)
\delta_{jk}
\delta_{il}
\end{Bmatrix}
\delta \Sigma_{kl}^{cd}(x_3,x_4)
.
\end{aligned}
\end{equation}
The propagator for $\delta \hSig$ is thus
\begin{equation}
\label{eqApp:deltaSig_contraction}
\left\langle 
\delta \Sigma_{ij}^{ab}(x_1,x_2)
\delta \Sigma_{kl}^{cd}(x_3,x_4)
\right\rangle 
=
\bra{x_2,x_1;ji;ba}
\hat{\cal K}^{-1}
\ket{x_4,x_3;lk;dc}
=
\bra{x_2,x_1;ji;ba}
\frac{-1}{
\hat{1} - \hat{Y}
}
\hat{\cal T}
\ket{x_4,x_3;lk;dc}
\end{equation}
where the matrix elements of $\hat{Y}$ and $\hat{\cal T}$ are respectively
\begin{equation}
\bra{x_2,x_1;ji;ba}
\hat{Y}
\ket{x_4,x_3;lk;dc}
=
{\cal T}_{ji;lk}^{ba;dc} (x_1,x_2)
G^{bc}_{\sfsp}(x_2,x_3)
G^{da}_{\sfsp}(x_4,x_1)
\end{equation}
and
\begin{equation}
\bra{x_2,x_1;ji;ba}
\hat{\cal T}
\ket{x_4,x_3;lk;dc}
=
{\cal T}_{ji;lk}^{ba;dc} (x_1,x_2)
\delta_{x_1,x_4}
\delta_{x_2,x_3}.
\end{equation}
This implies that contracting a pair of $\delta \hSig$ amounts to inserting an interaction ladder, 
as illustrated diagrammatically in Fig.~\ref{fig:delta_Sig_propagator}.
In practice, however, we only keep the first term in the series since higher-order terms are parametrically suppressed by $1/\gammael$, 
as we assumed impurity scattering is the dominating scattering mechanism.   

We are now ready to derive $S_{\sfint}$ and $S_{\sfintII}$. 
The interaction among $\delta \hq$ arises from the last two terms in the last line of Eq.~(\ref{eqApp:delta_S_Trln}), which read
\begin{equation}
\delta S_{\sfint}
=
-
i \gammael
\Tr
\left[
\hG_{\sfsp} \delta \hSig
\,
\hG_{\sfsp} \delta \hq
\right]
-
\frac{\gammael^2}{2}
\Tr
\left[
\hG_{\sfsp} \delta \hSig
\,
\hG_{\sfsp} \delta \hq
\,
\hG_{\sfsp} \delta \hSig
\,
\hG_{\sfsp} \delta \hq
\right].
\end{equation}
Consider the expansion
\begin{equation}
\label{eqApp:expSint}
\begin{aligned}
\left\langle 
\e^{-\delta S_{\sfint}}
\right\rangle_{\delta \Sigma} 
&=
\left\langle 
1
-\delta S_{\sfint}
+
\frac{1}{2!}
(-\delta S_{\sfint})^2
+
\ldots
\right\rangle_{\delta \Sigma} 
\\
&=
\left\langle 
1
+
\frac{\gammael^2}{2}
\Tr
\left[
\hG_{\sfsp} \delta \hSig
\,
\hG_{\sfsp} \delta \hq
\,
\hG_{\sfsp} \delta \hSig
\,
\hG_{\sfsp} \delta \hq
\right]
-
\frac{\gammael^2}{2!}
\left(
\Tr
\left[
\hG_{\sfsp} \delta \hSig
\,
\hG_{\sfsp} \delta \hq
\right]
\right)^2
+
\ldots
\right\rangle_{\delta \Sigma} 
\end{aligned}
\end{equation}
Consider the second term in Eq.~(\ref{eqApp:expSint})
\begin{equation}
\begin{aligned}
&
	\left\langle 
	\frac{\gammael^2}{2}
	\Tr
	\left[
	\hG_{\sfsp} \delta \hSig
	\,
	\hG_{\sfsp} \delta \hq
	\,
	\hG_{\sfsp} \delta \hSig
	\,
	\hG_{\sfsp} \delta \hq
	\right]
	\right\rangle_{\delta \Sigma} 
\\
&=
	\frac{\gammael^2}{2}
	\int_{}
	\left\langle 
	G_{\sfsp}^{aa}(x_1,x_2) 
	\delta \Sigma^{ab}_{ij}(x_2,x_3)
	\,
	G_{\sfsp}^{bb}(x_3,x_4)
	\delta q_{jk}^{bc}(x_4,x_5) 
	\delta_{\textbf{x}_4,\textbf{x}_5}
	\,
	G_{\sfsp}^{cc}(x_5,x_6)
	\delta \Sigma^{cd}_{kl}(x_6,x_7)
	\,
	G_{\sfsp}^{dd}(x_7,x_8)
	\delta q^{da}_{li}(x_8,x_1)
	\delta_{\textbf{x}_8,\textbf{x}_1}
	\right\rangle_{\delta \Sigma} 
\\
&\simeq
	-\frac{\gammael^2}{2}
	\int_{}
	G_{\sfsp}^{aa}(x_1,x_2) 
	\,
	G_{\sfsp}^{bb}(x_3,x_4)
	\delta q_{jk}^{bc}(x_4,x_5) 
	\delta_{\textbf{x}_4,\textbf{x}_5}
	\,
	G_{\sfsp}^{cc}(x_5,x_6)
	\,
	G_{\sfsp}^{dd}(x_7,x_8)
	\delta q^{da}_{li}(x_8,x_1)
	\delta_{\textbf{x}_8,\textbf{x}_1}
	\left[
	{\cal T}^{ba,dc}_{ji;lk}
	(x_2,x_3)
	\delta_{x_2,x_7}
	\delta_{x_3,x_6}
	\right]
\\
&=
	-\frac{i g^2 \gammael^2}{2N}
	\int_{}
	G_{\sfsp}^{aa}(x_1,x_2) 
	\,
	G_{\sfsp}^{bb}(x_3,x_4)
	\delta q_{jk}^{bc}(x_4,x_5) 
	\delta_{\textbf{x}_4,\textbf{x}_5}
	\,
	G_{\sfsp}^{cc}(x_5,x_3)
	\,
	G_{\sfsp}^{dd}(x_2,x_8)
	\delta q^{da}_{li}(x_8,x_1)
	\delta_{\textbf{x}_8,\textbf{x}_1}
\\
&\, 	\qquad\qquad 
	\times
	\left[
	\gamma^s_{bc}
	D^{ss'}_{\sfsp}(x_2,x_3)
	\gamma^{s'}_{da}
	\delta_{ij}
	\delta_{kl}
	\right]
\\
&=
	-\frac{i g^2 \gammael^2}{2N}
	\int_{}
	\left[
	G_{\sfsp}^{aa}(x_1,x_2) 
	G_{\sfsp}^{dd}(x_2,x_8)
	\delta q^{da}_{ji}(x_8,x_1)
	\delta_{\textbf{x}_8,\textbf{x}_1}
	\gamma^{s'}_{da}
	\right]
	\,
	D^{ss'}_{\sfsp}(x_2,x_3)
	\,
	\left[
	G_{\sfsp}^{bb}(x_3,x_4)
	G_{\sfsp}^{cc}(x_5,x_3)
	\delta q_{ij}^{bc}(x_4,x_5) 
	\delta_{\textbf{x}_4,\textbf{x}_5}
	\gamma^s_{bc}
	\right],
\end{aligned}
\end{equation}
where $\int$ integrates all $x_i$ appearing in the integrand. 
By switching to momentum space and performing integrals similar to those 
that appeared in Eq.~(\ref{eqApp:GqGq}), we have 
\begin{equation}
\label{eqApp:Sint1_trln}
\begin{aligned}
&
\left\langle 
\frac{\gammael^2}{2}
\Tr
\left[
\hG_{\sfsp} \delta \hSig
\,
\hG_{\sfsp} \delta \hq
\,
\hG_{\sfsp} \delta \hSig
\,
\hG_{\sfsp} \delta \hq
\right]
\right\rangle_{\delta \Sigma} 
\\
&\simeq
-
\Gamma_1
\int\limits_{\omega_{1,2,3,4},\textbf{q}}
\delta_{\omega_1+\omega_3,\omega_2+\omega_4}
\Tr
[
\hq_{ij;\omega_1,\omega_2}(-\textbf{q})
\hgamma^{s}_{\omega_2,\omega_1}
]
D^{ss'}_{\sfsp,\omega_2-\omega_1}(\textbf{q})
\Tr[
\hgamma^{s'}_{\omega_4,\omega_3}
 \hq_{ji;\omega_3,\omega_4}(\textbf{q})
],
\end{aligned}
\end{equation}
where $\Gamma_1$ is given in Eq.~(\ref{eq:def_Gamma_1_and_2}). 

For the third term in Eq.~(\ref{eqApp:expSint})
\begin{equation}
\begin{aligned}
&
	-
	\frac{\gammael^2}{2!}
	\left\langle 
	\left(
	\Tr
	\left[
	\hG_{\sfsp} \delta \hSig
	\,
	\hG_{\sfsp} \delta \hq
	\right]	
	\right)^2
	\right\rangle_{\delta \Sigma} 
\\
&	
	=
	-
	\frac{\gammael^2}{2}
	\int_{}
	\langle 
	\left[
	G_{\sfsp}^{aa}(x_1,x_2) 
	\delta \Sigma_{ij}^{ab}(x_2,x_3)
	\,
	G_{\sfsp}^{bb}(x_3,x_4)
	\delta q^{ba}_{ji}(x_4,x_1) 
	\delta_{\textbf{x}_4,\textbf{x}_1}
	\right]
\\
&
	\qquad \qquad
	\times
	\left[
	G_{\sfsp}^{cc}(x_5,x_6) 
	\delta \Sigma_{kl}^{cd}(x_6,x_7)
	\,
	G_{\sfsp}^{dd}(x_7,x_8)
	\delta q^{dc}_{lk}(x_8,x_{5}) 
	\delta_{\textbf{x}_8,\textbf{x}_{5}}
	\right]
	\rangle_{\delta \Sigma} 
\\
&
	\simeq
	\frac{\gammael^2}{2}
	\int_{}
	\left[
	G_{\sfsp}^{aa}(x_1,x_2) 
	G_{\sfsp}^{bb}(x_3,x_4)
	\delta q^{ba}_{ji}(x_4,x_1) 
	\delta_{\textbf{x}_4,\textbf{x}_1}
	\right]
\\
&
	\qquad \qquad
	\times
	\left[
	G_{\sfsp}^{cc}(x_5,x_6) 
	G_{\sfsp}^{dd}(x_7,x_8)
	\delta q^{dc}_{lk}(x_8,x_{5}) 
	\delta_{\textbf{x}_8,\textbf{x}_{5}}
	\right]
	\left[
	{\cal T}^{ba,dc}_{ji;lk}
	(x_2,x_3)
	\delta_{x_2,x_7}
	\delta_{x_3,x_6}
	\right]
\\
&
	=
	i\frac{\gammael^2 g^2}{2N}
	\int_{}
	\left[
	G_{\sfsp}^{aa}(x_1,x_2) 
	G_{\sfsp}^{bb}(x_3,x_4)
	\delta q^{ba}_{ii}(x_4,x_1) 
	\delta_{\textbf{x}_4,\textbf{x}_1}
	\right]
	\left[
	G_{\sfsp}^{cc}(x_5,x_3) 
	G_{\sfsp}^{dd}(x_2,x_8)
	\delta q^{dc}_{jj}(x_8,x_{5}) 
	\delta_{\textbf{x}_8,\textbf{x}_{5}}
	\right]
	\left[
	\gamma^s_{bc}
	D^{ss'}_{\sfsp}(x_2,x_3)
	\gamma^{s'}_{da}
	\right], 
\end{aligned}
\end{equation}
where $\int$ integrates all $x_i$ appearing in the integrand. 
By Fourier transforming the above expression and assuming small momentum transfer through the bosonic line, we have 
\begin{equation}
\label{eqApp:Sint2_trln}
\begin{aligned}
&
-
\frac{\gammael^2}{2!}
\left\langle 
\left(
\Tr
\left[
\hG_{\sfsp} \delta \hSig
\,
\hG_{\sfsp} \delta \hq
\right]
\right)^2
\right\rangle_{\delta \Sigma} 
\simeq
 \Gamma_2
\int\limits_{1-4,\textbf{k}}
\delta_{1+3,2+4}
\Tr
\bigg[
	\hq_{ii;1,2}(-\textbf{k}) 
	\,
	\hgamma^s_{2,3}
	\,
	\hq_{jj;3,4}(\textbf{k})
	\,
	\hgamma^{s'}_{4,1}
\bigg]
\left[
\int_{\textbf{k'}}
D^{ss'}_{\sfsp,1-4}(\textbf{k'})
\right]
,
\end{aligned}
\end{equation}
where $\Gamma_2$ is given in Eq.~(\ref{eq:def_Gamma_1_and_2}). 

Combining Eqs.~(\ref{eqApp:Sint1_trln}) and Eq.~(\ref{eqApp:Sint2_trln}), we have 
\begin{equation}
\left\langle
\e^{-\delta S_{\sfint}}
\right\rangle_{\delta \Sigma} 
\simeq
\e^{
-S_{\sfintI} - S_{\sfintII}
}
.
\end{equation}
Thus, we have recovered $S_{\sfintI}$ in Eq.~(\ref{eq:S_intI_q}) and $S_{\sfintII}$ in Eq.~(\ref{eq:S_intII_q}) of the main text. 

\section{Cancelation of MFL self-energy and boson vertex corrections in the semiclassical limit, $T > 0$}
\label{app:no_lin_T}

In this appendix, we demonstrate that the semiclassical conductivity is given in our MFL-FNLsM by 
the Drude result at finite temperature, despite the incorporation of the anomalous fermion 
MFL self-energy. This result arises via the cancelation with boson vertex corrections,
as exhibited at $T = 0$ in Sec.~\ref{sec:semi_classic}.

The density-density correlator is determined by Eqs.~(\ref{eq:dens_resp_exp}) and (\ref{eq:dens_resp_exp_dyn}).
Working to first order in $\gbar^2$, the dynamical part can be written as 
\begin{align}
	\pi^R_{\sfdyn}(\Omega,\textbf{q})
	\simeq&\,
	- 
	\frac{N i \nu_0 \Omega}{D \vex{q}^2 - i \Omega}
	+
	\delta\pi^R_{\sfdyn,A}(\Omega,\textbf{q})
	+
	\delta\pi^R_{\sfdyn,B}(\Omega,\textbf{q}).
\end{align}
The first correction arises from expanding the anomalous diffuson
[Eq.~(\ref{eq:free_W_propagator})]
to lowest-order in the MFL self-energy,
\begin{align}
	\delta\pi^R_{\sfdyn,A}(\Omega,\textbf{q})
	=
	\frac{N \pi \nu_0}{\left(D \vex{q}^2 - i \Omega\right)^2}
	\int\limits_{1}
	\left(
		F_{1 - \Omega}
		-
		F_1
	\right)
	\left[
		\Sigma^R_{\sfsp}(\omega_1)
		-
		\Sigma^A_{\sfsp}(\omega_1 - \Omega)
	\right].
\end{align}
The second correction incorporates the first vertex correction 
from $S_{\sfintII}$. 
Using Eq.~(\ref{sintii_quad}), we obtain
\begin{align}
	\delta\pi^R_{\sfdyn,B}(\Omega,\textbf{q})
	=&\,
	\pi \gbar^2
	\frac{N \pi \nu_0}{\left(D \vex{q}^2 - i \Omega\right)^2}
	\int\limits_{1,2}
	\left(F_{1 - \Omega} - F_1\right)
\left\{
\begin{aligned}
&\,
	\left(
		\frac{2}{F_{1 - 2}}
		+
		F_2
		+
		F_{2 - \Omega}
	\right)
	i
	\tan^{-1}\left(\frac{\alpha \, \omega_{12}}{\mb^2}\right)
\\&\,
+
	\left(
		F_2 - F_{2 - \Omega} 
	\right)
	\ln\left[\frac{q_{\sfmax}^2}{\sqrt{\mb^4 + \alpha^2 \omega_{12}^2}}\right]
\end{aligned}
\right\},
\end{align}
where $\omega_{12} \equiv \omega_1 - \omega_2$. 
Finally, employing the integral representation for the MFL self-energy 
[Eq.~(\ref{MFL-IntRep})], we find that 
\begin{align}
	\delta\pi^R_{\sfdyn,A}(\Omega,\textbf{q})
	+
	\delta\pi^R_{\sfdyn,B}(\Omega,\textbf{q})	
	=
	0,
\end{align}
consistent with the results of Ref.~\cite{SYK_Patel_linearT_arxiv_22}.

\section{
Expressions for the type-B diagrams
}
\label{app:typeB_diag}

In this appendix, we present the detailed expressions associated with the type-B diagrams shown in Fig.~\ref{fig:AA_WFR}. 
Let's write
\begin{equation}
\delta \pi^R_{\text{type B}}(\Omega,\textbf{q})
=
\sum_{i}
\delta \pi^R_{\text{type B},i}(\Omega,\textbf{q})
\end{equation}
where $i$ corresponds to the label of the subfigure. 
Specifically, 
\begin{equation}
\begin{aligned}
	\delta \pi^R_{\text{type B,a(i)}}(\Omega,\textbf{q})
	=
	\frac{C}{2i}
	\int\limits_{\omega,\xi,\textbf{k}}
&\,
	D_{\sfscr,\xi}^R(\textbf{k})
	(F_{\omega} - F_{\omega + \xi})
	\left(
	\frac{1}{F_{\xi}}
	+
	F_{\omega}
	\right)
	\Delta_{\omega+\xi,\omega-\Omega}^R(\textbf{k} + \textbf{q})
	\Delta^R_{\omega+\xi,\omega}(\textbf{k})
\\
&\,
	\times
	\left(
	\textbf{q}^2 - ih\lambda \gbar^2 \Omega \ln \frac{\omega_c}{|\omega|} 
	\right)
	[\Delta^R_{\sfFL,\Omega}(\textbf{q})]^2,
\end{aligned}
\end{equation}
\begin{equation}
\begin{aligned}
	\delta \pi^R_{\text{type B,a(ii)}}(\Omega,\textbf{q})
	=
	-\frac{C}{2i}
	\int\limits_{\omega,\xi,\textbf{k}}
&\,
	D_{\sfscr,\xi}^R(\textbf{k})
	(F_{\omega} - F_{\omega + \xi})
	\left(
	\frac{1}{F_{\xi}}
	+
	F_{\omega+ \Omega}
	\right)
	\Delta_{\omega+\xi + \Omega,\omega}^R(\textbf{k} + \textbf{q})
	\Delta^R_{\omega+\xi,\omega}(\textbf{k})
\\
&\,
	\times
	\left(
	\textbf{q}^2 - ih\lambda \gbar^2 \Omega \ln \frac{\omega_c}{|\omega + \xi|} 
	\right)
	[\Delta^R_{\sfFL,\Omega}(\textbf{q})]^2,
\end{aligned}
\end{equation}
\begin{equation}
\begin{aligned}
	\delta \pi^R_{\text{type B,b(i)}}(\Omega,\textbf{q})
	=
	\frac{C}{2i}
	\int\limits_{\omega,\xi,\textbf{k}}
&\,
	D_{\sfscr,\xi}^R(\textbf{k})
	(F_{\omega} - F_{\omega + \xi})
	\left(
	-\frac{1}{F_{\xi}}
	+
	F_{\omega + \xi -  \Omega}
	\right)
	\Delta_{\omega+\xi,\omega-\Omega}^R(\textbf{k} + \textbf{q})
	\Delta^R_{\omega+\xi,\omega}(\textbf{k})
\\
&\,
	\times
	\left(
	\textbf{q}^2 - ih\lambda \gbar^2 \Omega \ln \frac{\omega_c}{|\omega|} 
	\right)
	[\Delta^R_{\sfFL,\Omega}(\textbf{q})]^2,
\end{aligned}
\end{equation}
\begin{equation}
\begin{aligned}
	\delta \pi^R_{\text{type B,b(ii)}}(\Omega,\textbf{q})
	=
	-\frac{C}{2i}
	\int\limits_{\omega,\xi,\textbf{k}}
&\,
	D_{\sfscr,\xi}^R(\textbf{k})
	(F_{\omega} - F_{\omega + \xi})
	\left(
	-\frac{1}{F_{\xi}}
	+
	F_{\omega + \xi }
	\right)
	\Delta_{\omega+\xi+\Omega,\omega}^R(\textbf{k} + \textbf{q})
	\Delta^R_{\omega+\xi,\omega}(\textbf{k})
\\
&\,
	\times
	\left(
	\textbf{q}^2 - ih\lambda \gbar^2 \Omega \ln \frac{\omega_c}{|\omega + \xi|} 
	\right)
	[\Delta^R_{\sfFL,\Omega}(\textbf{q})]^2,
\end{aligned}
\end{equation}
\begin{equation}
\begin{aligned}
	\delta \pi^R_{\text{type B,c(i)}}(\Omega,\textbf{q})
	=
	-\frac{C}{2i}
	\int\limits_{\omega,\xi,\textbf{k}}
&\,
	D_{\sfscr,\xi}^R(\textbf{k})
	F_{\omega + \xi}
	(F_{\omega} - F_{\omega + \Omega})
	\Delta^R_{\omega+\xi,\omega}(\textbf{k})
	\Delta^R_{\omega+\Omega + \xi,\omega}(\textbf{k} + \textbf{q})
\\
&\,
	\times
	\left(
	\textbf{q}^2 - ih\lambda \gbar^2 \Omega \ln \frac{\omega_c}{|\omega + \xi|} 
	\right)
	[\Delta^R_{\sfFL,\Omega}(\textbf{q})]^2,
\end{aligned}
\end{equation}
\begin{equation}
\begin{aligned}
	\delta \pi^R_{\text{type B,c(ii)}}(\Omega,\textbf{q})
	=
	-\frac{C}{2i}
	\int\limits_{\omega,\xi,\textbf{k}}
&\,
	D_{\sfscr,\xi}^R(\textbf{k})
	F_{\omega + \Omega}
	(F_{\omega} - F_{\omega + \Omega})
	\Delta^R_{\omega + \Omega +\xi,\omega + \Omega}(\textbf{k})
	\Delta^R_{\omega+\Omega + \xi,\omega}(\textbf{k} + \textbf{q})
\\
&\,	\times
	\left(
	\textbf{q}^2 - ih\lambda \gbar^2 \Omega \ln \frac{\omega_c}{|\omega|} 
	\right)
	[\Delta^R_{\sfFL,\Omega}(\textbf{q})]^2,
\end{aligned}
\end{equation}
\begin{equation}
\begin{aligned}
	\delta \pi^R_{\text{type B,d(i)}}(\Omega,\textbf{q})
	=
	\frac{C}{2i}
	\int\limits_{\omega,\xi,\textbf{k}}
&\,
	D_{\sfscr,\xi}^R(\textbf{k})
	F_{\omega}
	(F_{\omega} - F_{\omega + \Omega})
	\Delta^R_{\omega,\omega-\xi}(\textbf{k})
	\Delta^R_{\omega+\Omega,\omega - \xi}(\textbf{k} + \textbf{q})
\\
&\,
	\times
	\left(
	\textbf{q}^2 - ih\lambda \gbar^2 \Omega \ln \frac{\omega_c}{|\omega|} 
	\right)
	[\Delta^R_{\sfFL,\Omega}(\textbf{q})]^2,
\end{aligned}
\end{equation}
\begin{equation}
\begin{aligned}
	\delta \pi^R_{\text{type B,d(ii)}}(\Omega,\textbf{q})
	=
	\frac{C}{2i}
	\int\limits_{\omega,\xi,\textbf{k}}
&\,
	D_{\sfscr,\xi}^R(\textbf{k})
	F_{\omega + \Omega - \xi}
	(F_{\omega} - F_{\omega + \Omega})
	\Delta^R_{\omega + \Omega,\omega + \Omega-\xi}(\textbf{k})
	\Delta^R_{\omega+\Omega,\omega - \xi}(\textbf{k} + \textbf{q})
\\
&\,
	\times
	\left(
	\textbf{q}^2 - ih\lambda \gbar^2 \Omega \ln \frac{\omega_c}{|\omega - \xi|} 
	\right)
	[\Delta^R_{\sfFL,\Omega}(\textbf{q})]^2,
\end{aligned}
\end{equation}
and
\begin{equation}
	\delta \pi^R_{\text{type B,c(iii)}}(\Omega,\textbf{q})
	+
	\delta \pi^R_{\text{type B,c(iv)}}(\Omega,\textbf{q})
	\simeq
	\delta \pi^R_{\text{type B,d(iii)}}(\Omega,\textbf{q})
	+
	\delta \pi^R_{\text{type B,d(iv)}}(\Omega,\textbf{q})
	\simeq
	0.
\end{equation}
At $T = 0$, by expanding to leading order of external frequency $\Omega$ and momentum $\textbf{q}$ and making use of the fact that the $\xi$ integral is dominated by the infrared regime at which $\xi \simeq \Omega$, one can cast $\delta \pi^R_{\text{type B}}(\Omega,\textbf{q})$ into the form of Eq.~(\ref{eq:AA_typeB}), up to logarithmic accuracy.

\begin{figure}[b!]
	\centering
	{\includegraphics[width=0.4\textwidth]{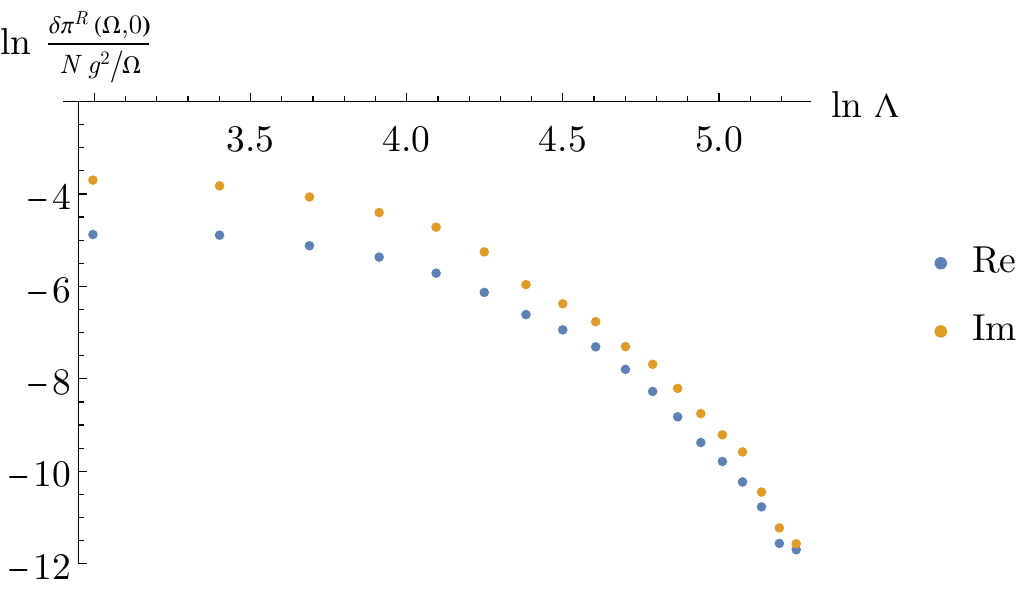} }%
	\caption{
	Log-log plot of the real part (blue) and imaginary part (orange) of the AA 
	density response correction $\delta \pi^R (\Omega,\textbf{0})$, 
	based on the numerical evaluation of the diagrams in Fig.~\ref{fig:AA_Hikami_Fock}--\ref{fig:AA_WFR} 
	when the external momentum $\textbf{q} = 0$. 
	We use the parameters $h = 1$, $\lambda = 0.1$,  $\gbar^2 = 0.25$, $h^2\lambda g^2 /\pi N = 10^{-3}$, $\Omega = 0.1$, $T = 0.005$, $\omega_c = \Lambda$. 
	As required by the U(1) Ward identity for electric charge conservation, $\delta \pi^R (\Omega,\textbf{0}) \rightarrow 0$
	as the UV cutoff $\Lambda \rightarrow \infty$. 
	}  	
	\label{fig:app_AA_sum}
\end{figure}

\section{Numerical verification of the Ward identity}
\label{app:numerical_Ward}

Here, we provide additional numerical evidence that the Ward identity is satisfied at $T \rightarrow 0$.  
We numerically evaluate the interaction corrections represented by the Feynman diagrams in Fig.~\ref{fig:AA_Hikami_Fock}--\ref{fig:AA_WFR}
when the external momentum $\textbf{q}$ is zero.
We perform the integrals by the method of Monte Carlo using \textit{Mathematica}. 
We employ $10^9$ for maximum number of sampling points, $0.15$ for ``BisectionDithering,'' and average over 100 configurations. 
We use the parameters $h = 1$, $\lambda = 0.1$,  $\gbar^2 = 0.25$, $h^2\lambda g^2/\pi N = 10^{-3}$, $\Omega = 0.1$, $T = 0.005$, and $\omega_c = \Lambda$. 
The frequency integral is cut in the infrared by $\Omega$. 
The dependence on the ultraviolet cutoff $\Lambda$ of the quantum correction $\delta \pi^R(\Omega,\textbf{0})$ when $\textbf{q} = 0$ is shown in Fig.~\ref{fig:app_AA_sum} 
in log scale. It is clear that for a sufficiently large cutoff $\Lambda$, the sum $\delta \pi^R(\Omega,\textbf{0})$ goes to zero, 
indicating that the Ward identity is satisfied, which is in line with our analytical analysis. 
Similar behaviors are observed in other parameter sets.

\section{AA integral kernels}
\label{app:AA_integral}

\subsection{Intermediate temperature, ignoring dynamical screening}

The general form of the kernel $\mathcal{I}_1$ that determines the finite-temperature
AA correction in Eqs.~(\ref{eq:AA-Eval2}) and (\ref{GAADef}) is given by 
\begin{align}
\label{AA--Ikernel}
\begin{aligned}
	\mathcal{I}_1
	=&\,
	\frac{i D \alpha_m}{2}
	\int_{-\infty}^\infty
	d y 
	\,
	\tanh\left(\frac{y}{3}\right)
	\,
	\left\{
		\frac{
			\Theta^2(y) 
			+ 
			(D \alpha)^2 \, \Xi^2(y) 
			+ 
			2 i (D \alpha) \, \Theta(y) \, \Xi(y) 
			\,
			\log\left[\frac{i}{D \alpha}\frac{\Theta(y)}{\Xi(y)}\right]
		}{
			\Theta(y) \left[\Theta(y) + i (D \alpha) \, \Xi(y)\right]^3
		}
	\right\},
\\
	\Theta(y)
	\equiv&\,
	y + \gbar^2 \left\{y \ln\left[\frac{\omega_c}{T \max(|y|,J)}\right] + i \frac{\pi}{2} \max(|y|,J)\right\},
\\
	\Xi(y)
	\equiv&\,
	i y - \frac{\alpha_m}{\alpha}.
\end{aligned}
\end{align}
The parameter $J = J(2 \alpha/\alpha_m)$ stabilizes the MFL self-energy at finite temperature,
and arises due to the finite thermal mass coefficient $\alpha_m$, Eqs.~(\ref{eq:MFL_J}) and (\ref{JAsym}).

For $\alpha = 0$ (static boson) and $\gbar^2 = 0$ (Fermi liquid, vanishing MFL self-energy),
Eq.~(\ref{AA--Ikernel}) reduces to 
Eq.~(\ref{AA--FLKernel}). 
Comparisons between this and the general case are exhibited in Figs.~\ref{fig:GAA_z_plot1}--\ref{fig:GAA_T_plot}.

\subsection{Incorporating dynamical screening}

The AA correction in Eq.~(\ref{eq:sigma_AA1}) of the main text involves the kernel $\calM$ defined in Eq.~(\ref{eq:AA_kernel_M}) as
\begin{equation}
\calM(\fraka,\frakb,\mb^2)
=
\im
\int_0^{\infty} dx
\;
\frac{x}{
\left(
x - i\fraka 
\right)^2
[
x^2
+
(
\mb^2
-
i\fraka
) 
x
-
i \frakb
]
}.
\end{equation}
This integral can be done exactly, the result is 
\begin{equation}
\label{eqApp:kernel_M}
\calM(\fraka,\frakb,\mb^2)
=
\frac{-1}{4
(-\mb^2 \fraka + \frakb)^2
}
\begin{Bmatrix}
\pi \fraka^2
+
4(-\mb^2 \fraka + \frakb)
+
2
\frakb 
\ln \dfrac{\fraka^2}{\frakb}
\\
-
2\im 
\left[
\dfrac{
\fraka^2(\mb^2-i\fraka)
+
\frakb(i\mb^2 - 3\fraka)
}{
\zeta
}
\ln
\dfrac{
\mb^2-i\fraka - \zeta
}{
\mb^2-i\fraka + \zeta
}
\right]
\end{Bmatrix},
\end{equation}
where
\begin{equation}
\zeta(\fraka,\frakb,\mb^2) \equiv
\sqrt{
(\mb^2 - i\fraka)^2
+
4i \frakb
}
.
\end{equation}
The remaining frequency integral in the AA correction [Eq.~(\ref{eq:sigma_AA2})] involves the kernel $\calI_2$ defined as
\begin{equation}
\label{eqApp:kernel_I}
{\calI_2}(a,b,\mb^2,T)
=
{\calI}_{2,1}(a,b,\mb^2,T)
+
{\calI}_{2,2}(a,b,\mb^2,T),
\end{equation}
where
\begin{eqnarray}
\begin{aligned}
{\calI}_{2,1}(a,b,\mb^2,T)
=&\,
\int_0^{3T} 
d\xi
\frac{\xi}{3T}
\calM(a\xi ,b\xi,\mb^2)
,
\\
{\calI}_{2,2}(a,b,\mb^2,T)
=&\,
\int_{3T}^{\infty}
d\xi
\,
\calM(a\xi ,b\xi, \mb^2)
.
\end{aligned}
\end{eqnarray}
The integrals can again be done exactly. 
The full expressions are
\begin{equation}
\label{eq:IntI1}
	{\calI}_{2,1}(a,b,\mb^2,T)
	=
	\frac{1}{24T (b - a \mb^2)^2}
	\begin{Bmatrix}
	\mb^4 \pi 
	+
	12a \mb^2 T
	-
	3T(4b + 3\pi a^2 T)
	-
	12b T
	\ln
	\dfrac{3a^2 T}{b}
\\
	+
	2\im 
	\left[
	(\mb^2 + 3ia T)
	\tilzeta
	\ln 
	\dfrac{
	\mb^2 - 3ia T - \tilzeta
	}{
	\mb^2 - 3ia T + \tilzeta
	}
	\right]
	\end{Bmatrix}
\end{equation}
and
\begin{equation}
\label{eq:IntI2}
	{\calI}_{2,2}(a,b,\mb^2,T)
	=
	\frac{1}{
	16(b - a \mb^2)^2
	}
	\begin{bmatrix}
	(16-\pi^2)b
	+
	4a(-4\mb^2 + 3\pi a T)
	+
	4 \ln \dfrac{3a^2 T}{b}
	\left(
	4b - 2a \mb^2
	+
	b \ln \dfrac{3a^2 T}{b}
	\right)
\\
	-8
	\re
	\left(
	a \tilzeta 
	\ln 
	\dfrac{
	\mb^2 - 3iaT - \tilzeta
	}{
	\mb^2 - 3iaT + \tilzeta
	}
	\right)
	-4
	\re
	\left(
	b
	\ln^2
	\dfrac{
	\mb^2 - 3iaT - \tilzeta
	}{
	\mb^2 - 3iaT + \tilzeta
	}
	\right)
	\end{bmatrix}
\end{equation}
\end{widetext}
with
\begin{equation}
\tilzeta 
= 
\sqrt{
12i b T
+
(\mb^2 - 3 ia T)^2
}
\end{equation}
Although the expressions of $\calI_{2,1}$ and $\calI_{2,2}$ are complicated, their asymptotic behaviors are quite simple. In particular, for $T \gg g^2/N$, we have 
\begin{equation}
{\calI}_{2,1}(a,b,\mb^2,T)
\simeq
\frac{\pi}{8 a^2 T},
\end{equation}
\begin{equation}
{\calI}_{2,2}(a,b,\mb^2,T)
\simeq
-
\frac{\pi}{12 a^2 T}
\end{equation}
On the other hand, in the very low temperature limit  $T \ll g^2/N$, we have 
\begin{equation}
{\calI}_{2,1}(a,b,\mb^2,T)
\simeq
-
\frac{1}{2b}
\ln 
\left(
\frac{a^2 T}{b}
\right),
\end{equation}
$\phantom{0}$
\begin{equation}
{\calI}_{2,2}(a,b,\mb^2,T)
\simeq 
\frac{1}{4b}
\ln^2 
\left(
\frac{a^2 T}{b}
\right)
.
\end{equation}
Combining the above expressions gives Eq.~(\ref{eq:AA_kernelI}).

\end{document}